%% file: PDF4LHC_Run2_recommendations.tex
\documentclass[11pt]{article}

\usepackage{graphicx,epsfig}
\usepackage{multicol,url,hyperref,multirow}
\usepackage{amsmath,amssymb,cite,color,hyperref}
\newcommand{\be}{\begin{equation}}
\newcommand{\ee}{\end{equation}}
\newcommand{\bea}{\begin{eqnarray}}
\newcommand{\eea}{\end{eqnarray}}
\newcommand{\bi}{\begin{itemize}}
\newcommand{\ei}{\end{itemize}}
\newcommand{\ben}{\begin{enumerate}}
\newcommand{\een}{\end{enumerate}}
\newcommand{\la}{\left\langle}
\newcommand{\ra}{\right\rangle}
\newcommand{\lc}{\left[}
\newcommand{\rc}{\right]}
\newcommand{\lp}{\left(}
\newcommand{\rp}{\right)}

\def\frac#1#2{{{#1}\over {#2}}}
\def\gsim{\mathrel{\rlap{\lower4pt\hbox{\hskip1pt$\sim$}}
    \raise1pt\hbox{$>$}}}         
\def\lsim{\mathrel{\rlap{\lower4pt\hbox{\hskip1pt$\sim$}}
    \raise1pt\hbox{$<$}}}         

\newcommand{\rep}{\mathrm{rep}}

\newcommand{\draft}[1]{}

\definecolor{grey}{rgb}{0.5,0.5,0.5}

\begin{document}
\begin{flushright}
  OUTP-15-17P \\
  SMU-HEP-15-12\\
  TIF-UNIMI-2015-14\\
  LCTS/2015-27\\
   CERN-PH-TH-2015-249 \\
\end{flushright}

\begin{center}
{\large\bf PDF4LHC recommendations for LHC Run II}
\vspace{0.3cm}

Jon Butterworth$^1$, Stefano Carrazza$^{2,4}$, Amanda Cooper-Sarkar$^3$, Albert
De Roeck$^{4,5}$, Jo\"el Feltesse$^6$, Stefano~Forte$^2$, Jun Gao$^7$,
Sasha Glazov$^{8}$, Joey Huston$^9$, Zahari Kassabov$^{2,10}$, Ronan McNulty$^{11}$, Andreas Morsch$^{4}$, 
Pavel Nadolsky$^{12}$, Voica Radescu$^{13}$, Juan~Rojo$^{14}$ and
Robert Thorne$^{1}$.

\vspace{.3cm}
       {\it \small
         ~$^1$Department of Physics and Astronomy, University College London, \\Gower Street, London WC1E 6BT, UK.\\
          ~$^2$ TIF Lab, Dipartimento di Fisica, Universit\`a di Milano and
         INFN, Sezione di Milano,\\ Via Celoria 16, I-20133 Milano, Italy\\
         ~$^3$ Particle Physics, Department of Physics,
University of Oxford,\\ 1 Keble Road, Oxford OX1 3NP, UK.\\
         ~$^4$PH Department, CERN, CH-1211 Geneva 23, Switzerland \\
         ~$^5$Antwerp University, B–2610 Wilrijk, Belgium \\
~$^6$ CEA, DSM/IRFU, CE-Saclay, Gif-sur-Yvette, France \\
 ~$^7$ High Energy Physics Division, Argonne National Laboratory,\\
Argonne, Illinois 60439, U.S.A.\\
~$^{8}$ Deutsches Elektronen-Synchrotron (DESY), \\
Notkestrasse 85, D-22607 Hamburg,
Germany.\\
  ~$^9$ Department of Physics and Astronomy, Michigan State University,\\
East Lansing, MI 48824 U.S.A.\\
~$^{10}$ Dipartimento di Fisica, Universit\`a di Torino and
INFN, Sezione di Torino,\\ Via Pietro Giuria 1, I-10125 Torino, Italy\\
~$^{11}$ School of Physics, University College Dublin Science Centre North,\\ UCD Belfeld, Dublin 4, Ireland \\
  ~$^{12}$ Department of Physics, Southern Methodist University,
Dallas, TX 75275-0181, U.S.A.\\
~$^{13}$ Physikalisches Institut, Universit\"at Heidelberg, Heidelberg, Germany. \\
~$^{14}$ Rudolf Peierls Centre for Theoretical Physics, 1 Keble Road,\\ University of Oxford, OX1 3NP Oxford, UK\\
}

\end{center}

\vspace{0.05cm}

\begin{center}
{\bf \large Abstract:}
\end{center}

We provide an updated recommendation for the usage of 
sets of parton distribution functions (PDFs)
and the assessment of PDF and PDF+$\alpha_s$ uncertainties
suitable for applications at the LHC Run II.
We review developments since the previous PDF4LHC recommendation, and
discuss and compare the
 new generation of PDFs, which  include
substantial information from experimental data from the Run I of the
LHC.
We then  propose a new prescription for the combination of a
suitable subset of the available PDF sets, which is presented
in terms of
 a single combined PDF set. We finally discuss tools which allow for the
 delivery of this combined set in terms of optimized sets of Hessian
 eigenvectors or Monte Carlo replicas, and their usage, and provide some
 examples of their application to LHC phenomenology.

 \vspace{0.2cm}
 \noindent
{\it This paper is dedicated to the memory
of Guido Altarelli (1941-2015), whose seminal work made
possible the quantitative study of parton distribution functions.}

\clearpage

\tableofcontents

\input{sec-introduction}

\input{sec-updates}
\input{sec-comparisons}
\input{sec-prescription}

\input{sec-frameworks}
\input{sec-recommendations}
\input{sec-future}

\input{PDF4LHC_Run2_recommendations.bbl}

\end{document}

%% file: sec-introduction.tex
\clearpage

\section{Introduction}
\label{sec:introduction}

In this first section we introduce the general context for the updated
PDF4LHC 2015 recommendations, and describe the layout of
this document.
Users whose main interest is the application of
the PDF4LHC15 recommendations to their specific analysis
can move directly to Sect.~\ref{sec:recommendations}.

\subsection{Parton distributions at the LHC}

The accurate determination of the parton distribution functions (PDFs)
of the proton
is crucial for precision predictions at the Large Hadron Collider
(LHC)~\cite{Rojo:2015acz,Forte:2010dt,Forte:2013wc}.
Almost all cross-sections of interest are now available at next-to-leading order (NLO), a rapidly
increasing number also at next-to-next-to-leading order (NNLO),
and even one, inclusive
Higgs production in gluon fusion, at NNNLO~\cite{Anastasiou:2015ema}.
The resulting improvements in theoretical uncertainties from
the inclusion of higher order matrix elements demand a
correspondingly improved control of  PDF uncertainties.

A number of groups have recently produced updates
of their PDFs fits~\cite{Alekhin:2013nda,Dulat:2015mca,Owens:2012bv,Jimenez-Delgado:2014twa,Abramowicz:2015mha,Harland-Lang:2014zoa,Ball:2014uwa} that have been compared
to LHC
data.
Even for PDFs based on
similar data
sets, there are still variations in both the ensuing central values
and
uncertainties.
This suggests that use
of the PDFs from one group alone might underestimate the true
uncertainties, and   a combination of
individual PDF sets is required for a robust uncertainty estimate in LHC
cross-sections.

\subsection{The PDF4LHC Working Group and the 2010 recommendations}

The PDF4LHC Working Group has been tasked with:
\begin{enumerate}
\item performing
    benchmark studies of PDFs and of predictions at the LHC, and
    \item making recommendations for a standard method of estimating PDF and PDF+$\alpha_s(m_Z^2)$ uncertainties at the LHC through a combination of the results from different
      individual groups.
\end{enumerate}
      This mandate has led to
 several benchmarking papers~\cite{Alekhin:2011sk,Ball:2012wy} and to the 2010
PDF4LHC recommendation~\cite{Botje:2011sn} which has undergone several
intermediate updates, with the last version available (along with a summary
of PDF4LHC activities) from the PDF4LHC Working Group website: 
\begin{center}
\url{http://www.hep.ucl.ac.uk/pdf4lhc/}. 
\end{center}
  
In 2010 the PDF4LHC Working Group carried out an exercise to which all PDF groups were invited to participate\cite{Alekhin:2011sk}.
Benchmark comparisons were made at NLO for LHC cross-section predictions at 7 TeV using {\tt MCFM}~\cite{Campbell:2002tg}
as a common framework using carefully prescribed input files.
The benchmark processes included $W/Z$ total cross-sections and rapidity distributions,
$t\bar{t}$ cross-sections and Higgs boson production through $gg$ fusion for masses of 
120, 180 and 240 GeV.
The PDFs used in this comparison
included ABKM/ABM09~\cite{Alekhin:2009ni}, CTEQ6.6/CT10~\cite{Nadolsky:2008zw,Lai:2010vv}, GJR08~\cite{Gluck:2007ck,Gluck:2008gs}, HERAPDF1.0~\cite{Aaron:2009aa}, MSTW2008~\cite{Martin:2009iq} and NNPDF2.0~\cite{Ball:2010de}.
The results were summarized in a PDF4LHC report~\cite{Alekhin:2011sk} and in the
LHC Higgs Cross Section Working Group Yellow Reports 1~\cite{Dittmaier:2011ti} and
2~\cite{Dittmaier:2012vm}.
In this study, each group used their native value of $\alpha_s(m_Z^2)$ along with its
corresponding uncertainty. 

This 2010 PDF4LHC
prescription~\cite{Botje:2011sn} for the estimation of combined
PDF+$\alpha_s(m_Z^2)$ uncertainties was based on PDFs from 
the three fitting groups performing a global analysis with a variable
flavor number scheme, namely CTEQ, MSTW and NNPDF.
Here,  ``global analysis'' is meant to signify that the widest
available  set  of data from a variety of experiments and processes
was used, 
including deep-inelastic scattering, gauge boson Drell-Yan production,
and inclusive jet production at hadron colliders. 

The recommendation at NLO was to use the envelope provided by the central values and PDF+$\alpha_s(m_Z^2)$ errors from CTEQ6.6, MSTW2008 and NNPDF2.0,
using each group's prescription for combining the PDF with the $\alpha_s$
uncertainties~\cite{Demartin:2010er,Lai:2010nw,Martin:2009bu}, and with 
the central value
given by the midpoint of the envelope.
By definition, the extent of the  envelope is determined by the
extreme PDFs at the upper and lower edges.
A drawback of this procedure
was that it assigned a higher weight to these outlier PDF sets than
what would be statistically correct
when all the individual input sets have equal prior likelihood.

In addition, as each PDF uses its own native value of $\alpha_s(m_Z^2)$ and its own PDF+$\alpha_s(m_Z^2)$ uncertainties about that 
central value, the resultant envelope inflates the impact of the
$\alpha_s(m_Z^2)$ uncertainty.
This very conservative prescription was
adopted partly because of ill-understood disagreements between the PDF
sets entering the combination (with other sets differing even more),
and it was considered suitable specifically for the search of the Higgs
boson at the LHC.
Furthermore, it was suggested that in
case of updates of the various PDFs set by the respective groups, the
most recent sets should always be used.

At the time of this first recommendation, only MSTW2008 (of the three global fitting groups) had produced PDFs at NNLO.
Since PDF errors are determined primarily by the
experimental errors in the data sets, as well as by the methodology
used in the corresponding PDF extraction (such as tolerance),
it was known that PDF uncertainties were relatively unchanged
when going from NLO to NNLO.
The 
NNLO recommendation was thus 
to use MSTW2008 NNLO as the central result,
and to take the same
percentage error on that NNLO prediction as was found at NLO using the
uncertainty prescription described above. 

\subsection{Intermediate updates}

A follow-up benchmarking study in 2012 was carried out with the  NNLO versions
of the most up-to-date PDF sets~~\cite{Ball:2012wy}.
By that time, CTEQ6.6 had been replaced by CT10~\cite{Lai:2010vv,Gao:2013xoa} and
NNPDF2.0 by NNPDF2.3~\cite{Ball:2012cx}, both at NNLO. In addition, HERAPDF1.0 had been replaced by HERAPDF1.5~\cite{CooperSarkar:2011aa}, and
ABM09 by ABM11~\cite{Alekhin:2012ig}.
In that 
study, a common value of $\alpha_s(m_Z^2)=0.118$ was used by all PDFs, 
with a variation between 0.117 and 0.119 
to account for the uncertainty on $\alpha_s(m_Z^2)$.

\begin{figure}[t]
  \centering
  \includegraphics[scale=0.38]{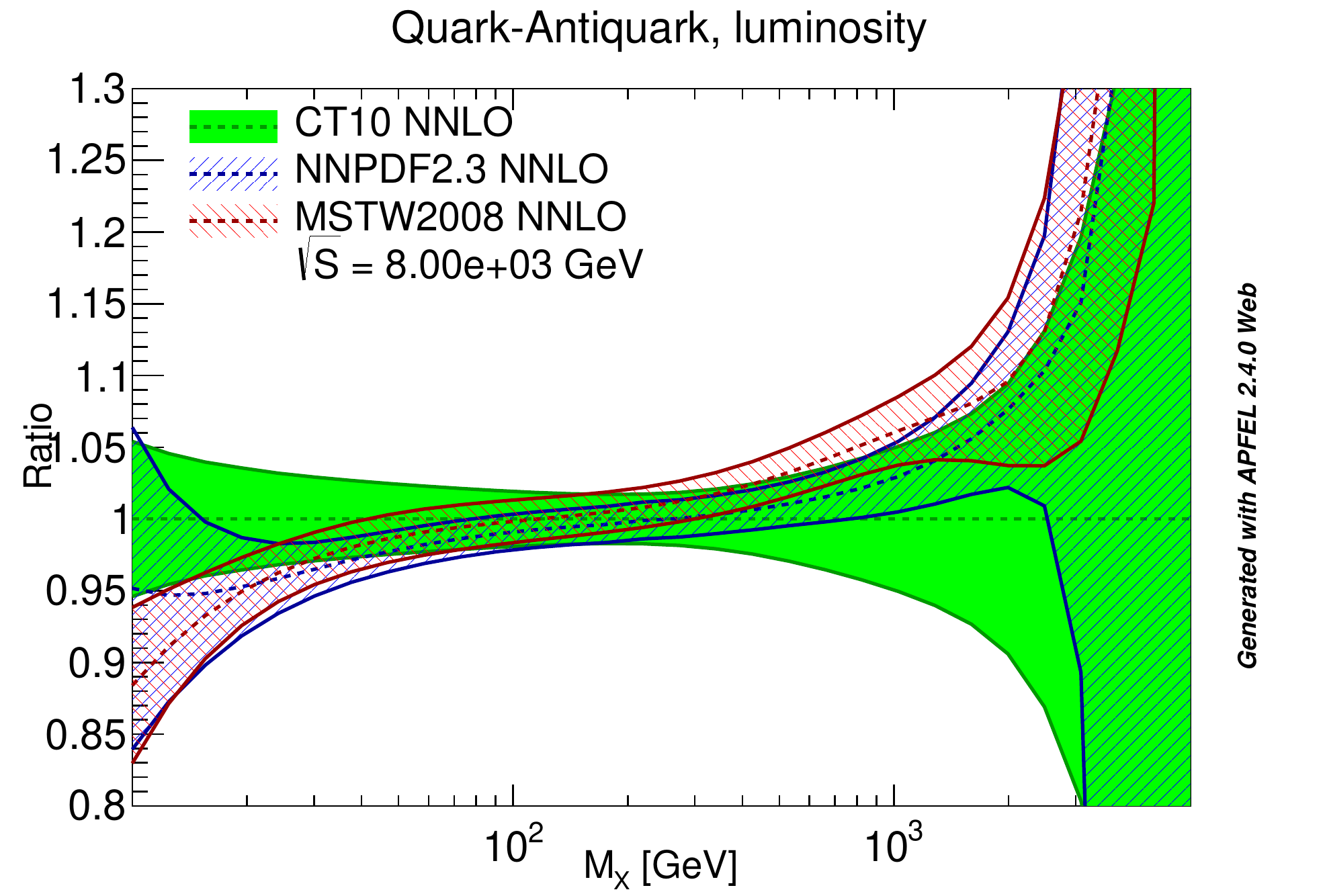}
  \includegraphics[scale=0.38]{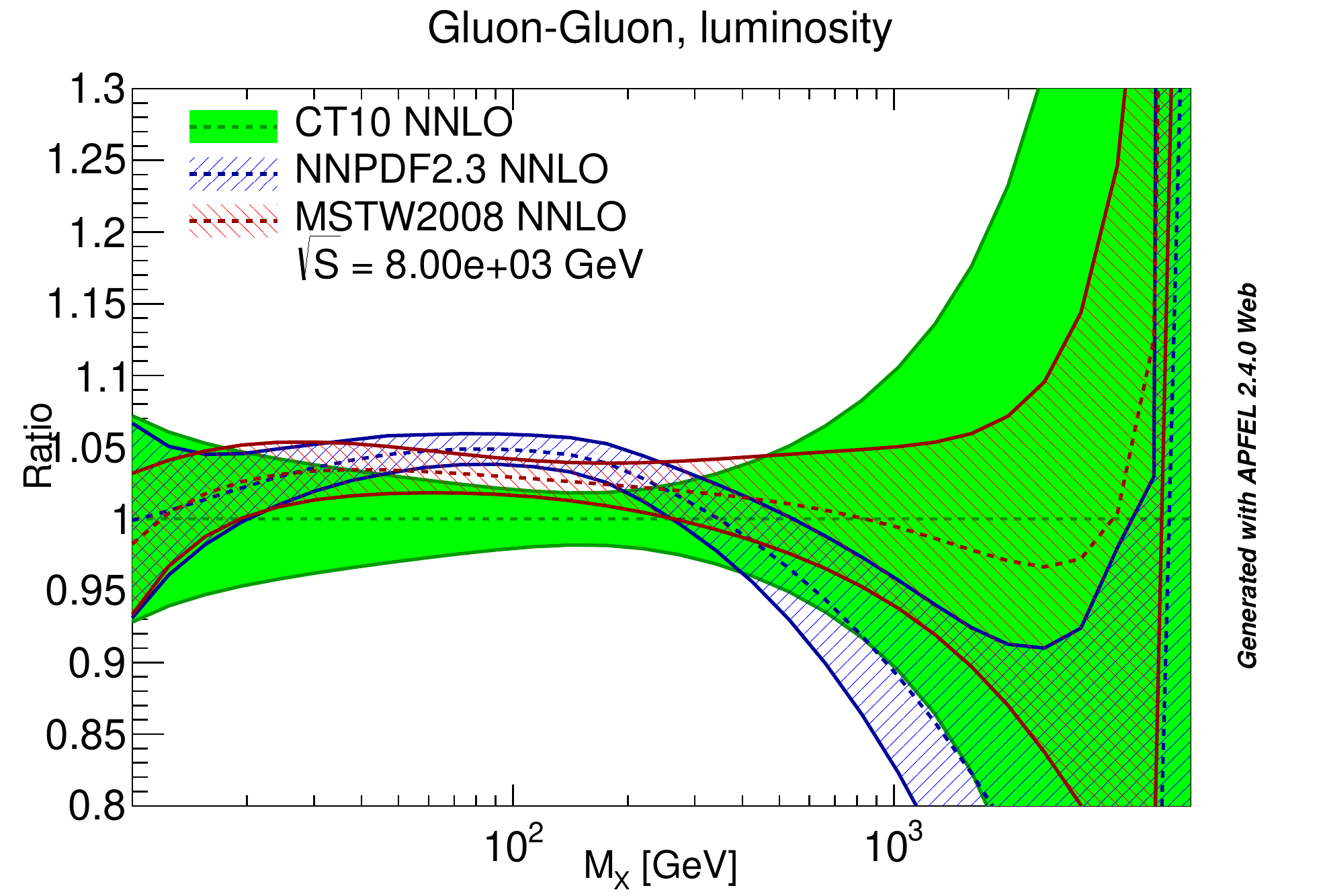}\\
  \caption{\small Comparison of the $q\bar{q}$ (left) and $gg$ (right)
  PDF luminosities at the LHC 8 TeV
 for
CT10, MSTW2008 and NNPDF2.3.
Results are shown normalized to the central value of CT10.
}  
\label{fig:CT10_MSTW08_NNPDF2.3}
\end{figure}

Data from the LHC Run I was then already available,
and detailed comparisons were made to data for inclusive jet and electroweak
boson production from ATLAS~\cite{Aad:2011fc,Aad:2011dm},
CMS~\cite{Chatrchyan:2012xt} and LHCb~\cite{Aaij:2012vn}.
Much better agreement was observed between the
updated versions of the three PDF sets used in the first study,
CT10, MSTW2008, and NNPDF2.3, 
for example for the quark-antiquark PDF luminosity in the mass region of the $W/Z$ bosons,
see Fig.~\ref{fig:CT10_MSTW08_NNPDF2.3}.
However some disagreements remained: Fig.~\ref{fig:CT10_MSTW08_NNPDF2.3} also shows
the gluon-gluon PDF luminosity for $M_X$ values in the
region of the Higgs boson mass.
In this case, the resulting envelope
 was more than twice
the size of the uncertainty band for any of the 
individual PDFs.

\begin{figure}[t]
  \centering
  \includegraphics[scale=.74]{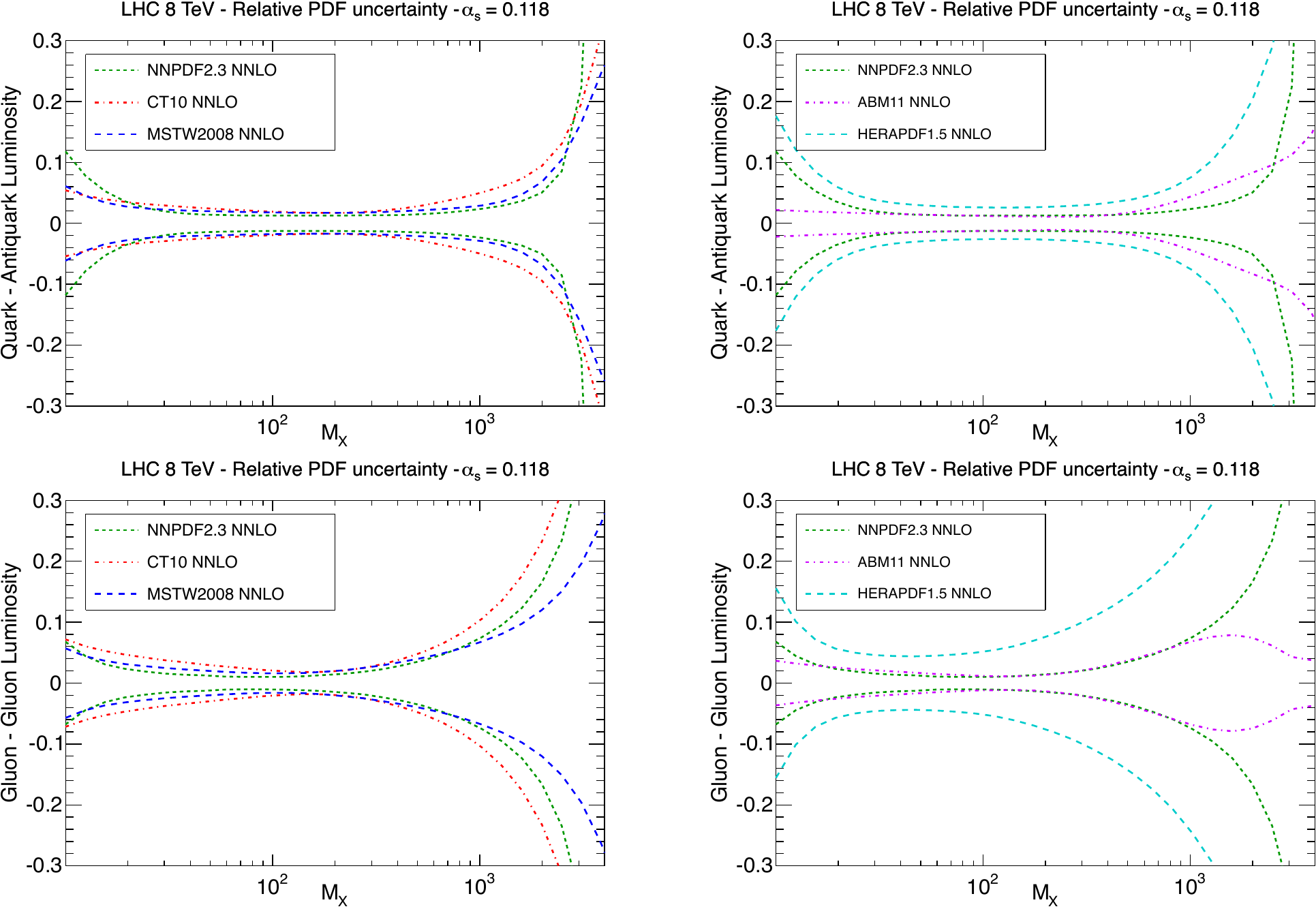}
  \caption{\small A comparison of PDF uncertainties in the $q\bar{q}$
    (upper plots) and $gg$ luminosity (lower plots)
at the LHC 8 TeV for the 
    ABM11, CT10, HERAPDF1.5,  MSTW2008 and NNPDF2.3 NNLO PDFs, for
  a common value of the strong coupling $\alpha_s(m_Z^2)=0.118$.
  PDF uncertainties correspond to 68\% CL intervals.
}  
\label{fig:gg_uncert}
\end{figure}

It is interesting to observe that 
the uncertainty bands for CT10, MSTW2008 and NNPDF2.3 are all reasonably 
similar, as shown in Fig.~\ref{fig:gg_uncert},  even though the
methodologies used to determine them
differ.
The uncertainties for ABM11 are smaller, 
despite using a more limited data set, and also 
including sources of uncertainties due to $\alpha_s$ and heavy-quark masses; we will come
back to this issue in Sect.~\ref{sec:currentPDFsets} below.
The uncertainties for HERAPDF1.5 were substantially larger, as expected from the
more limited data set used.
%

Based on the 2012 study, the 2010 PDF4LHC recommendation was
updated in 2013.
First, it was  recommended that the most up-to-date versions
of the  PDF sets from the three groups included  
in the previous recommendation be used, namely CT10, MSTW2008 and
NNPDF2.3. Furthermore, as all sets now had both NLO and NNLO sets, it
was recommended that the same procedure should be used both at NLO and
NNLO.
Finally, a somewhat simpler way of combining the PDF and
$\alpha_s$ uncertainties was suggested. Namely,
 the central value of $\alpha_s(m_Z^2)$  was fixed for all PDFs to be
 $\alpha_s(m_Z^2)=0.118$, obtained by 
rounding off  the then-current PDG world average 
$0.1184\pm0.0007$~\cite{Beringer:1900zz} (and near the preferred value
 of each group anyway).
An uncertainty range for $\alpha_s(m_Z^2)$ was taken to be $\pm0.002$ at the
90\% Confidence Level (CL) around the central value of 0.118.
This corresponds to a 68\% CL uncertainty of $\pm0.0012$, somewhat more conservative than the PDG estimate.

The total PDF$+\alpha_s(m_Z^2)$ uncertainty was then 
determined for each group  by adding in quadrature 
the PDF  uncertainty and $\alpha_s$ uncertainty, with the latter determined as 
the difference in the results found using the best-fit PDFs for
the upper and lower $\alpha_s$ values of the 68\% CL range.
Indeed, it can be
shown~\cite{Lai:2010nw} that
addition of PDF and $\alpha_s$ uncertainties in quadrature automatically
accounts for the correlation between $\alpha_s$ and PDF uncertainties,
assuming gaussianity and linear error propagation.
As in the 2010 recommendation, it was then suggested to determine
the PDF+$\alpha_s$ uncertainty for each of the three groups, at 
the upper and lower $\alpha_s$ value of the chosen range
$\alpha_s(m_Z^2)=0.118\pm0.0012$, and finally take the envelope of the results.

\subsection{Scope of this document}

The scope of the current document is to provide an update of the PDF4LHC
recommendation suitable for its use at the LHC Run II.
Similar to previous
recommendations, the new recommendation will be based on a discussion
and comparison of existing PDF sets.
Specifically, we will argue that,
based on more recent results, an envelope procedure is no longer
necessary and a purely statistical combination is appropriate.
In addition,
we will specify criteria for the inclusion of PDF sets in this
combination.
We
will then discuss dedicated tools which allow for a streamlined
construction and delivery of this statistical combination, and finally
discuss the main feature of a combined PDF4LHC PDF set constructed in
this way.

The outline of the paper is as follows.
In Sect.~\ref{sec:updates} we review developments in PDF determination
since the original PDF4LHC recommendation, and in
Sect.~\ref{sec:comparisons} we compare the most
recent PDF sets from all groups.
In Sect.~\ref{sec:prescription} we motivate the new prescription for the
combination of PDF sets and enumerate criteria for the inclusion of PDF sets
in this combination.
In Sect.~\ref{sec:frameworks}  we discuss
the practical
implementation of the 2015 PDF4LHC prescription, based on the combination
of the CT14, MMHT14 and NNPDF3.0 PDF sets.
First we
 introduce the 
Monte Carlo (MC) statistical combination, and we then present various methods
suitable for the production and delivery of a manageable reduced
combined PDF set, based on either MC or Hessian methodologies.
We also present the resulting 
PDF4LHC15 combined sets, both at the level of
individual PDFs and parton luminosities, and use them to compute
several representative LHC cross-sections, specifically comparing the various
delivery forms.

For the majority of the users of the new PDF4LHC15 recommendations,
the only section which is essential is Sect.~\ref{sec:recommendations}.
This section presents a self-contained summary
with
the general guidelines for the usage of the PDF4LHC15 combined sets,
including the formulae for the computations of PDF and PDF+$\alpha_s$
uncertainties, and the corresponding citation policy.
Finally, conclusions are drawn and directions for
future developments are presented in 
Sect.~\ref{sec:future}.

%% file: sec-updates.tex
\section{Recent developments in PDF determination}
\label{sec:updates}

Since the 2010 PDF4LHC recommendation document~\cite{Botje:2011sn}
there has been very considerable progress in the determination of PDFs.
This includes both the usage of an increasingly large and diverse set
of data, exploiting new 
measurements from the LHC, Tevatron and HERA experiments, and
various methodological developments, which have been incorporated in
various new PDF sets released by different groups.
In this section we will give a brief summary of these developments.

\subsection{Intermediate PDF updates}
\label{sec:intermediate}

Some of the updates in PDF sets were fully expected at the time of the last 
recommendation document, 
and, as  mentioned in Sect.~\ref{sec:introduction}, 
have been already 
implemented in updated recommendations,  available 
at the PDF4LHC website~\cite{PDF4LHCrecom}.

Specifically, at the time of the 
previous recommendation the most updated set from NNPDF was NNPDF2.0~\cite{Ball:2010de},
obtained using the zero-mass variable-flavour-number (ZM-VFN)
scheme, which is known to
miss important mass-dependent terms
for scales near the quark mass and to induce inaccuracies in the PDFs.
However, soon after the recommendation this set was updated to  NNPDF2.1~\cite{Ball:2011mu}
which used the FONLL general-mass variable flavour number scheme (GM-VFN scheme) described
in~\cite{Forte:2010ta}, and this automatically improved general agreement 
with CT and MSTW PDFs.
These PDFs were later also made available at NNLO~\cite{Ball:2011uy}.
Subsequently NNPDF updated their PDFs to the NNPDF2.3
set~\cite{Ball:2012cx}, the main change being the inclusion of early LHC 
data on rapidity-dependent vector boson production and asymmetries and on
inclusive jet production.
In addition, the theoretical treatment of charged-current structure functions
in neutrino DIS was improved in NNPDF2.3, leading to a somewhat
enhanced strangeness around $x\simeq 10^{-2}$.
Even though changes from  NNPDF2.1 to NNPDF2.3 are moderate, 
the new data and improved fitting methodology reduced uncertainties 
a little. 

In 2013 the CT10 NNLO PDFs were released, specifically using the  
extension to NNLO~\cite{Guzzi:2011ew} of the S-ACOT-$\chi$ heavy
flavour scheme used by this group. This enabled an intermediate
improved prescription based on the envelope of CT10, MSTW08 and
NNPDF2.3 at NLO and NNLO, as discussed in Sect.~\ref{sec:introduction}.

Coming to MSTW08, it was soon clear that despite good 
predictions for most LHC observables, the MSTW2008 PDFs did not describe low 
rapidity lepton asymmetry data, which is sensitive to small-$x$ ($x\sim 0.01$)
valence quarks.
This was rectified in an intermediate update, the MMSTWW
(or sometimes MSTWCPdeut) PDFs~\cite{Martin:2012da}, based on 
 the same data sets as MSTW2008, but with an extended PDF parametrization 
based on Chebyshev polynomials in $(1-2\sqrt{x})$ rather than just powers of 
$\sqrt{x}$,  and more flexible deuteron correction (given as a 
function with four free parameters that are fit to data).
This included a study
of the number of terms in the parametrization
of quarks and gluons
required to reproduce functions to an accuracy of 
small fractions of a percent: seven parameters were found to be
sufficient.
These PDFs agree well with LHC $W$ asymmetry data, even
though  the data itself is not actually included in the fit.

There have also been
intermediate updates from groups not included in the 2010
PDF4LHC combination.
 PDF
sets  from the HERAPDF group 
are based on fitting to HERA data only. 
While the HERA structure function data is undoubtedly the
most constraining single data set available, particularly for the gluon and the
total quark singlet 
distribution at moderately and small $x$, it provides fewer 
constraints in some kinematic 
regions (such as the gluon at large $x$) and some PDF combinations,
and specifically it does not allow a separation of the down and
strange distribution.
At the time of the last recommendation,
the most recent  available  PDF
set was HERAPDF1.0, which was obtained from a fit
to the HERA Run-I combined
total cross-section data.
This set was in good broad agreement with CT, MSTW and NNPDF sets, but 
with larger uncertainties for many PDF regions and a much softer high-$x$ gluon and harder 
high-$x$ sea quarks.
Since then, there 
have been a number of updates.
HERAPDF1.5~\cite{CooperSarkar:2011aa} included some preliminary data on inclusive cross 
sections from HERA-II running.
This reduced uncertainties slightly, and also led to
a harder high-$x$ gluon and sea-quark distributions 
at NNLO, though not as much at NLO.
These differences between NLO and NNLO can be partly explained
by the fact that the NNLO sets appeared some time after the
NLO ones, and
also that the NNLO HERAPDF1.5 set
was based on a more flexible PDF parametrisation that its
NLO counterpart.

There are also two other PDF sets that are based on smaller total sets of data than the CT, MMHT
and NNPDF fits.
These PDFs differ from all those discussed so far in
that they adopt a fixed-flavour-number (FFN) scheme
for the treatment of heavy quarks:
{\it i.e.}, they do not introduce PDFs for
charm and bottom quarks and
antiquarks.
The most recent sets from these groups 
at the time of the previous  recommendation were 
ABKM09~\cite{Alekhin:2009ni} and
(G)JR09~\cite{Gluck:2007ck,JimenezDelgado:2008hf}.
The group responsible for the former, 
now the ABM collaboration, has since produced two updated sets.
The first one was the
ABM11 PDFs~\cite{Alekhin:2012ig}, which
included the combined HERA Run-I cross-section 
data for the first time, and also some H1 data at different beam energies in addition to 
the fixed target DIS and Drell-Yan data already used in the ABKM analysis.
The PDFs were mostly unchanged,
but with a slightly larger small-$x$ gluon. As in the ABKM analysis a
rather low 
value of $\alpha_s(m_Z^2)$, i.e. $\alpha_s(m_Z^2)=0.1134$ at NNLO,
was obtained.

In Ref.~\cite{Alekhin:2012ce},
the impact on the ABM11 analysis 
of fitting different jet datasets from D0, CDF and the 2010 ATLAS data was studied.
This was done both at NLO and at NLO$^{*}$ (including the effects
of threshold resummation) in jet production.
As compared to the default fit, once the Tevatron jet data sets were included,
the fitted value of $\alpha_s(m_Z^2)$ was found to vary by an amount $^{+0.001}_{-0.002}$, 
depending on the particular data set and jet algorithm.
These jet measurements have not been included in the posterior ABM PDF releases.

All these PDF fitting groups have
presented major updates recently:
MMHT14~\cite{Harland-Lang:2014zoa}, CT14~\cite{Dulat:2015mca}
NNPDF3.0~\cite{Ball:2014uwa}, ABM12~\cite{Alekhin:2013nda} and
HERAPDF2.0~\cite{Abramowicz:2015mha}, which will be discussed
in more detail in Sect.~\ref{sec:currentPDFsets} below.

\subsection{New experimental measurements}
\label{sec:newexp}

One obvious reason for updates of the various
PDF sets is the availability of new measurements
from a 
variety of collider and fixed-target experiments.
Recently,  an up-to-date overview of 
all PDF-sensitive measurements at 
LHC Run-I has been presented in the ``PDF4LHC report on PDFs and LHC
data''~\cite{Rojo:2015acz}. We refer to this document for an extensive
discussion (including full references) and here we only provide a
very brief summary.

First of all, HERA has provided combined H1 and ZEUS data on 
the charm structure functions, in addition to
the already available combined inclusive Run-I structure function
data. Run-II ZEUS and H1 data have also been
published, and 
a  
legacy HERA combination of all inclusive structure function measurements
from Run-I and Run-II has been presented very
recently~\cite{Abramowicz:2015mha}; it is only included in the
HERAPDF2.0 set.
Also,  Tevatron collaborations are   
still releasing  data on lepton asymmetry
and top-pair cross-sections and differential distributions.

A wide variety 
of data sets which help constrain the parton 
distributions has been made available at the LHC  from the ATLAS, CMS and LHCb collaborations.
This includes
data sets on rapidity- and mass-dependent $W^{\pm}, Z$ and $\gamma^{\star}$
production, inclusive jet cross-sections, inclusive top-pair production 
and $W +$ charm quark production, 
differential top pair production and dijet cross 
sections.

Many of these measurements have been
included in recent global
PDF sets.
Therefore, for all PDF 
groups the  fitted dataset
is now considerably
wider than that at the time of the 2010 PDF4LHC recommendation.

\subsection{Current PDF sets and  methodological improvements}
\label{sec:currentPDFsets}

We now discuss in turn all the presently available PDF sets.
These
differ  from previous releases not only because of the inclusion of new
data, but also because of various methodological and
theoretical improvements, which
we also review.

We  first consider the most updated
PDF sets from the three groups represented in 
the original recommendation.
An important observation is that the most
recent sets from these groups,  CT14, MMHT2014 and NNPDF3.0 sets are
all  made available  with full PDF uncertainty 
information available at the common value of $\alpha_s(m_Z^2)=0.118$,
both  at NLO and NNLO. This is to be contrasted to the situation
 at the time of the
previous recommendation, when full information of PDF uncertainties
was only available in each set for
fixed (different) values of $\alpha_s$, spanning a
range of 
approximately $\Delta\alpha_s(m_Z^2)\approx 0.002$.
We now discuss each set in turn.

The CT14 PDF sets~\cite{Dulat:2015mca} have been made recently available 
at NLO, NNLO, and also at LO.
These sets include a variety 
of LHC data sets as well as the most recent D0 data on electron charge asymmetry
\cite{D0:2014kma} and combined H1+ZEUS data on charm production \cite{Abramowicz:1900rp}.
The PDFs also use an updated parametrization based on 
Bernstein polynomials, which reduces parameter correlation. The PDF
sets contain 28 pairs of eigenvectors. LHC inclusive jet data
are included at NLO also in the NNLO fit, despite the lack of a full NNLO calculation, in the expectation that the NNLO corrections are relatively small for single-inclusive cross-sections.
The main change in the PDFs as
compared to CT10 is a softer high-$x$ gluon, a smaller strange quark (partially due to 
correction of the charged current DIS cross-section code) and the 
details of the flavour decomposition, e.g. $\bar u /\bar d$ and the high-$x$ valence quarks,
due both to the parametrization choices and new data.
In the CT14 publication~\cite{Dulat:2015mca} a number
of additional comparisons were made to data not
included in the fit, such as $W$+charm and $\sigma_{\bar t t}$ from the LHC,
and the predictions 
are found to be in good agreement.
One should also mention that
the CT14 PDF uncertainties are provided as 90\% confidence level
intervals, that need to be rescaled by a factor 1.642 to compare
with other PDF sets, for which uncertainties are provided
as 68\% confidence level intervals.

We now consider PDFs from the  MSTW group, currently renamed MMHT due
to a change in personnel.
The MMHT14 PDFs~\cite{Harland-Lang:2014zoa}
incorporate the improved parametrization and deuteron 
corrections in the MMSTWW study~\cite{Martin:2012xx}, and also a change in the GM-VFN scheme to the
``optimal''
scheme in~\cite{Thorne:2012az}, and a change in the branching fraction 
$B_{\mu} = B(D \to \mu)$ used in the determination of the strange quark from 
$\nu N \to \mu\mu X$ data.
Charged-current structure functions were also updated in the MMHT14 analysis,
in particular by including the NLO gluon coefficient functions.
The updated analysis  includes new data: the combined HERA-I inclusive 
structure function data~\cite{Aaron:2009aa} (which post-dates the MSTW2008 PDFs) and charm structure function 
data, updated Tevatron lepton asymmetry data, vector boson and inclusive jet data from the 
LHC (though LHC jet data is not included at NNLO),  and top pair cross-section data from the 
Tevatron and LHC.
No PDFs change dramatically in comparison to
 MSTW2008, with the most significant changes being the shift in 
the small-$x$ valence quarks already observed 
in the MMSTWW study, a slight increase in the central value of the strange quark to help 
the fit to LHC data, and a much expanded uncertainty on the strange distribution due to the 
inclusion of a conservative uncertainty on $B_{\mu}$.
The MMHT14 PDFs provide a much better description
to the LHC $W$ lepton asymmetry data~\cite{Chatrchyan:2013mza,Aad:2011dm}
than its predecessor
MSTW08, thanks to the various improvements already implemented
in the MMSTWW (also known as MSTW08CPdeut) intermediate release.
Deuteron nuclear corrections are fitted to the data, and
their uncertainty propagates into the total PDF
uncertainty.

The MMHT14
PDFs are made available with 25 eigenvector pairs for $\alpha_s(m_Z^2) =0.118$ and 
0.120 at NLO and 0.118 at NNLO, as well as at LO (for $\alpha_s(m_Z^2) =0.135$). However, 
$\alpha_s(m_Z^2)$ is also determined by the NLO and NNLO fits and values of 
$\alpha_s(m_Z^2)=0.1201$ and $0.1172$ respectively are found, or 0.1195 and 0.1178 if the 
world average of $\alpha_s(m_Z^2)$ is included as a data point. These are in good agreement 
with the PDG world average without DIS
data of $\alpha_s(m_Z^2)=0.1187\pm 0.0007$~\cite{Agashe:2014kda}.
A dedicated study of the uncertainties in the determination of $\alpha_s(m_Z^2)$
in the MMHT14 analysis has been presented in~\cite{Harland-Lang:2015nxa}.

The NNPDF3.0 PDF sets~\cite{Ball:2014uwa} are the recent major
update within the NNPDF framework.
In comparison to NNPDF2.3, NNPDF3.0 includes  HERA inclusive
structure function Run-II data from H1 and ZEUS (before their combination),
more  recent ATLAS, CMS and LHCb data on gauge boson production and
inclusive jets, and  $W+$charm 
 and top quark pair production.
A subset of jet data was included at NNLO using an approximate NNLO treatment,
based on a study~\cite{Carrazza:2014hra} of the region where the
threshold approximation~\cite{deFlorian:2013qia} is reliable
by comparing with the exact $gg$ channel calculation~\cite{Ridder:2013mf}.
Heavy quarks are treated using the FONLL-B scheme (FONLL-A was used in
NNPDF2.1 and NNPDF2.3), which includes mass corrections to an extra
order in $\alpha_s$, thereby achieving a 
better description of low-$Q^2$ data on the charm 
structure function.
The compatibility of the NNPDF3.0 NLO set with the LHCb forward charm production
cross-sections has been demonstrated in~\cite{Gauld:2015yia}, where it
was also shown that these data could be useful in reducing
the uncertainty on the small-$x$ gluon.

The NNPDF3.0 fitting procedure has been tuned by means  of a 
closure test, namely, by generating pseudo-data based on an assumed
underlying set of PDFs. One verifies in this case that the output of the
fitting procedure is consistent with the a priori known answer.
As a byproduct, one can investigate directly the origin
of PDF uncertainties, and specifically how much of it is due to the
uncertainty in the data, how much it is due to the need to interpolate
between data points, and how much is due to the fact that there is an
infinite number of functions which produce equally good fits to a
given finite number of data-points.
The minimization has been
optimized based on the closure test, specifically by choice of a more
efficient genetic algorithm, and the choice of optimal stopping point
based on cross-validation and the search for the 
absolute minimum of the validation $\chi^2$ (look-back fitting). 
The NNPDF3.0 PDFs display moderate changes in comparison to NNPDF2.3:
specifically somewhat smaller  uncertainties and a 
noticeable change in the gluon-gluon luminosity.

The HERAPDF2.0 PDFs has also become recently
available~\cite{Abramowicz:2015mha}.
HERAPDF2.0
is the only PDF set to include the full  legacy Run-I and Run-II
combined HERA structure function
data, based on 
runs at beam energies of 
$920, 820, 575$ and 460 GeV.
This PDF set has considerably reduced uncertainties on PDFs compared to 
HERAPDF1.0. In particular there is a much improved constraint on flavour 
decomposition
at moderate and high $x$ due to the difference between neutral current $e^+$ and $e^-$ 
cross-sections, and due to much more precise charged current data, which provides a genuine 
constraint on the $d_V$ distribution.
The running at different energies gives sensitivity to 
$F_L(x,Q^2)$ which provides some new information on the gluon. The HERAPDF2.0 sets are made 
available also with a default of  $\alpha_s(m_Z^2)=0.118$ and with 14 eigenvector sets
along with further variations to cover uncertainties due to model
assumptions and 
changes in the form of the parametrization; for CT, MMHT and NNPDF
this is not necessary because the
input form of the PDFs is flexible enough.
Sets 
including HERA jet and charm data have also been
made available.
The HERAPDF2.0 PDFs agree fairly well in general with the CT, MMHT and NNPDF sets, but 
there are some important differences in central values, particularly for high-$x$ quarks.
Overall, the uncertainties on HERAPDF2.0 PDFs are markedly smaller than for previous 
versions, but there are still some regions where the PDFs have 
significantly larger uncertainties than those in the global fits.

Let us also mention that the three general-mass variable-flavour
number schemes used by these four PDF sets (S-ACOT-$\chi$, FONLL and
RT/RT$^{\rm opt}$) exhibit a remarkable convergence, especially
at NNLO~\cite{LHhq}.
All these four sets fit the HERA charm cross-sections
with comparable quality, and residual
differences between the three GM-VFN schemes translate into
rather small differences in the resulting
PDFs.~\cite{Butterworth:2014efa}

The most recent update
for the ABM collaboration,  ABM12~\cite{Alekhin:2013nda},
now includes the 
HERA combined charm cross-section data, an extension of the HERA inclusive data to higher $Q^2$ than 
previously used, and vector 
boson production data from ATLAS, CMS and LHCb. The heavy flavour contributions to structure 
function data are now calculated using the $\overline{MS}$ renormalization scheme~\cite{Alekhin:2010sv} rather 
than the pole mass scheme.
The main change compared to ABM11 PDFs is in the 
details of the decomposition into up and down quarks and antiquarks, this being affected by the 
LHC vector boson data.
The ABM12 PDF sets are determined together with 
$\alpha_s$, whose value  comes out to be rather lower
than the PDG average, namely  $\alpha_s(m_Z^2)=0.1132\pm 0.0011$ 
at NNLO.
Top quark pair production data from the LHC is investigated, but not included in the 
default PDFs.
Its inclusion tend to raise the high-$x$ gluon and $\alpha_s(m_Z^2)$ a little;
the precise details depend on the precise value of the
top quark mass (and mass renormalization scheme) used.
The ABM12
PDFs are available with 28 eigenvector pairs for the default $\alpha_s(m_Z^2)$ and at NNLO only.
Within the same framework there has been a recent specific investigation of the strange sea~\cite{Alekhin:2014sya} in the light of LHC data and NOMAD and CHORUS fixed-target data, but this
is not accompanied by a PDF set release.
More recently, an update of the ABM12 with additional data
from gauge boson production
at the Tevatron and the LHC has been presented~\cite{Alekhin:2015cza},
though this study is not accompanied
by the release of a new PDF set.

There has also been an update of the (G)JR PDFs:
JR14~\cite{Jimenez-Delgado:2014twa}. The default sets from this group
make the assumption that PDFs must be valence-like at  low
$Q_0^2=0.8~$GeV$^2$.
This analysis  
extends and updates the fixed-target and collider DIS data used, in particular including 
CCFR and NuTeV dimuon data as a constraint on strange quarks, and includes both 
HERA and Tevatron jet data at NLO, but does not include any LHC data.
The strong coupling is also determined along with the PDF, yielding 
a  value $\alpha_s(m_Z^2)=0.1136$ 
at NNLO, and a large small-$x$ gluon distribution. There are some significant changes compared 
to the JR09 NNLO PDFs.
These are mainly at high $x$ values, where there is an increase in 
down and strange quarks and antiquarks, and a decrease in the gluon. 
At low $x$ values, the features
are largely determined by the assumption of a valence-like input, 
and are largely unchanged.
A determination without the valence-like
assumption (and thus a more flexible parametrization) is also
performed, leading to  PDFs and a best fit value of $\alpha_s(m_Z^2)$
more in line with other groups, though it is not adopted as a default.

In addition, there has  been a number of more specific PDF
studies.
The CJ12 PDFs~\cite{Owens:2012bv} emphasize the
 description of the high-$x$ region, with   particular care paid to higher-twist
corrections and the modeling the deuteron corrections to nucleon PDFs,
thereby allowing for  the inclusion of
high-$x$ and low $Q^2$ data which is often 
cut from other  fits, most notably Jefferson Lab data.
Several  
analyses have been focused  on the possible intrinsic
charm component of the proton.
Specifically,  
Refs.~\cite{Dulat:2013hea,Jimenez-Delgado:2014zga} have presented 
investigations
of the evidence for, and limits on, intrinsic charm in the proton.
Work in progress towards the determination of intrinsic charm is also
ongoing within NNPDF, and the required modifications of the FONLL GM-VFN
scheme have been presented in~\cite{Ball:2015tna}.
The
determination of  
mass of the charm quark from PDF fits  has been discussed in
Refs.~\cite{Alekhin:2012vu,Gao:2013wwa}.
PDFs including QED corrections, which were presented for the first time 
in Ref.~\cite{Martin:2004dh} by the MRST group, have been updated in Ref.~\cite{Martin:2014nqa};  a determination of the photon
PDF from the data  has been presented for the first time
in~\cite{Ball:2013hta}, with the corresponding study
in the CT14 framework found in 
 Ref.~\cite{Schmidt:2015zda}.

It is also important to mention that
a number of PDF-related studies based on
the open-source {\tt HERAfitter} framework~\cite{Alekhin:2014irh}
have also become available.
From the {\tt HERAfitter} developers team, two studies have
been presented: PDFs with correlations between
different perturbative orders~\cite{::2014uva} and the impact
on PDFs of the Tevatron legacy Drell-Yan measurements~\cite{Camarda:2015zba}.
Within the LHC experiments, various studies of the constraints
of new measurements on PDFs have been performed using
{\tt HERAfitter},
including the ATLAS $W,Z$ analysis~\cite{Aad:2012sb} that lead to a strangeness
determination and the CMS studies constraining the gluon
and $\alpha_s(m_Z^2)$ from inclusive
jet production~\cite{Chatrchyan:2012bja,Khachatryan:2014waa}.
The impact of the LHCb forward charm production data
on the small-$x$ gluon has been explored by the
{\tt PROSA} collaboration in~\cite{Zenaiev:2015rfa}.
Finally, {\tt HERAfitter} has also been used to quantify the
impact on PDFs of data from a future Large Hadron Electron
Collider (LHeC)~\cite{AbelleiraFernandez:2012cc}.

\subsection{Origin of the differences between PDFs}
\label{sec:developementsTheory}

There  have been many developments in understanding the 
differences between the PDF sets obtained by different groups. This
understanding inevitably leads to an improvement of 
the best procedures for combining these different PDFs.
In the benchmark document 
\cite{Alekhin:2011sk}, 
on which the first  PDF4LHC recommendation was based, the differences
between the PDFs and the consequent differences in predictions for LHC cross-sections 
were discussed; subsequent  benchmarking exercises were published in
Refs.~\cite{Watt:2011kp,Alekhin:2012ig}; and then
in~\cite{Ball:2012wy} including comparisons with published LHC
measurements.
See also~\cite{Alekhin:2010dd} for an earlier benchmark exercise.
Here we concentrate on three topics on which
some progress has been made recently: the
impact of the choice of heavy-quark scheme; the
origin of the remaining differences between global PDF fits;
and the origin of the differences between the sizes
of PDF uncertainties.
Graphical comparisons of the PDFs
illustrating their present level of agreement
are presented in Sect.~\ref{sec:comparisons}.

\subsubsection{Dependence on the heavy-quark scheme}

The results of various benchmark studies~\cite{Watt:2011kp,Alekhin:2012ig,Ball:2012wy} illustrate how the
softer high-$x$ gluon and smaller $\alpha_s(m_Z^2)$ 
present in PDF sets like ABM12 and JR09 leads to a smaller prediction 
for top pair cross 
sections and Higgs bosons as compared to the global
PDF sets.
In both cases, the magnitude of the difference
 depends on 
 whether the default $\alpha_s(m_Z^2)$ is obtained directly
 from the fit or if it is treated as an external input.
 In these sets also
 the quark PDFs seem to lead to slightly larger vector boson production
 cross-sections as compared to the global fits.
On the other hand
 the CT, MSTW/MMHT, and NNPDF, and HERAPDF PDFs at NNLO are
 in generally good agreement, though HERAPDF has associated
 rather larger uncertainties from the reduced dataset.

In the previous recommendation it was stated that the systematic feature of a lower
high-$x$ gluon and lower value of $\alpha_s(m_Z^2)$ in some fits was due to the omission
of Tevatron jet data, which constrains the high-$x$ gluon. 
The understanding of this issue has improved. While it is true
that Tevatron and LHC jet data directly constrain the gluon distribution, and 
that PDF sets with a particularly small high-$x$ gluon  
distribution and/or small value of $\alpha_s(m_Z^2)$ do not provide the 
best fit to these data (see~\cite{Thorne:2011kq,Alekhin:2012ce} for
discussions on this issue with  
similar results, though not the same conclusions), 
it is unclear whether their omission 
automatically leads to a small high-$x$ gluon distribution or small
$\alpha_s(m_Z^2)$.
The results of including the jet data in the fit
also show some dependence on the treatment of
correlated systematic errors (additive or multiplicative) 
assumed, though these effects are way too
small~\cite{Gao:2013xoa,Ball:2014uwa}
to explain the aforementioned differences.

The behavior of the
gluon is constrained directly by Tevatron/LHC jet
production at high $x$, and indirectly by scaling violations in DIS data at
very small and high $x$, with the latter influencing the gluon at
moderate $x$ via the momentum sum rule. The gluon PDF is strongly
correlated with $\alpha_s(m_Z^2)$ at high $x$ and anti-correlated at low $x$. 
Due to this feature of correlation between the gluon and $\alpha_s(m_Z^2)$, 
there is no automatic strong pull to
lower $\alpha_s(m_Z^2)$ or small high-$x$ gluon in the fits that omit
hadron collider data~\cite{Thorne:2011kq,Ball:2011us}.
It appears that the DIS and Drell-Yan data provide only 
a weak constraint on  $\alpha_s(m_Z^2)$ and the gluon, if both are left free, 
a conclusion also evident in those HERAPDF1.6 studies
that added HERA jet data to the DIS data and 
more strongly constrain $\alpha_s(m_Z^2)$.

An analysis in the ABM framework~\cite{Alekhin:2011ey} suggested
that this issue is related to whether one makes a conservative
cut on low-$Q^2$ structure function data, or
includes it and includes higher-twist corrections as
well as higher-order QCD corrections to $F_L(x,Q^2)$
in fits of the NMC cross-section data. 
However, their findings were not corroborated 
by CTEQ, MSTW and NNPDF groups
~\cite{Thorne:2011kq,NNPDF:2011aa,Ball:2013gsa,Thorne:2014toa}, who
did not  observe clear-cut sensitivity of the PDFs to  
the details of the NMC treatment either at fixed 
$\alpha_s(m_Z^2)$~\cite{NNPDF:2011aa, Gao:2013xoa},  
or varying $\alpha_s(m_Z^2)$~\cite{Thorne:2011kq, Gao:2013xoa}.
Also,  no significant variation in PDFs was seen when including a fixed 
higher twist in the NNPDF2.3 analysis in~\cite{Ball:2013gsa}.
In~\cite{Thorne:2014toa}, when the higher-twist 
contribution is fit within the standard 
MSTW framework, there was fairly small impact on the PDFs, 
and less on $\alpha_s(m_Z^2)$. 

However, different schemes for heavy-quark treatment,
notably the usage of the fixed-flavor scheme 
as opposed to general-mass variable flavor number scheme,
may explain the most pronounced disagreements observed in the gluon
distribution between ABM/JR and the global sets.
When a fit was performed to DIS data using the 
3-flavor FFN scheme at NLO using the MSTW and NNPDF
frameworks~\cite{Thorne:2012az,Ball:2013gsa}, 
the gluon became softer at high $x$ and larger at small $x$;
the light quarks became larger at small $x$ 
(a feature previously noted to some extent 
in~\cite{CooperSarkar:2007ny}); the value of $\alpha_s(m_Z^2)$
decreased noticeably. This was 
confirmed in more detail in~\cite{Thorne:2014toa}, using an approximate 
NNLO FFN scheme.
At NNLO the decrease in
$\alpha_s(m_Z^2)$ was more marked that at NLO, and
these conclusions did not change after the 
Drell-Yan data and higher-twist corrections were added.
It was also observed that the quality of the fit to the DIS data was clearly 
worse when using the FFN scheme, as opposed to the GM-VFN scheme. At 
a fixed $\alpha_s(m_Z^2)$ the same conclusions about the PDFs,
more detailed confirmation of a worse fit were found,
while it was also observed
that the deterioration is mainly in high-$Q^2$ 
DIS data~\cite{Ball:2013gsa}.

In~\cite{Thorne:2014toa} a theoretical investigation of the difference 
between the speed of evolution of $F_2^c(x,Q^2)$ at high $Q^2$ was performed, and it was shown that,        
for the region $x\sim 0.05$, 
there is slow convergence of the $\ln(Q^2/m_c^2)$ terms included 
at finite order in the FFNS to the fully resummed result in a GM-VFNS,
leading to the worse fit to HERA data for this range of $x$ at high $Q^2$.
It
was also shown that  this slower evolution is partially 
compensated by an increase of the gluon in this region of $x$, which feeds
down to lower $x$ with evolution.
This, however, also results in a
smaller  gluon at high $x$, so that 
the fit to the fixed-target data then requires a lower value of 
$\alpha_s(m_Z^2)$.
This feedback between $\alpha_s(m_Z^2)$ and the gluon is somewhat
enhanced  if 
higher twist contributions are included, 
as the freedom in these provides more flexibility to 
change the PDFs and $\alpha_s(m_Z^2)$ without worsening the fit
quality. 

The general conclusion is thus that 
smaller values of $\alpha_s(m_Z^2)$ (particularly at NNLO), 
softer high-$x$ gluons, and slightly enhanced 
light quarks at high $x$ are found if a FFNS, rather than GM-VFNS, 
is adopted  in the PDF fit.
In an extremely conservative approach, these differences
 might be viewed as a ``theoretical uncertainty'', 
but the markedly worse fit quality for the 
FFNS motivates instead that a GM-VFNS should be adopted, 
especially for calculating quantities sensitive to high-$Q^2$ PDFs.  

\subsubsection{Differences within the global PDF fits}

While the above discussion
sheds light on
the differences between the sets based on a FFN scheme
and those which adopt a GM-VFN scheme, the fact remains
that CT,  HERAPDF, MSTW and NNPDF also show some smaller 
differences, even though they all adopt a GM-VFN scheme.
As mentioned in
the introduction, for the previous families of each
group, NNPDF2.3, CT10 and MSTW08, these
differences were especially noticeable for the gluon PDF and led to
a spread of predictions for Higgs production via gluon fusion, producing an
envelope uncertainty that is more than twice the size of individual PDF
uncertainties. 

An attempt to understand the origin of these moderate differences was 
made in Ref.~\cite{Butterworth:2014efa}. In this study, 
all groups produced fits to the HERA combined Run-I inclusive 
cross-section data only. Furthermore, all structure function calculation codes from the groups were 
benchmarked using a toy set of PDFs. The agreement was
good in general, with all deviations in 
neutral current processes being due only to
the differing choices in GM-VFNS choice, which were much smaller at NNLO than NLO due to the convergence 
of such schemes, and differing treatments of electroweak corrections at high $Q^2$.     
A difference in charged current cross-sections was noted, and corrected, but it did not affect the 
PDFs extracted in any significant manner.
Fits to the data were performed adopting a common choice of 
kinematic cuts, $\chi^2$ definition (i.e. treatment of correlated
uncertainties) and charm mass, all of 
which differ to some extent in the default fits.

This study found that the PDFs and predictions for Higgs 
cross-sections at NNLO still followed the same pattern: in particular, 
$\sigma_{gg\to h}$ for CT remained smaller than the  MSTW result,
which in turn was
smaller than the NNPDF prediction, as in the global fits, even though
all
predictions were compatible within the large uncertainties
of a HERA-only fit.
An investigation of variations of the 
heavy flavour scheme was also performed, but this turned out to have a
comparatively small impact. 
Some evidence of parametrization dependence between CT and MSTW was
noted (and  new parametrization forms were adopted by CT), but it was not 
conclusive.

Concurrent to this study with DIS data, NLO theoretical
calculations for fitting collider jet data, based on
programs {\tt MEKS}~\cite{Ellis:1992en} and
{\tt NLOJET++}~\cite{Nagy:2001fj,Nagy:2003tz}, and their
fast interfaces
{\tt APPLgrid} and
{\tt FastNLO}~\cite{Carli:2010rw, Wobisch:2011ij,Kluge:2006xs,Britzger:2012bs}, were
validated~\cite{Gao:2012he,LesHouches1,Ball:2012wy}, providing greater confidence 
in the NLO theoretical codes and their fast interfaces, and the fits to the jet data.

Therefore, the details of the reasons for the disagreement at, or around, 
the one-$\sigma$ level,
between the global fits for the gluon luminosity in the region of the
Higgs boson mass were still only partially understood.
This was a situation that
conservatively could only be handled by taking an envelope of results,
which amounts to saying that one or more of the results are affected
by an unknown source of systematics (or bias).

This state of affairs has changed completely
with the 
release of the
most recent CT14, MMHT14 and NNPDF3.0 PDF sets.
Indeed, due to increased data
constraints and methodological improvements in each of the fitting
procedures, these turn out 
to be in
better agreement for most PDFs in a wide
range of $x$, as will be illustrated
in Sect.~\ref{sec:comparisons}.

\subsubsection{Differences in the size of the PDF
  uncertainties}

In general the sizes of the PDF uncertainties between
the three global sets, NNPDF, CT and MMHT, turn out
to be in rather reasonable agreement.
On the other hand, the sizes of the uncertainties for
PDF sets based on a HERA-only dataset, such as
HERAPDF, are be expected to be
larger, and in some cases much larger, than those of the global
fits.
This is indeed what happens for the HERAPDF2.0 PDFs, despite
including the legacy HERA combination
data~\cite{Abramowicz:2015mha}, since deep-inelastic
scattering cannot constrain a number of quark
flavor combinations and the medium and large-$x$
gluon.
The much larger PDF uncertainties in HERA-only fits
as compared to global fits have also been verified
in the NNPDF framework.
The NNPDF3.0 HERA-only fit~\cite{Ball:2014uwa} indeed
shows substantially larger uncertainties for most PDF
combinations (with the exception of the quark singlet
and the small-$x$ gluon) than the corresponding
global analysis.

Whereas HERAPDF2.0 includes the legacy HERA combination
data~\cite{Abramowicz:2015mha}, which are not included in any other
fit,
it has been shown~\cite{Thorne:2015caa,Rojo:2015nxa} 
that the impact of these data on a global fit that already
included
the HERA-I data was moderate.
In addition, it has been checked~\cite{Rojo:2015nxa} that
their impact on a
global fit like NNPDF3.0 which already included both the HERA-I combination
and the individual HERA-II measurements is very small.
Therefore, the availability of the legacy HERA inclusive dataset
does not modify the general
anticipation
that HERA-only fits
(or more general, DIS-only fits) will have
larger PDF errors than the global fits.
However, it is worth emphasizing that
the uncertainties of HERAPDF2.0 are in
many cases comparable (for the $u$ quark) 
or even smaller (for the gluon) to the global
fits, as will be
illustrated in
the comparisons of Sect.~\ref{sec:comparisons},

The PDF uncertainties in ABM12 are systematically smaller, 
and often much smaller, that those of the global fits.
Therefore the PDF uncertainty estimates of the three global fits, on one
hand, and those of ABM12, on the other, 
do not appear to be consistent.
The most likely
explanation
for the disparate sizes of the uncertainties is their varied statistical interpretations adopted by the PDF analysis groups.
In the global sets, they
are determined according to complex
procedures, and include more sources of 
uncertainties  than the purely experimental uncertainty obtained
from the  
$\Delta \chi^2=1$ criterion,
for example, arising from partial inconsistencies
of the input datasets, using
a variety of methods~\cite{Harland-Lang:2014zoa,Pumplin:2002vw,
  Pumplin:2001ct,Collins:2001es,Watt:2012tq,Ball:2014uwa}.

%% file: sec-comparisons.tex
\section{Comparisons of PDF sets}
\label{sec:comparisons}

Let us now
present a comparison of PDFs and parton
luminosities, using  the most
up-to-date versions of the PDF sets.
All the PDFs and luminosity comparison plots shown in this document have been obtained
with the APFEL on-line PDF plotting tool~\cite{Carrazza:2014gfa,Bertone:2013vaa}.
We will show comparisons for the ABM12, CT14, HERAPDF2.0, MMHT2014 and NNPDF3.0
NNLO PDF sets.
For HERAPDF2.0, the uncertainty shown includes experimental, model,
and parametrization uncertainties.
For ABM12, we use the $N_f=5$ set and the native value of $\alpha_s(m_Z^2)=0.1132$,
since sets for different values of the strong coupling are not available.

\subsection{Parton distributions}

\begin{figure}[t]
  \centering
  \includegraphics[scale=.37]{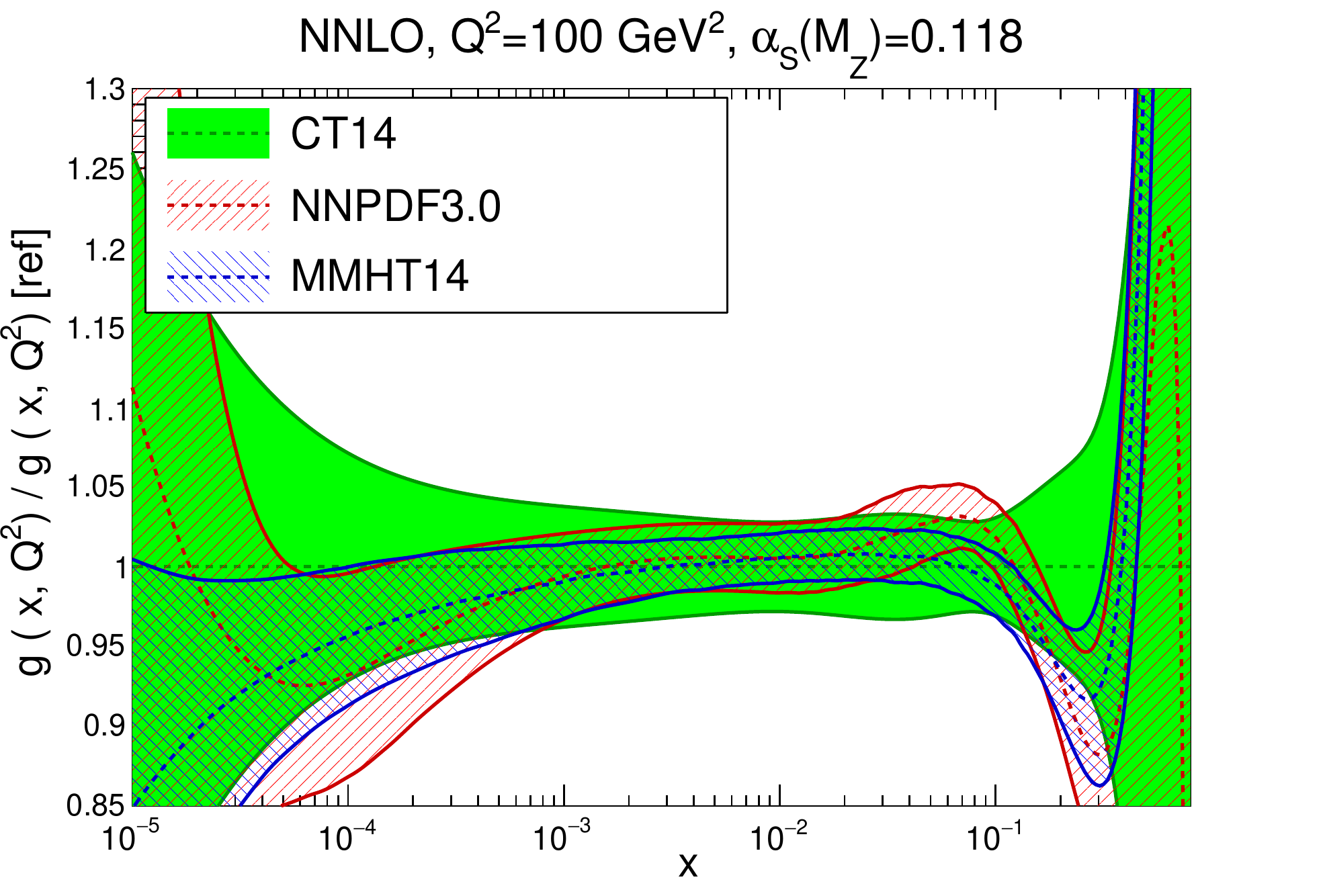}
  \includegraphics[scale=.37]{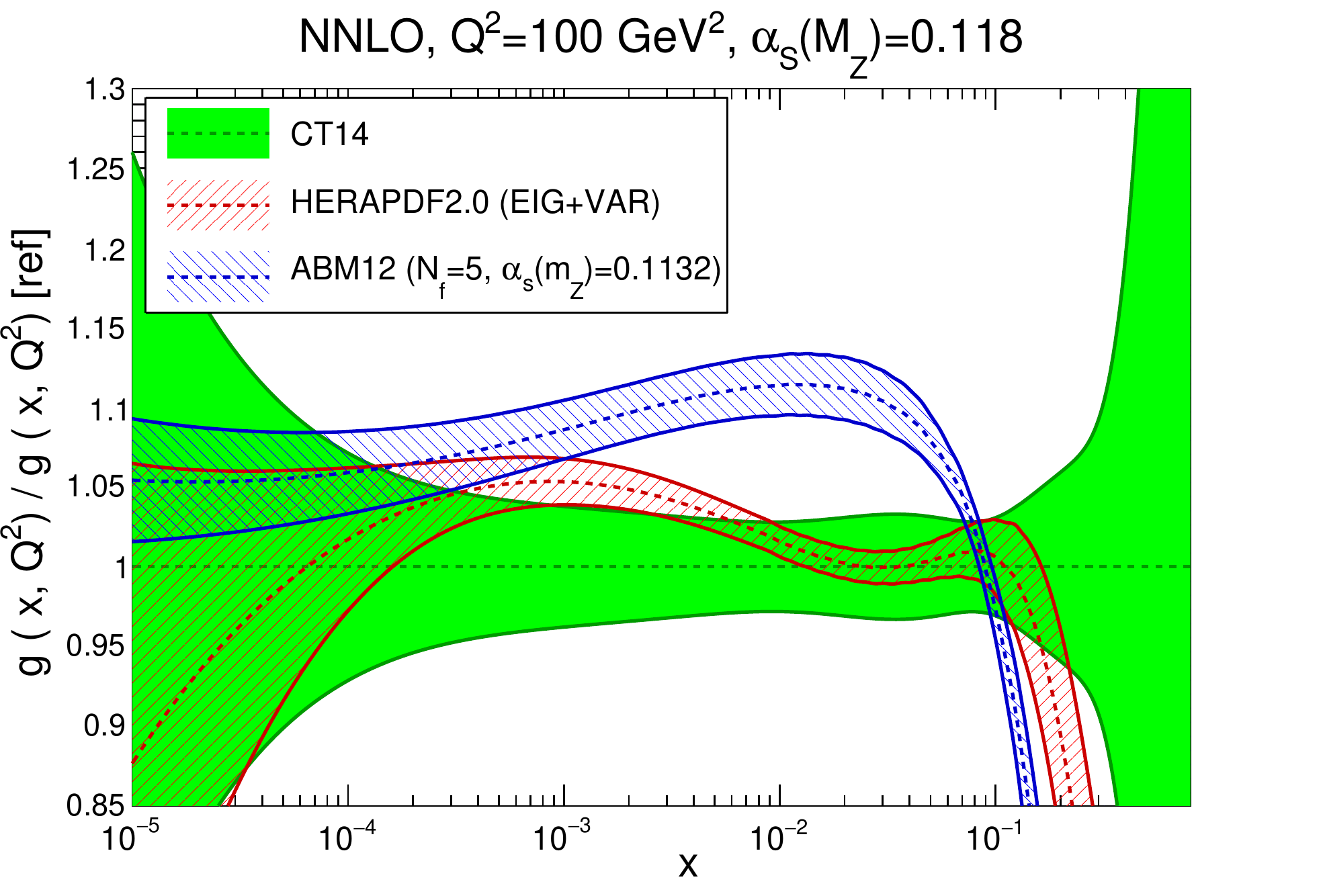}\\
  \includegraphics[scale=.37]{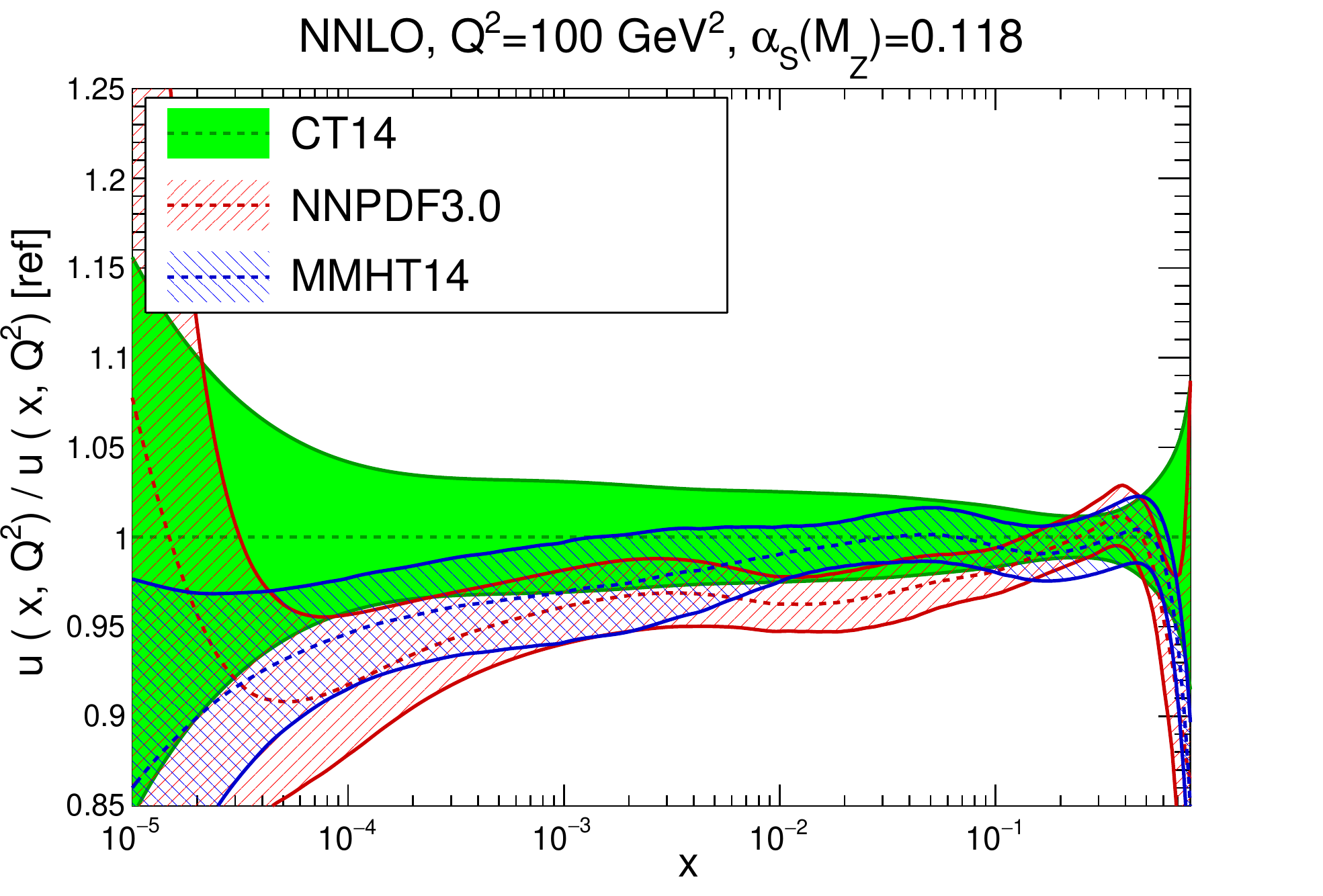}
  \includegraphics[scale=.37]{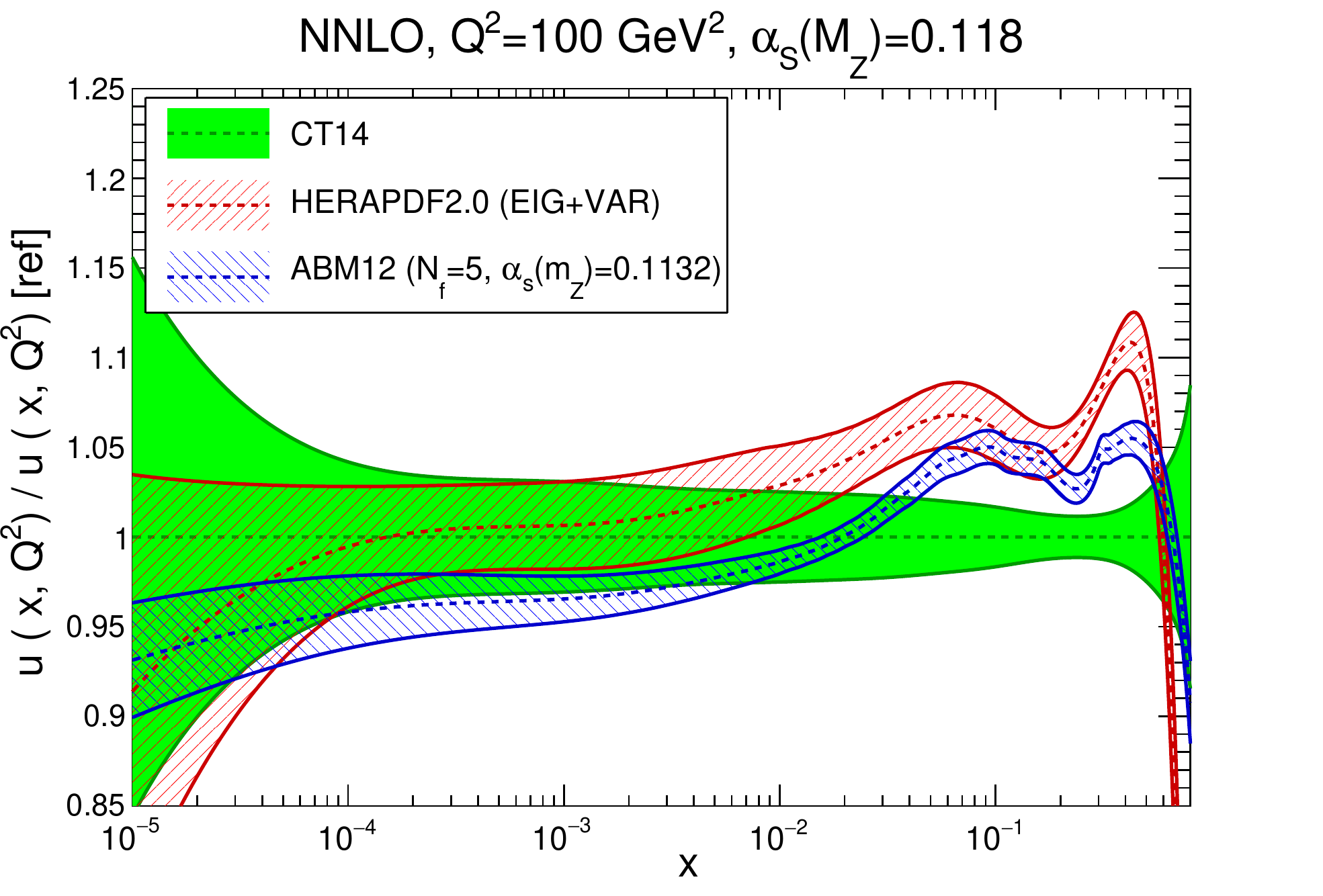}\\
  \caption{\small Comparison of the gluon (upper plots) and
    up quark (lower plots) PDFs from the CT14,  MMHT14 and NNNPDF3.0 NNLO sets
    (left plots) and from the  CT14,  ABM12 and HERAPDF2.0 sets
    (lower plots).
    The comparison is performed at a scale of $Q^2=100$ GeV$^2$,
    and results are shown normalized
    to the central value of CT14.
  }  
\label{fig:gluon_up_quark}
\end{figure}

\begin{figure}[t]
  \centering
   \includegraphics[scale=.37]{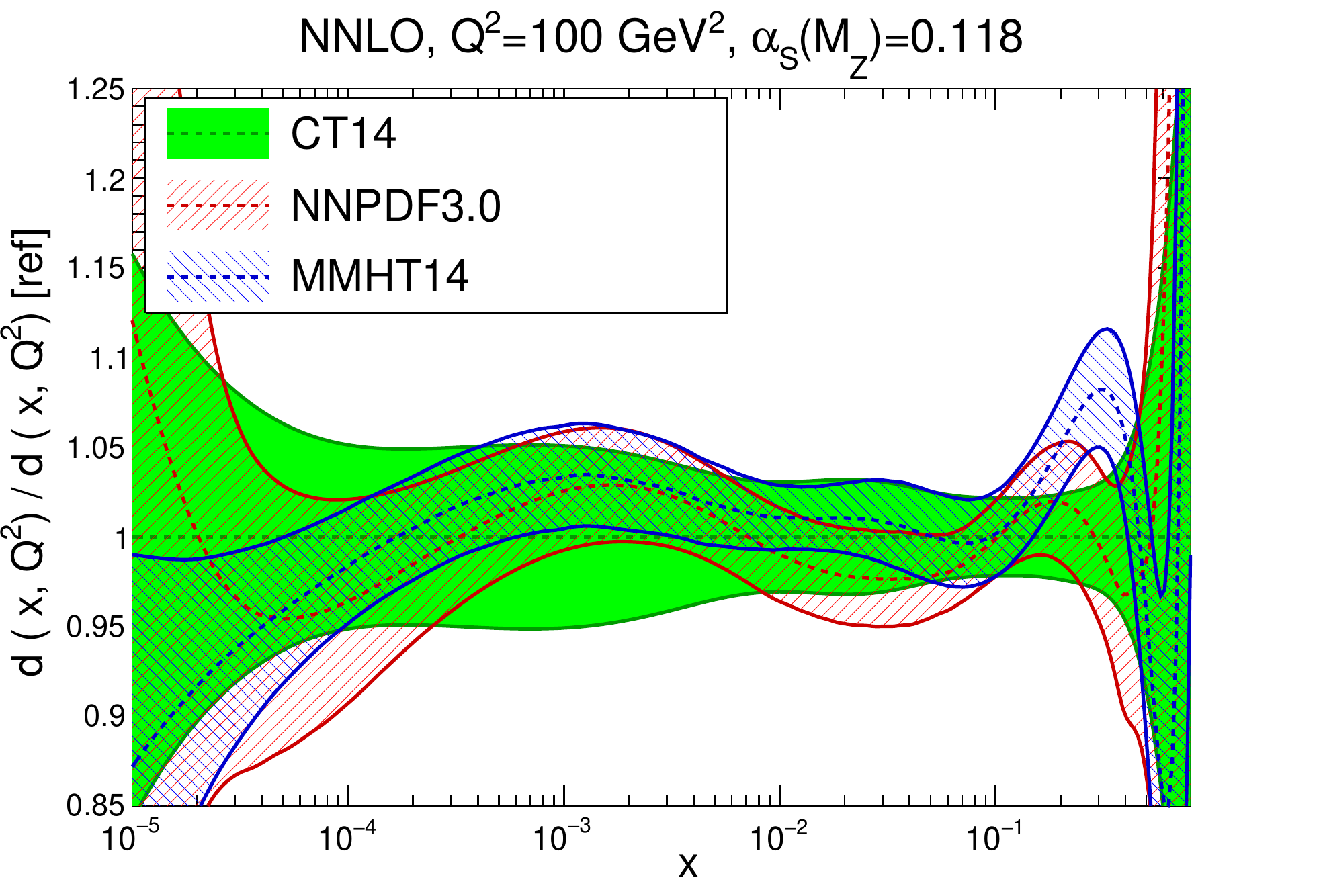}
  \includegraphics[scale=.37]{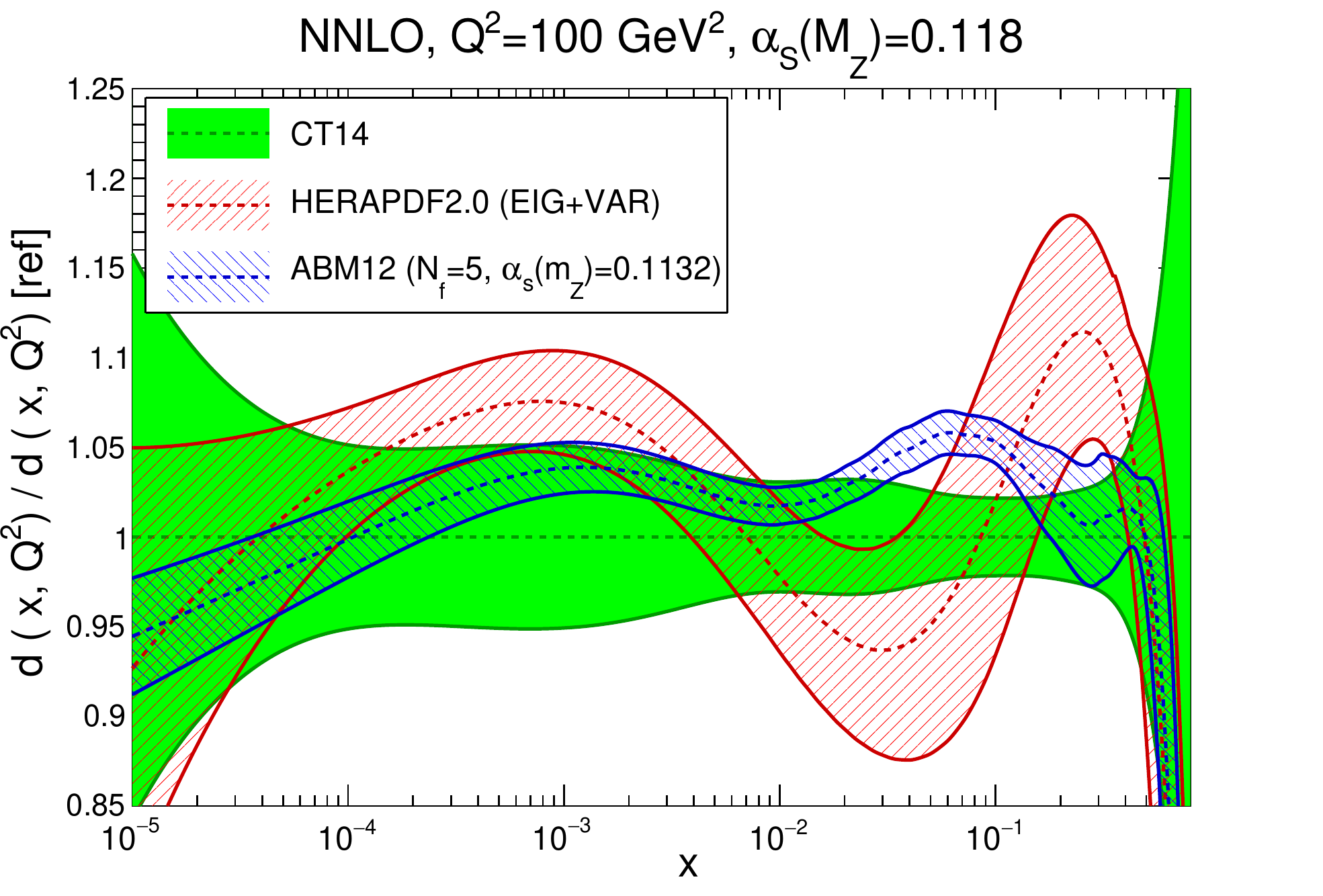}\\
   \includegraphics[scale=.37]{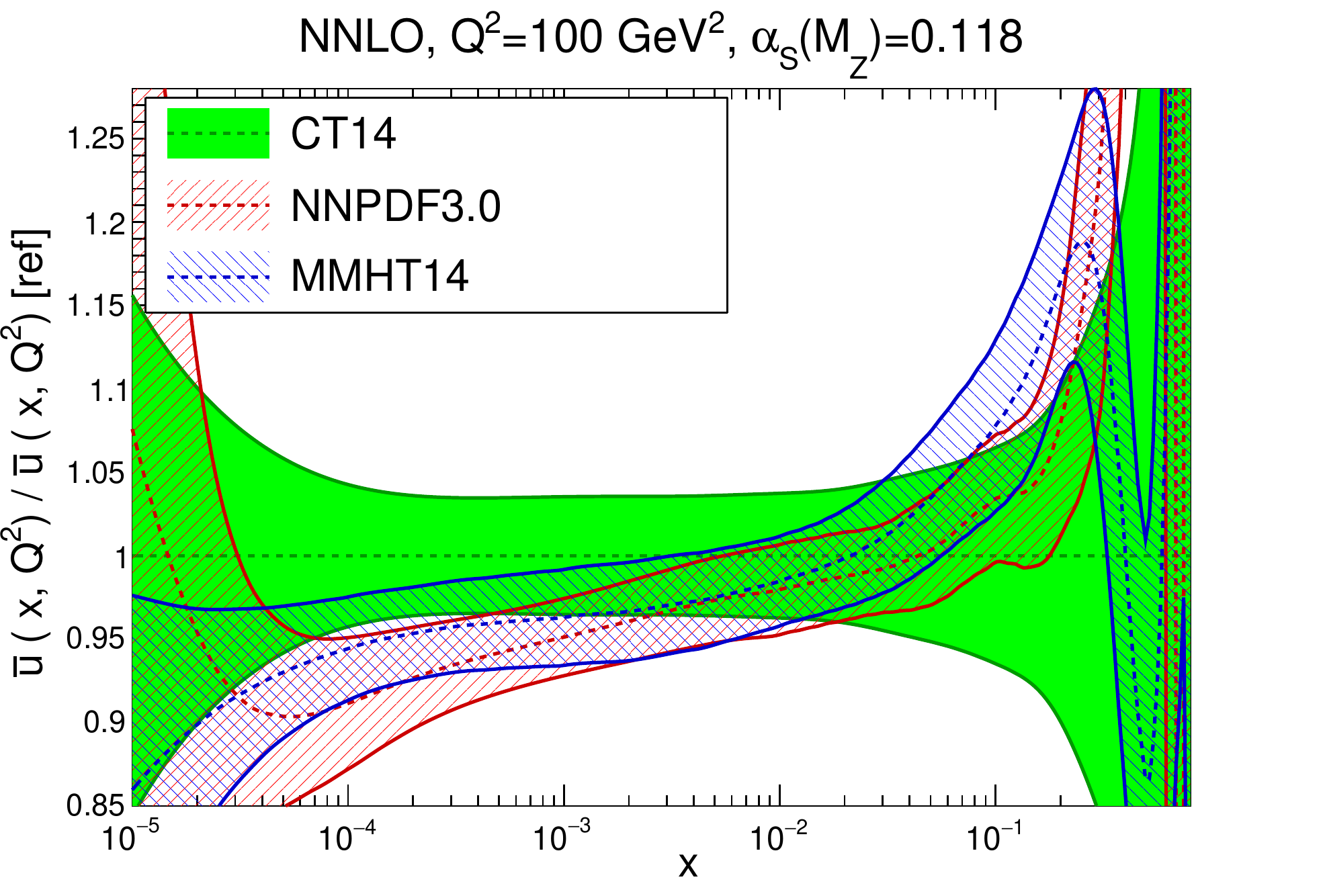}
   \includegraphics[scale=.37]{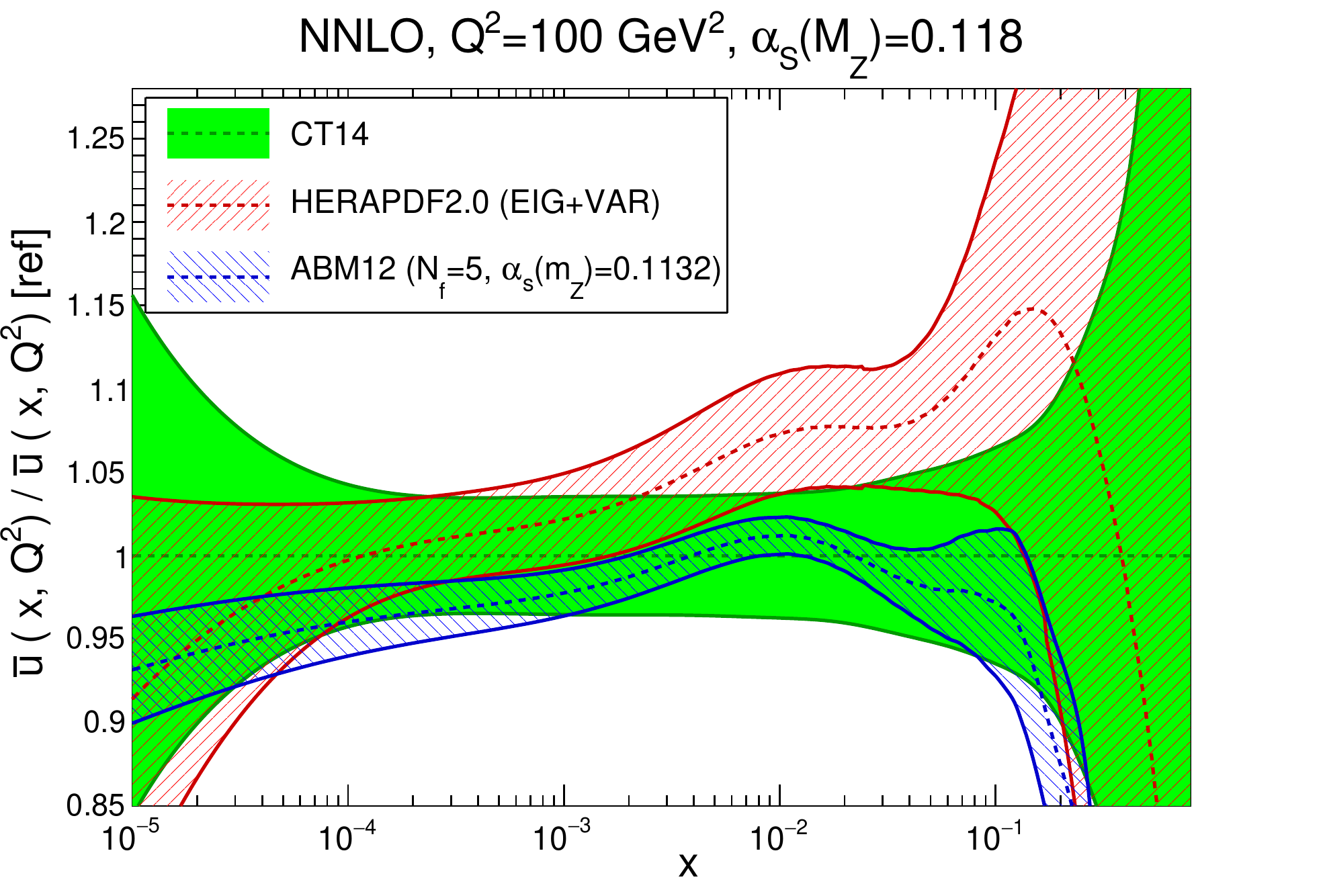}\\
    \caption{\small Same as Fig.~\ref{fig:gluon_up_quark} for the down quark
(upper plots) and the anti-up quark (lower plots).
  }  
\label{fig:down_upbar_quark}
\end{figure}

We  begin by comparing the
gluon and the up quark PDFs, with
the corresponding one-sigma uncertainty bands.
These are shown
in Fig.~\ref{fig:gluon_up_quark}
for $Q=100$ GeV, normalized to the
central value of the CT14 distribution.
Similar plots are shown for the $d$  and $\bar{u}$ quarks in Fig.~\ref{fig:down_upbar_quark}.
For each flavor, the left plots compare the three global sets among them,
while the right plots compare CT14 with the two sets based on reduced
datasets, HERAPDF2.0 and ABM12.

In general, the gluon 
distributions for CT14, MMHT2014 and NNPDF3.0 agree well, but not perfectly,  both in central value and in uncertainty, except at very low $x$ and at high $x$.
This level of agreement is reassuring, given the importance of the gluon
distribution in the dominant Higgs boson production mechanism.
Although each group uses similar, but not identical, data sets,
the fitting procedures and tolerances are not the same, and  for this reason, exact agreement is not expected.
As compared to the global fits,
the HERAPDF2.0 gluon PDF is somewhat larger
in the $x$ range from $10^{-4}$ to $10^{-2}$ and is
substantially smaller for
$x\ge 0.1$.
Similar considerations apply to ABM12, where their differences
as compared to the global sets are due only in part to the difference
$\alpha_s$ value, with the bulk of the differences due to the use
of a FFNS, as discussed in Sect.~\ref{sec:developementsTheory} above.

The overlap for the up quark distribution is not as good as for the
gluon distribution, so an envelope of 
CT14, MMHT2014 and NNPDF3.0 would be somewhat broader than the
individual predictions.
The HERAPDF2.0 and ABM12 up quark distribution are higher than the global sets,
especially in the $x$ range from $10^{-2}$ to 0.5, while HERAPDF2.0 agrees
well with CT14 for smaller values of $x$.
The CT14, MMHT2014 and NNPDF3.0 down quark distributions,
see Fig.~\ref{fig:down_upbar_quark}, are in
reasonable agreement with each other over most of the $x$
range, but have slightly different behaviors at high $x$.
The HERAPDF2.0 down quark has a different shape, though 
its uncertainty bands partially overlap with those of the others.
ABM12 agrees with the global fits except around $x\sim 0.07$, where
it is systematically larger.

Similar conclusions as those drawn
from the comparison of $u$ quark PDFs between
the three global fits.
apply to 
the $\bar{u}$ quark distributions.
Note the blow-up of PDF uncertainties at large $x$ due to the reduced
experimental information available in this region.
The
HERAPDF2.0 $\bar{u}$ quark distribution is generally higher than in the
global fits for $x>10^{-4}$  by a substantial amount (around 10\%).
Here ABM12 agrees reasonably with CT14, except for $x\ge 0.1$, where
is substantially smaller.
%

\begin{figure}[t]
  \centering
  \includegraphics[scale=.37]{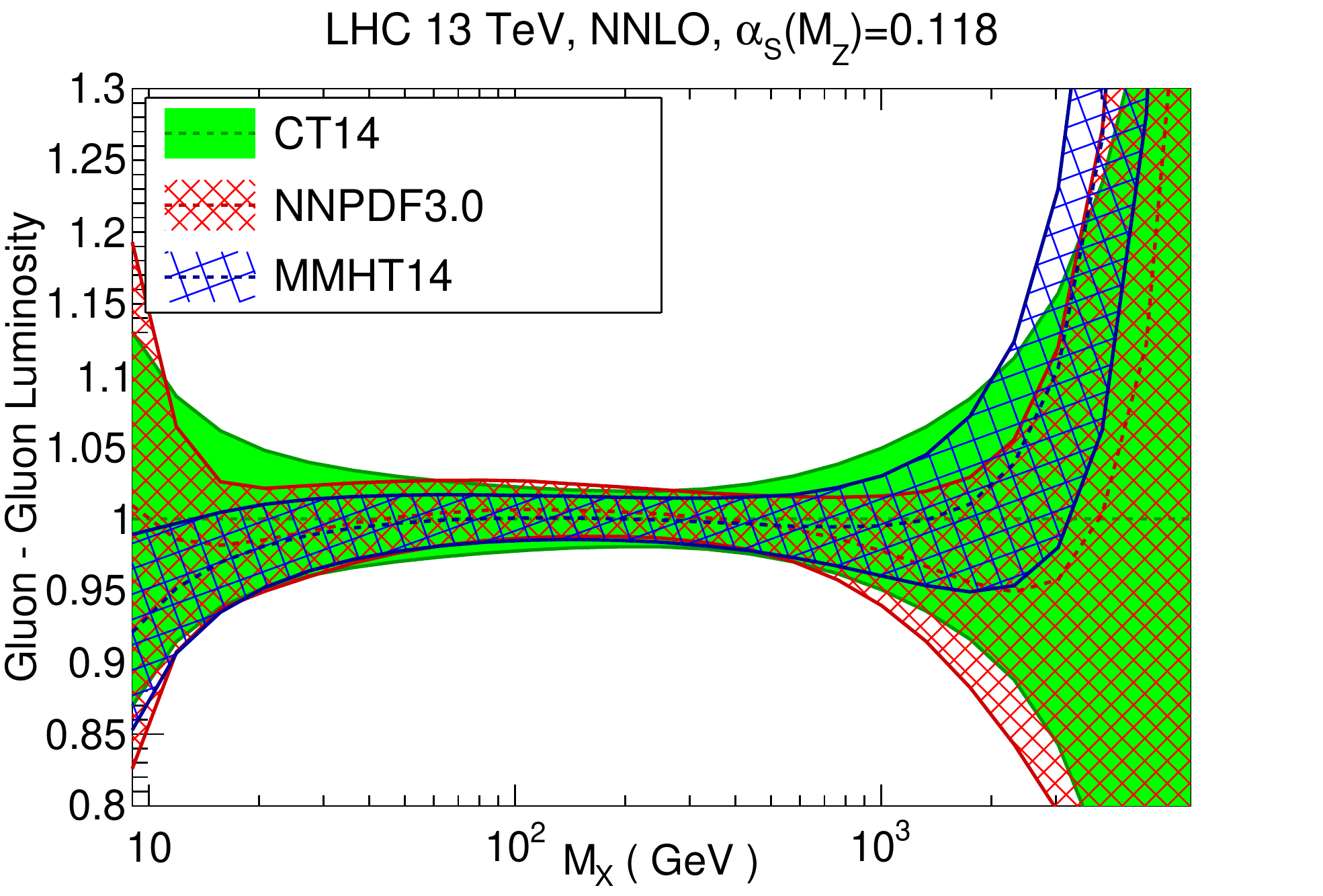}
  \includegraphics[scale=.37]{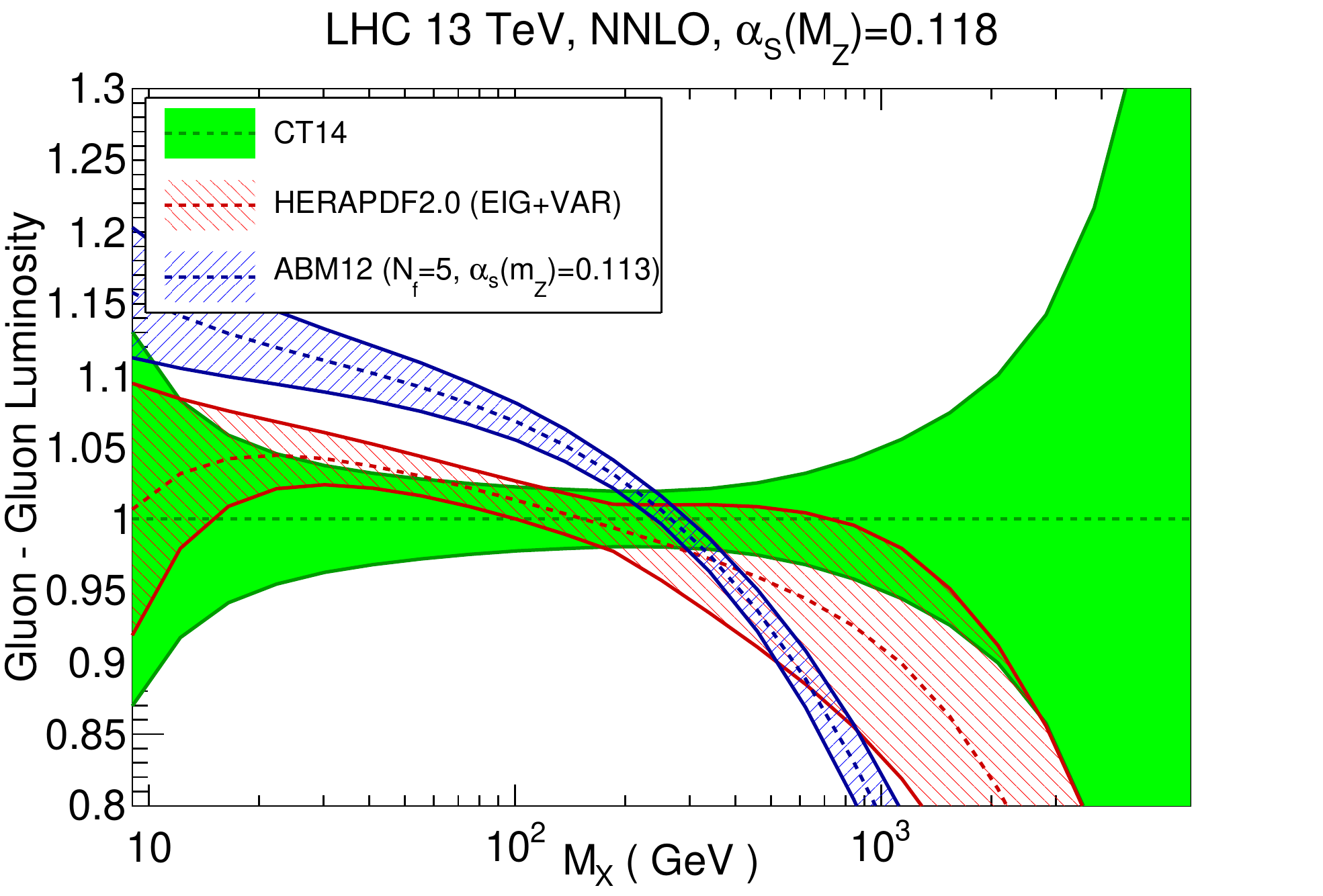}\\
  \includegraphics[scale=.37]{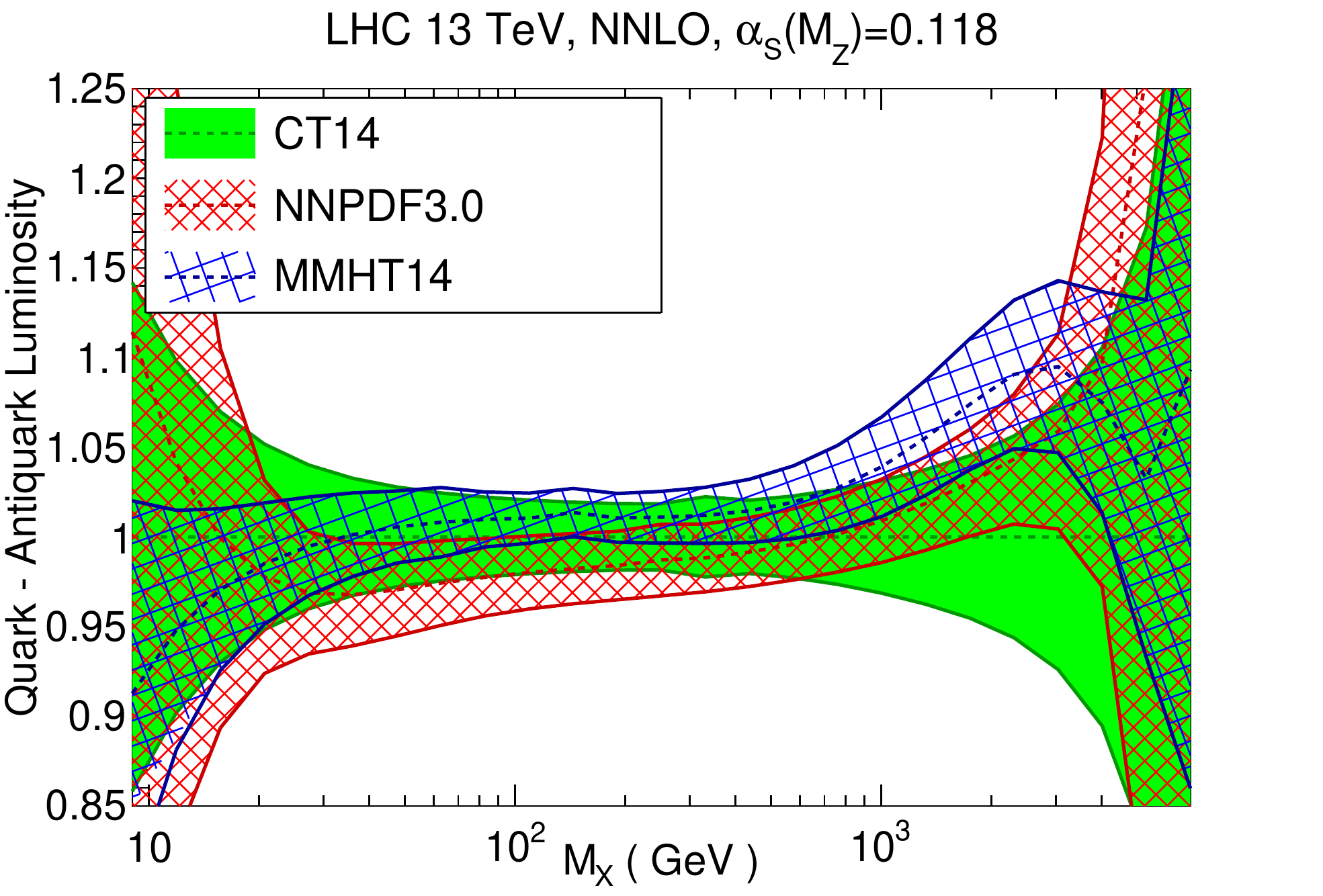}
  \includegraphics[scale=.37]{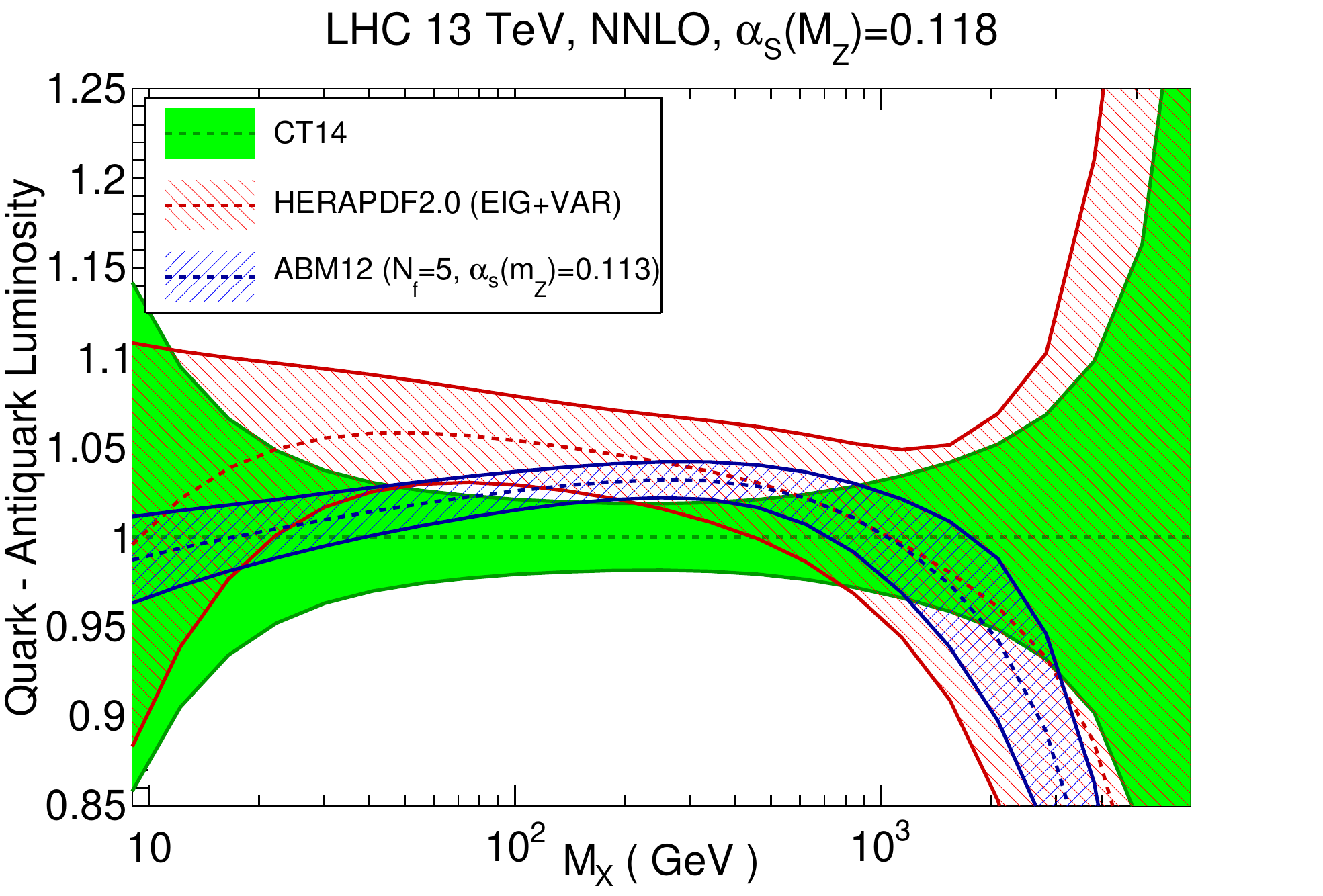}\\
  \caption{\small Comparison of the gluon-gluon (upper plots) and
    quark-antiquark (lower plots)
    PDF luminosities from the CT14, MMHT14 and NNNPDF3.0 NNLO sets
    (left plots) and from the NNPDF3.0, ABM12 and HERAPDF2.0 NNLO sets
    (right plots), for a center-of-mass energy of 13 TeV, as a function
  of the invariant mass of the final state $M_X$.}  
\label{fig:gg_qQ_lum}
\end{figure}

\begin{figure}[t]
  \centering
  \includegraphics[scale=.37]{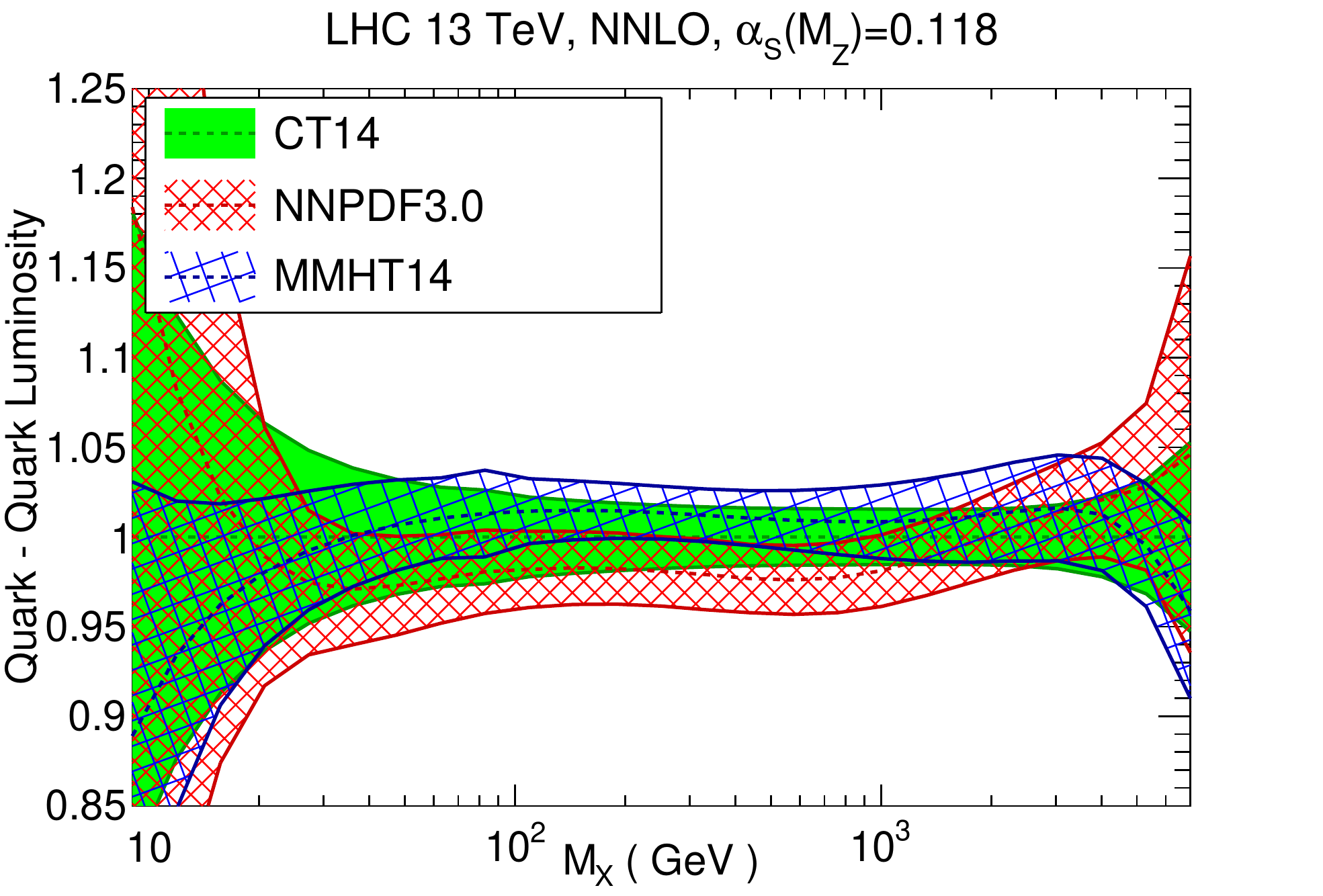}
  \includegraphics[scale=.37]{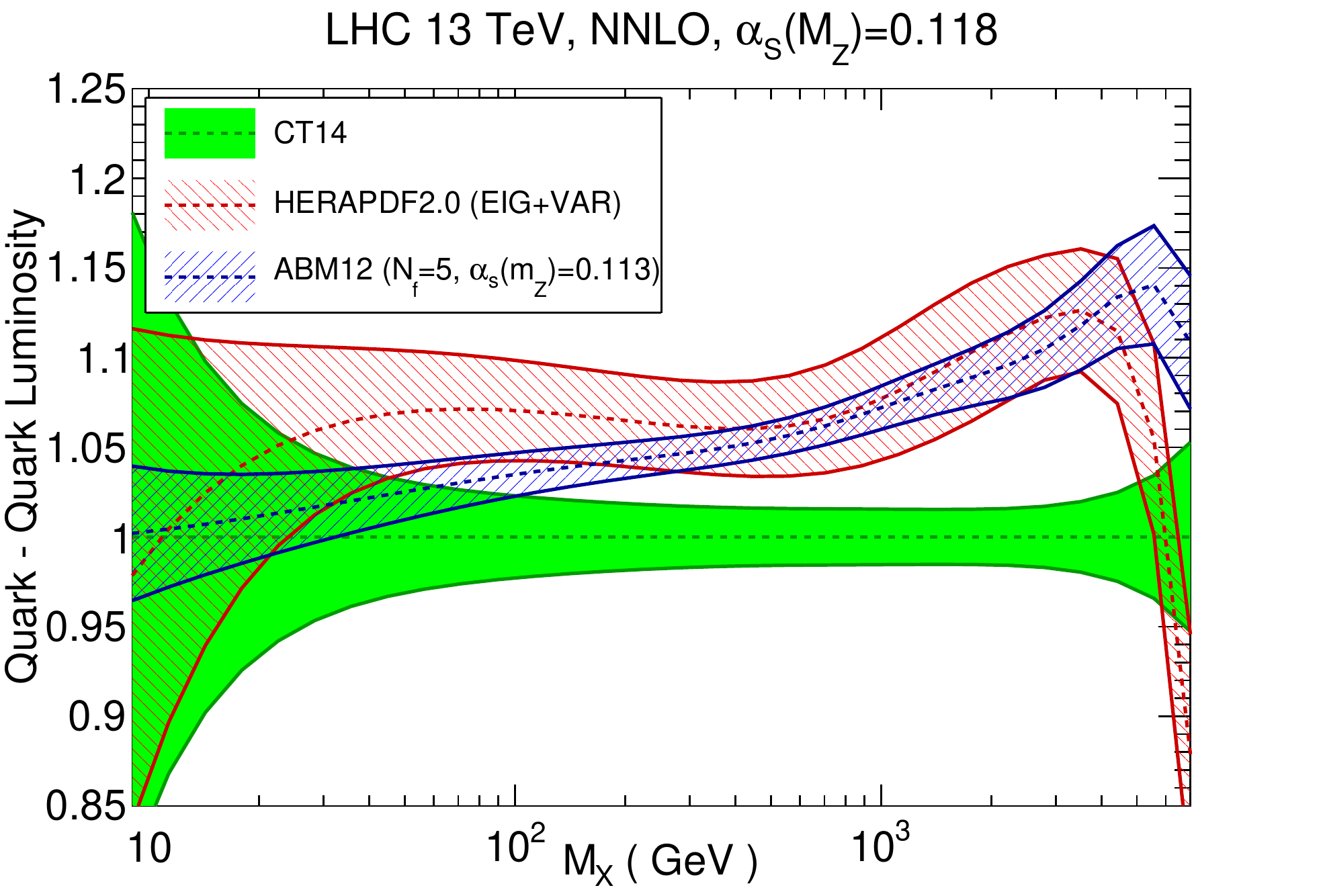}\\
  \includegraphics[scale=.37]{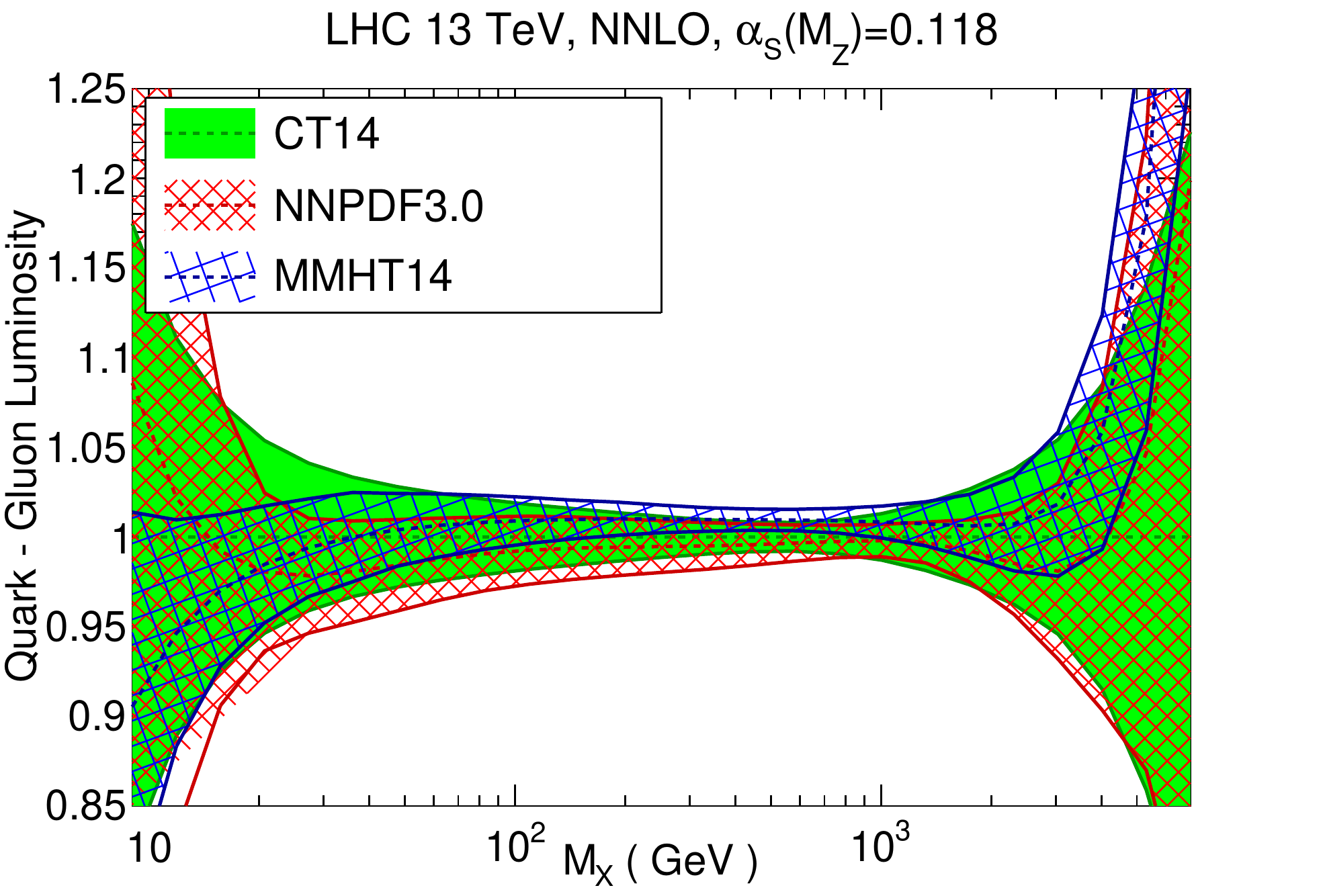}
  \includegraphics[scale=.37]{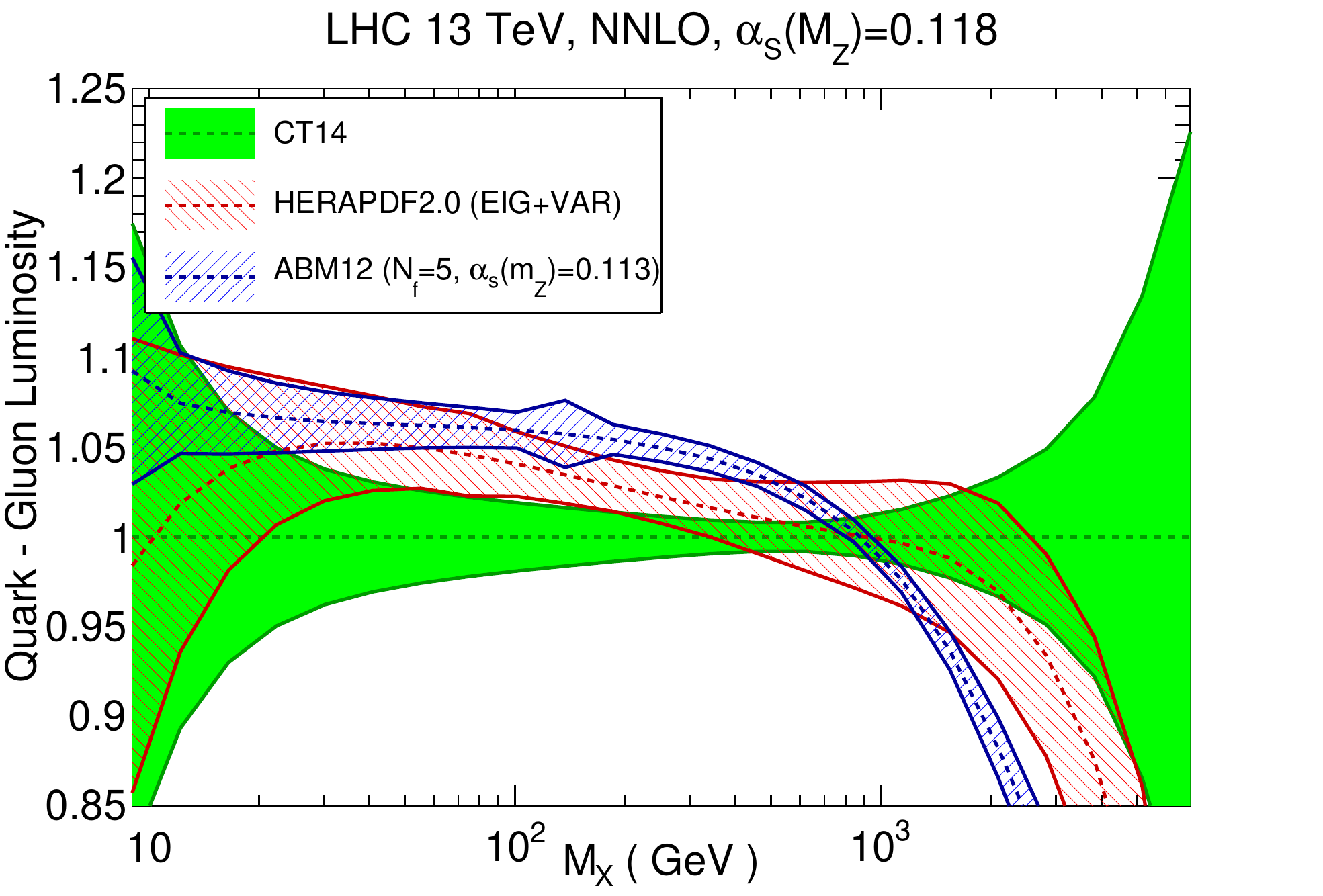}\\
  \caption{\small Same as Fig.~\ref{fig:gg_qQ_lum} for the quark-quark (upper plots) and the
  quark-gluon (lower plots) PDF luminosities.}  
\label{fig:qq_gq_lum}
\end{figure}

\subsection{PDF luminosities}

It is also instructive to examine the
parton-parton luminosities~\cite{Campbell:2006wx}, which are more closely related to the predictions for LHC cross-sections.
The gluon-gluon and
quark-antiquark luminosities, as a function of the invariant
mass of the final state $M_X$, for a center-of-mass energy of 13 TeV
are shown in Fig.~\ref{fig:gg_qQ_lum}, where we compare,
on the left, the three global fits, NNPDF3.0, CT14 and MMHT14,
and, on the right, CT14 with the fits based on reduced datasets,
HERAPDF2.0 and ABM12, using for the latter exactly the same settings as
in the PDF comparison plots.
All results are shown normalized to the central value of CT14, as before.
The corresponding comparison for the  quark-quark
and gluon-quark PDF luminosities is then shown in Fig.~\ref{fig:qq_gq_lum}.

The luminosity uncertainty ranges tend to blow-up at low
invariant masses ($M_{X} \le 50$ GeV) and high masses ($M_X \ge 500$
GeV for $gg$, $M_X \ge 1$ TeV for $q\bar{q}$ and $M_X \ge 5$ TeV for $qq$),
that is, in the 
regions that are relatively unconstrained in current global PDF fits.
The region of intermediate final-state invariant masses can be thought
of as the domain for precision physics measurements, where the PDF
luminosity uncertainties are less than 5\% (at 68\% CL).
There is good
agreement among the predictions of CT14, MMHT2014 and NNPDF3.0 for the
$gg$ PDF luminosities for this mass range, and in particular for the
production of a Higgs boson at $m_h=125$ GeV.  

This is an improvement over
the situation with the previous generation of PDFs (CT10, MSTW2008 and
NNPDF2.3), where, as mentioned previously, 
the total PDF luminosity uncertainty in the Higgs
mass range was more than a factor of 2 larger  
the uncertainty for any individual PDFs, as a result of differences in
central PDFs.
Clearly, this leads to a
corresponding improvement in agreement between predictions for the
cross section $\sigma(gg\to h)$ computed using these PDF sets.
Indeed, as compared to NNPDF2.3, the new  NNPDF3.0 prediction decreases by about 2\%, the CT14 prediction  increases by 1.1\%, compared to CT10 while
the MMHT14 predictions, as compared to MSTW08, decrease by
a small amount  (roughly 0.5\%).
As discussed in Sect.~\ref{sec:currentPDFsets}, it is difficult 
to pinpoint a single reason for this improvement in agreement, which
most likely arises from a combination of methodological advancements
and new experimental constraints in the global fits.

Still considering the $gg$ luminosities, HERAPDF2.0 is in good agreement
with the global fits for $M_X\le 0.5$ TeV, becoming rather softer above
this mass, while ABM12 shows strong differences, with a much harder
luminosity for $M_X\le 0.5$, and a much softer one above this $M_X$ value.
For the $q\bar{q}$ luminosity, there is a reasonable agreement
between the three global fits; note the blow-up of PDF uncertainties
at small and large values of $M_X$ because of absence of data.
HERAPDF2.0 best-fit luminosity for $q\bar q$ is above CT14 by about 5\% in the region of intermediate $M_X$,
with better agreement at smaller and larger masses, but markedly less certain. 
ABM12 is consistent with the global fits except for $M_X \ge $ 2 TeV when
it becomes rather softer.

Turning to the  quark-quark
and gluon-quark PDF luminosities shown in Fig.~\ref{fig:qq_gq_lum},
we see that for the global fits we get consistent results within uncertainties,
though the agreement is not quite as good as for the gluon-gluon
luminosity, especially in the region between 100~GeV
and 1~TeV.
The luminosities for HERAPDF2.0 and ABM12 in the $qq$ case are
harder than those of the global fits, by around 5\% at low
masses to up 15\% in the TeV region, with important phenomenological
implications.
For the $qg$ luminosity, there is good agreement for the global fits
(similar to the $gg$ luminosity) and slightly worse for the
PDF sets based on reduced datasets.

%% file: sec-prescription.tex
\section{Constructing the PDF4LHC15 combination}
\label{sec:prescription}

As discussed in Sect.~\ref{sec:comparisons}, 
we are now in the rather satisfactory position where differences
between PDF sets are either better understood or much reduced.
However, there is still the question of how best to combine PDF sets even if 
they are essentially compatible. 
In this section we motivate and present the updated PDF4LHC prescription for
the evaluation of PDF  uncertainties at the
LHC Run II,  discuss the
general conditions that  PDF sets should satisfy in order to be
included in the combination and its future updates, and list the
PDF sets which will enter the current prescription.
We then describe how the construction of the combined
sets based on the Monte Carlo method.

\subsection{Usage of PDF sets and their combinations}
\label{sec:combination}

We would like first to state that there are three main cases in which
PDFs are used in LHC applications:
\begin{enumerate} 
\item Assessment of the {\it total uncertainty on a
  cross section} based on
the available knowledge of PDFs, {\it e.g.}, when
  computing the cross section for a process that has not been measured
  yet (such as supersymmetric particle production cross-sections),
  or for estimating acceptance corrections on a given observable.
  This is
  also the case of the measurements that aim to verify overall,
  but not detailed, consistency with Standard Model expectations,
  such as when comparing theory with Higgs measurements.
\item Assessment of the accuracy of the {\it PDF sets themselves} 
  or of related Standard Model parameters, typically done by comparing
  theoretical predictions using individual PDF sets
  to the  most precise data available.
\item Input to the
  {\it Monte Carlo event generators} used to generate large MC
  samples for LHC data analysis.
\end{enumerate}

In the second case, it is important to always
use the PDF sets from the individual groups
for predictions.
This is especially true in comparisons 
that involve PDF-sensitive measurements,
providing information about the PDFs for individual fits. As a rule of thumb,
comparisons between  QCD or electroweak calculations and
unfolded data for Standard Model production processes should be done with 
individual PDF sets.
Similar considerations hold for the third case, since
MC event generators require to be carefully tuned to experimental
data using as input specific PDF sets.

However, in the first case above
a robust estimate of the PDF uncertainty must accommodate the 
fact that the individual PDF sets are not identical 
either in their central values or in their
uncertainties.
Consequently, an uncertainty based 
on the totality of available PDF sets must be estimated. 
Besides acceptance calculations, a contemporary example of this
situation is the extraction of the Higgs couplings from LHC data, in which
a robust estimate of the overall theoretical uncertainty
is essential for probing the nature of the Higgs boson
and for identifying possible
deviations from the Standard Model expectations.
In  searches for New Physics particles, such
as supersymmetric partners or new heavy $W'$ and $Z'$ bosons, 
estimates of the combined uncertainty are also necessary 
in order to derive robust exclusion limits,  or to be able to claim
a discovery.

For the applications of type 1, instead of
using an envelope,
the 2015 PDF4LHC recommendation proposes
to take a statistical
combination, to be described below,
of those PDF sets that satisfy a set of
compatibility requirements.
While these requirements may evolve with time, for the current combination
we select the individual PDF sets that satisfy the following properties:
\begin{enumerate}
\item The PDF sets to be combined should be {\it based on a global dataset},
  including
a large number of datasets of diverse types (deep-inelastic scattering, 
vector boson  and jet production, ...) 
from
fixed-target  and colliders experiments (HERA, LHC, Tevatron). 
\item Theoretical hard cross sections for DIS and hadron
  collider processes
  should be evaluated up to {\it two 
QCD loops in $\alpha_s$},
  in a {\it general-mass variable-flavor number scheme} with up to
  $n_f^{\rm max}=5$ active quark
flavors.\footnote{The combination can also be performed
  for sets with $n_f^{\rm max}=3$ or $n_f^{\rm max}=4$, provided
  that these are computed using the $n_f^{\rm max}=5$ sets as boundary
conditions for the evolution, see Sect.~\ref{sec:guidelines}.}
Evolution of $\alpha_s$ and PDFs should be performed up to
three loops, using public
codes such as {\tt HOPPET}~\cite{Salam:2008qg}
or
{\tt QCDNUM}~\cite{Botje:2010ay}, or a code benchmarked to
these.

\item The central value of
  {\it $\alpha_s(m_Z^2)$ should be fixed at an agreed common value},
  consistent with the PDG world-average~\cite{Agashe:2014kda}.
  This value is currently chosen to be $\alpha_s(m_Z^2) =0.118$ at both NLO and NNLO.\footnote{There are  arguments that a slightly higher value should be taken at NLO than at NNLO, but the majority request from PDF users is to have a common value for both orders.}
For the computation of $\alpha_s$ uncertainties, two
additional PDF members
corresponding to  agreed upper and lower values
of $\alpha_s(m_Z^2)$ should also be provided.
This uncertainty on $\alpha_s(m_Z^2)$ is currently assumed to be
$\delta \alpha_s=0.0015$, again the same at NLO and NNLO.

The input values of $m_c$ and $m_b$ should be compatible with their
world-average values; either pole or $\overline{\rm MS}$ masses are accepted.

\item
{\it All known experimental and procedural sources of uncertainty should be
properly accounted for}.
Specifically, it is now recognized that
the PDF uncertainty receives several contributions of comparable
importance: the measurement
uncertainty propagated from the experimental data,  uncertainties
associated with incompatibility of the fitted experiments, procedural
uncertainties such as those
related to the functional form of PDFs, the
handling of systematic errors, etc.
Sets entering the combination must
account for these through suitable methods, such as 
separate estimates for additional model and
parametrization components of the
PDF uncertainty~\cite{Abramowicz:2015mha},
tolerance~\cite{Harland-Lang:2014zoa,Dulat:2015mca}, or closure
tests~\cite{Ball:2014uwa}.  
\end{enumerate}

Following the needs of precision physics at the LHC,
future updates of the PDF4LHC recommendations might be based on a different
set of conditions.
For instance, in addition to the above PDF sets, one
might be required to provide fits in a range of $m_c$ and $m_b$
values, or to provide 
direct evidence of the ability to describe, reasonably accurately, a
wide variety of different data types that constrain PDFs of different
flavour and in different kinematic regions.
In the future, with
the progress in combination techniques, it might be possible to relax
the first requirement, and also include in the combination PDF sets
based on datasets of a different size by using suitable weighted
averaging techniques --- note that such a weighted averaging would have
to account for the fact that different data affect different PDFs and
kinematic regions, and thus the weights chosen cannot  be the same
for all PDF flavours.

The existing PDF sets which satisfy all of these requirements at
present have been identified as {\bf CT14, MMHT2014} and {\bf NNPDF3.0};
no other publicly available PDF sets currently appears to satisfy all
conditions.

\subsection{Statistical combination of PDF sets}
\label{sec:statistical}

Currently, two different representations of PDF uncertainties are
being used~\cite{Alekhin:2011sk}: 
the Monte Carlo representation~\cite{DelDebbio:2004qj,
  DelDebbio:2007ee,Ball:2008by}, in which the probability
distribution of PDFs is given as an ensemble of replicas, whose mean
and standard deviation provide respectively central value and
uncertainty, and the Hessian representation~\cite{Pumplin:2001ct}, 
in which a central PDF is given,
along with error sets, each of which corresponds to an eigenvector of the
covariance matrix in parameter space.

In Ref.~\cite{Watt:2012tq}, a comparison of the Hessian and the Monte
Carlo  methods was made within the MSTW framework, showing that
they provide compatible representations of PDF uncertainties, and in
particular lead to the same uncertainties when used to determine PDFs
from known pseudo-data.
Also, a way of obtaining a Monte Carlo
representation from starting Hessian representation
was presented,  and the accuracy of the Hessian to Monte Carlo
conversion was explicitly demonstrated.
As will be discussed in the next section, recently two methods
to perform the inverse conversion, namely transforming MC sets
into a Hessian representation, have also been
developed~\cite{Gao:2013bia,Carrazza:2015aoa}.

These developments make possible a statistical
combination of different PDF sets and their predictions,
as originally 
outlined in Ref.~\cite{Forte:2010dt}: if different PDF sets are
assumed to be equally likely representations of an underlying PDF
probability distribution, they can be combined by simply taking their
unweighted average. This in turn can be done simply by generating
equal numbers of Monte Carlo replicas from each input PDF set, 
and then merging these replica sets of equal sizes.
For the Hessian PDF sets
such as CT or MMHT, the Monte-Carlo replicas are generated by sampling along 
each eigenvector direction, assuming a Gaussian distribution.
Alternatively, a weighted average would correspond to taking different number
of replicas from the various sets sets entering the combination.
 First examples of this application for LHC cross sections were given in 
 Ref.~\cite{Forte:2013wc}.
 A detailed overview of how the MC  method
 has been used to construct the PDF4LHC 2015 combined sets
 will be presented in the next section.

Clearly, results obtained in this way are
less conservative than those obtained from an envelope:
they
correspond to the assumption that different PDF determinations are
statistically distributed  instances  of an underlying probability
distribution, rather than instances of a probability
distribution affected  by unknown underlying sources of
systematics.
Such a combination method appears therefore to be
adequate when results are compatible, or differences are understood,
as is the case now.


%% file: sec-frameworks.tex
\section{Implementation and delivery of the PDF4LHC15 PDFs}
\label{sec:frameworks}

In this section we discuss the technical construction of the
 new
PDF4LHC prescription, presented in Sect.~\ref{sec:prescription}, and
the delivery of the
combined PDFs based on it.
Readers interested in the more practical question of how
to use the PDF4LHC15 combination in their specific analysis
can jump directly to Sect.~\ref{sec:recommendations}.

Unlike  the previous PDF4LHC prescription, in which the task of
combining the PDF uncertainties was left to the user, with the current 
PDF4LHC prescription, several pre-packaged PDF sets that already combine the
uncertainties will be
constructed and delivered.
The combined sets are based on a statistical
combination of the CT14, MMHT2014 and NNPDF3.0 PDF sets, as discussed
in Sect.~\ref{sec:statistical}.
This combination leads to a ``prior'' Monte Carlo set with 
$N_{\rm rep}=900$ replicas, to be referred to as either {\tt MC900} or
{\tt PDF4LHC15\_prior} in the following discussion. 
Such a large replica set would be
unmanageable; however, various methods have been proposed recently to
deal with it, which we collectively refer to as {\it reduction
  methods}.
Usage of these methods will allow for a compact delivery of
the combined PDF sets.

The idea of producing a unified PDF set by combining various individual sets
was first suggested in Ref.~\cite{Gao:2013bia},
based on the idea of refitting a suitable functional form to a combined
set of Monte Carlo replicas, thereby leading to a representation
of the starting Monte Carlo probability distribution in terms of  Hessian
error sets in parameter space (META-PDFs).
This
then realizes the dual goal of producing a combination, and then
reducing the number of PDF error members to a manageable size.
More recently, other reduction methods
were suggested, with the similar
goal of turning the starting combined Monte Carlo
sample into a more manageable representation.
The Monte Carlo
compression method~\cite{Carrazza:2015hva} selects a subset of the original
replica sample which reproduces its statistical features with minimal
information loss, thereby keeping the MC representation, but
with a smaller number of replicas (CMC-PDFs).
The  Monte Carlo to
Hessian conversion method~\cite{Carrazza:2015aoa} 
turns a set of Monte Carlo replicas into Hessian
error sets by representing the covariance matrix in the space of PDFs
on a discrete set of points in $x$ as a linear combination of the
replicas, by means of a Singular Value Decomposition followed by
Principal Component Analysis (MCH-PDFs).

In this section, 
we will first present the Monte Carlo combination of the PDF sets and, in particular,
determine the number of replicas that is necessary in order to
achieve a faithful description of the combined set.
We will then
review each of the three reduction and delivery methods: CMC (Monte
Carlo), META (Hessian) and MCH (Hessian), and, in each case, identify
the size of the corresponding reduced PDF
sets that optimizes the specific features of the method.
These three sets will be adopted for delivery of the combined
PDF4LHC15 PDF sets.
Finally, we will compare and benchmark the PDF sets obtained with the three
different reduction strategies, both at
the level of parton distributions and of LHC observables.

\subsection{The Monte Carlo combination of PDF sets}
\label{sec:prior}

The first step in the construction of a Monte Carlo statistical
combination~\cite{Forte:2010dt,Watt:2012tq,Forte:2013wc,Watt:2013,Gao:2013bia,Carrazza:2015hva}
is the transformation 
of the Hessian PDF sets into
Monte Carlo PDF sets using the Watt-Thorne method~\cite{Watt:2012tq}.
Once all sets have been turned into a Monte Carlo representation,
the combination is simply built by adding together an equal number of
replicas from each set.
The generation of the Monte Carlo replicas
is based on the symmetric Thorne-Watt
formula, see Eq.~(22) in Ref.~\cite{Buckley:2014ana},
as implemented in the {\tt LHAPDF6} code~\cite{Buckley:2014ana}.
This has been cross-checked with the independent code
of the {\tt MP4LHC} package, finding excellent agreement.

\begin{figure}[t]
  \centering
  \includegraphics[width=.48\textwidth]{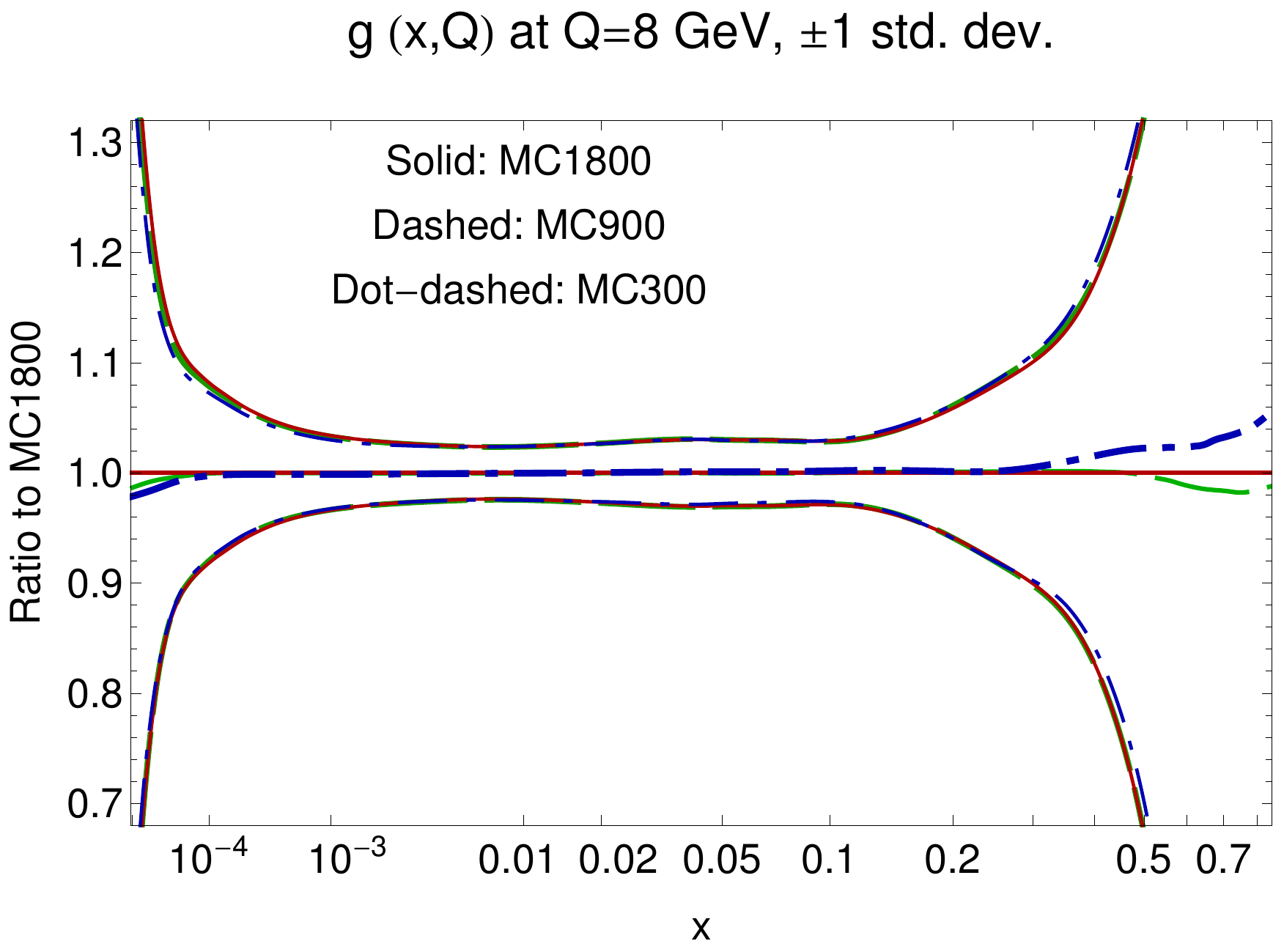}
  \includegraphics[width=.48\textwidth]{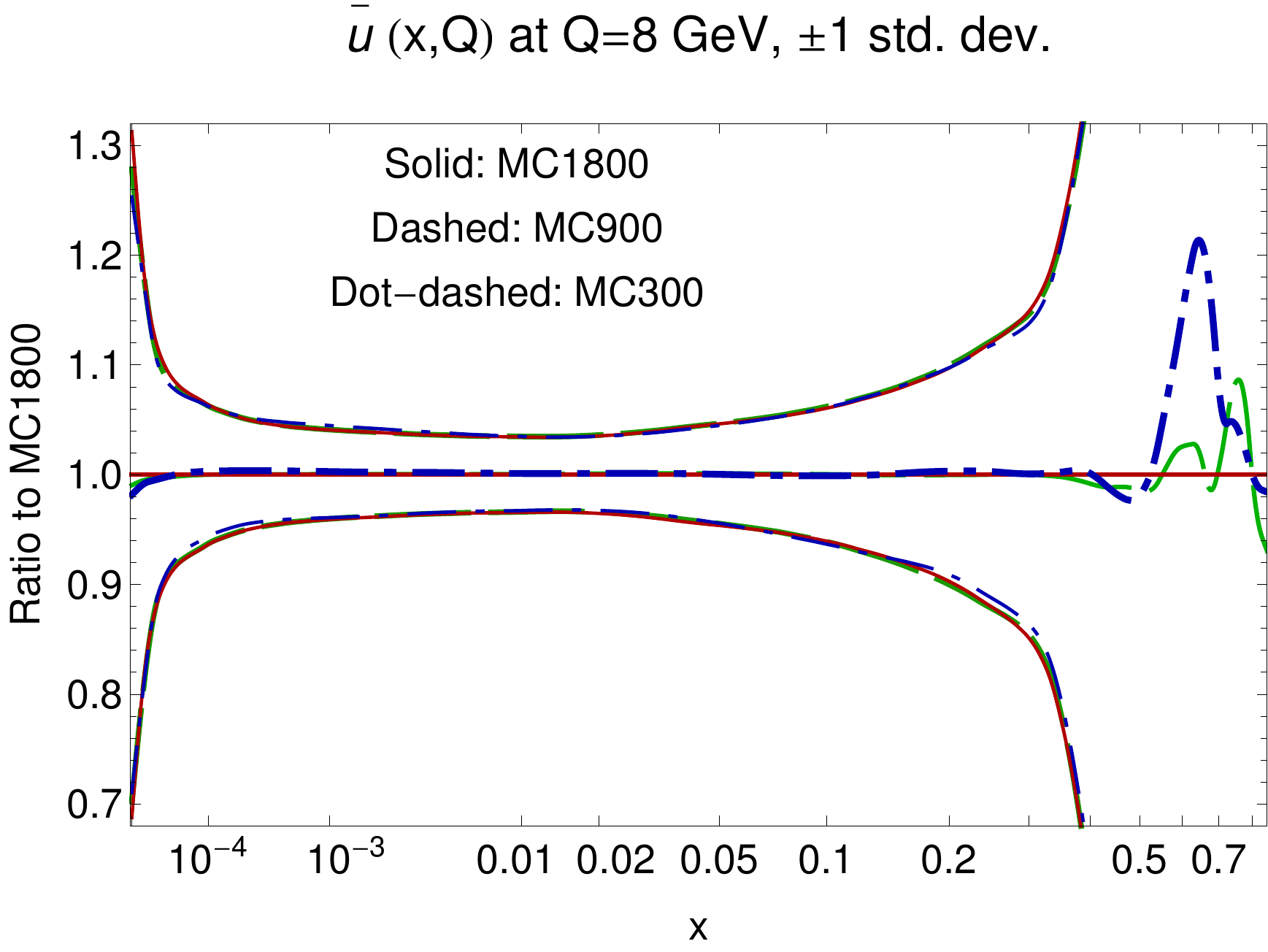}
  \includegraphics[width=.48\textwidth]{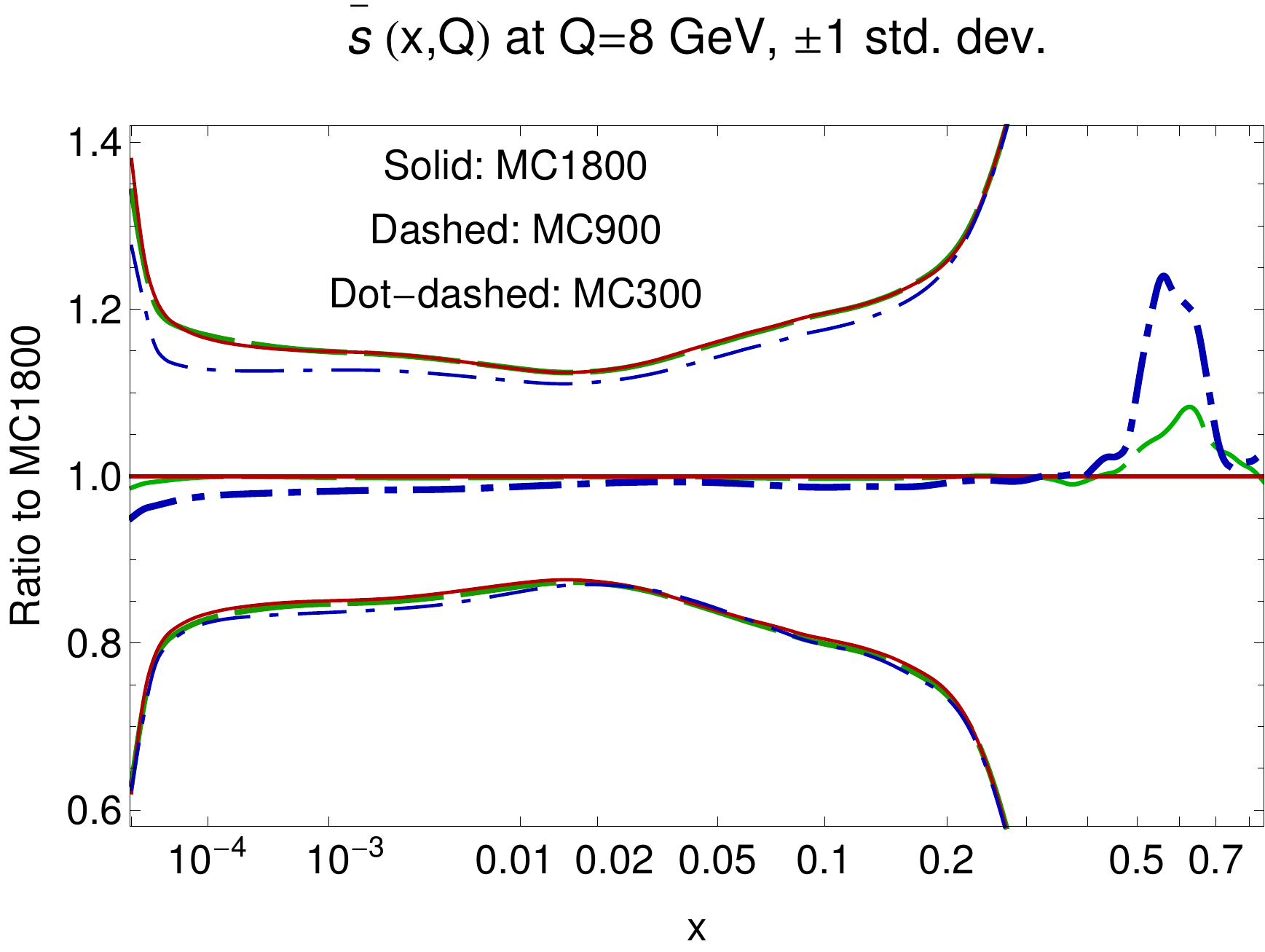}
 \caption{\small Comparison of central values and uncertainties for the MC combination of CT14, MMHT14 and NNPDF3.0 for different values
   of $N_{\rm rep}$, 300, 600 and 900, denoted by MC300, MC900 and MC1800 respectively. 
}  
\label{fig:mcrepdef}
\end{figure}

In Fig.~\ref{fig:mcrepdef} we show the
 comparison
    of the PDFs from the Monte Carlo combination of
    CT14, MMHT14 and NNPDF3.0 for a different
    number of MC replicas, $N_{\rm rep}=300$, 900 and 1800,
    referred in the following as MC300, MC900 and MC1800.
    The error bands correspond to one standard deviation from the mean
    value. 
    We see that, while between  $N_{\rm rep}=300$ and $N_{\rm
      rep}=900$ some visible, albeit small, changes are observed,
    results stabilize if a yet larger
    number of replicas is used.
    We conclude that $N_{\rm rep}=900$ (i.e. $300$ replicas from
    each of three individual sets) is an adequate number
    for the combination,
    and this is the value that we will adopt
    henceforth as a default for our prior.
    Therefore, the prior combined set {\tt PDF4LHC15\_prior} coincides
    with the MC900 set.

\begin{figure}[t]
  \centering
 \includegraphics[width=.48\textwidth]{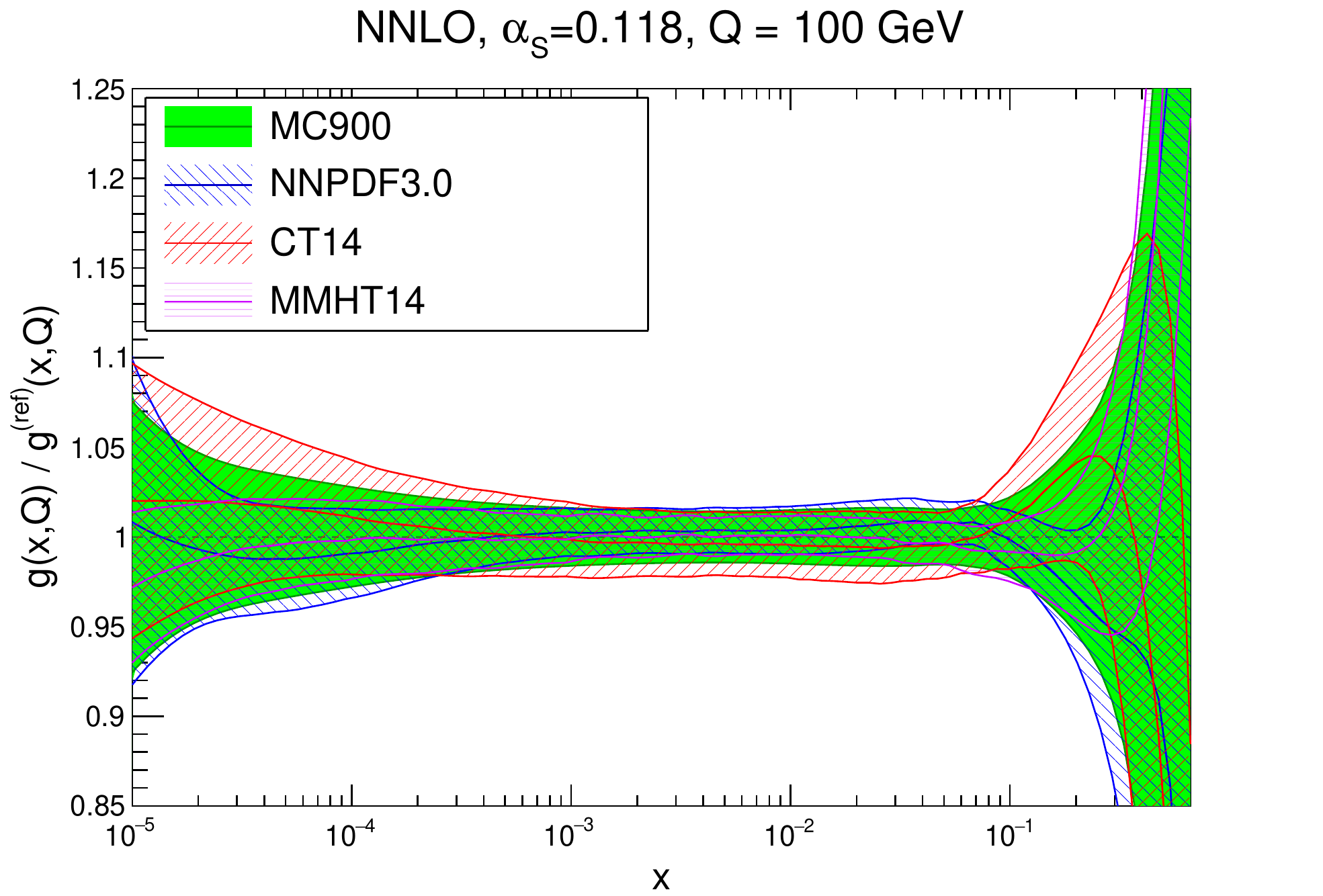}
  \includegraphics[width=.48\textwidth]{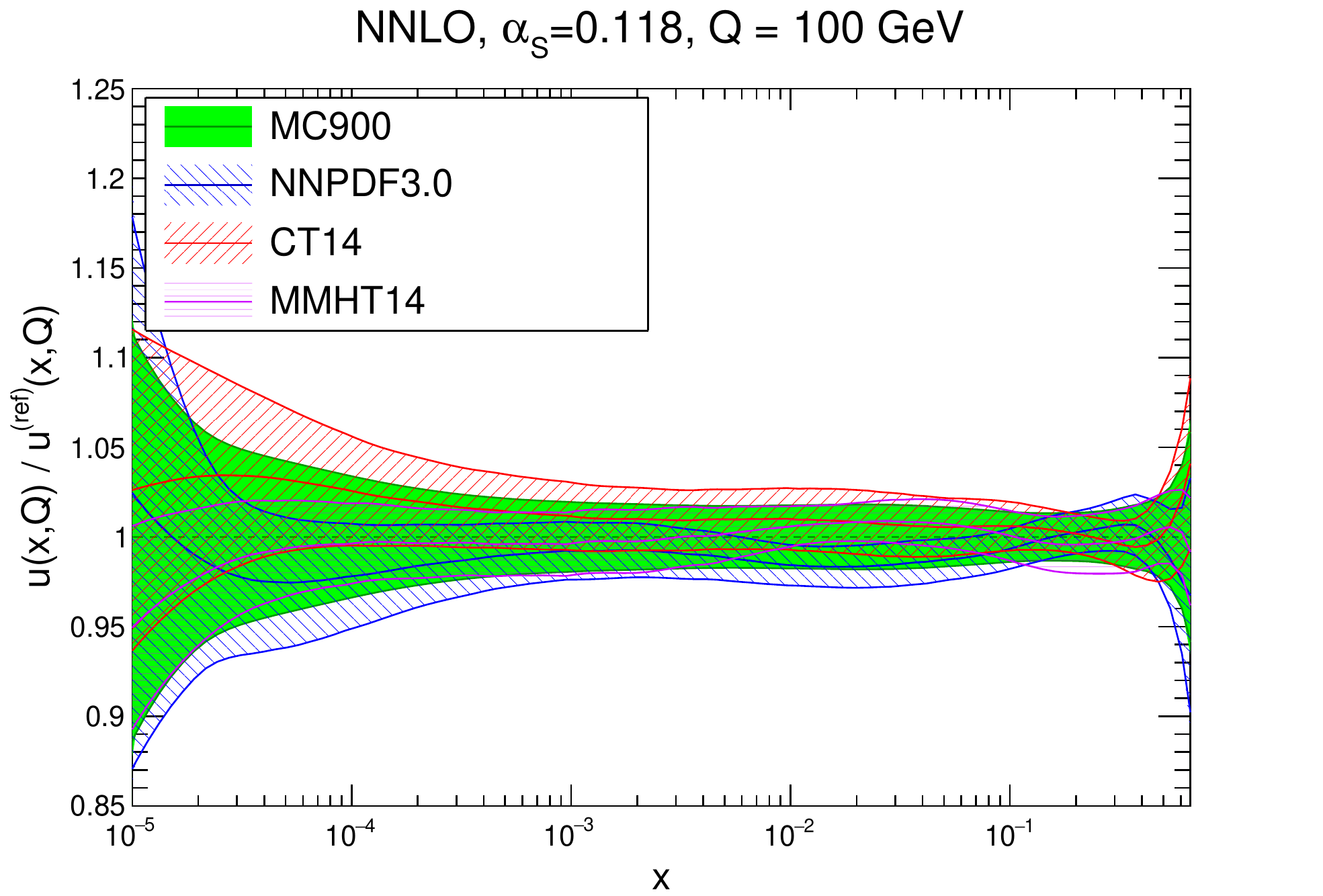}
  \includegraphics[width=.48\textwidth]{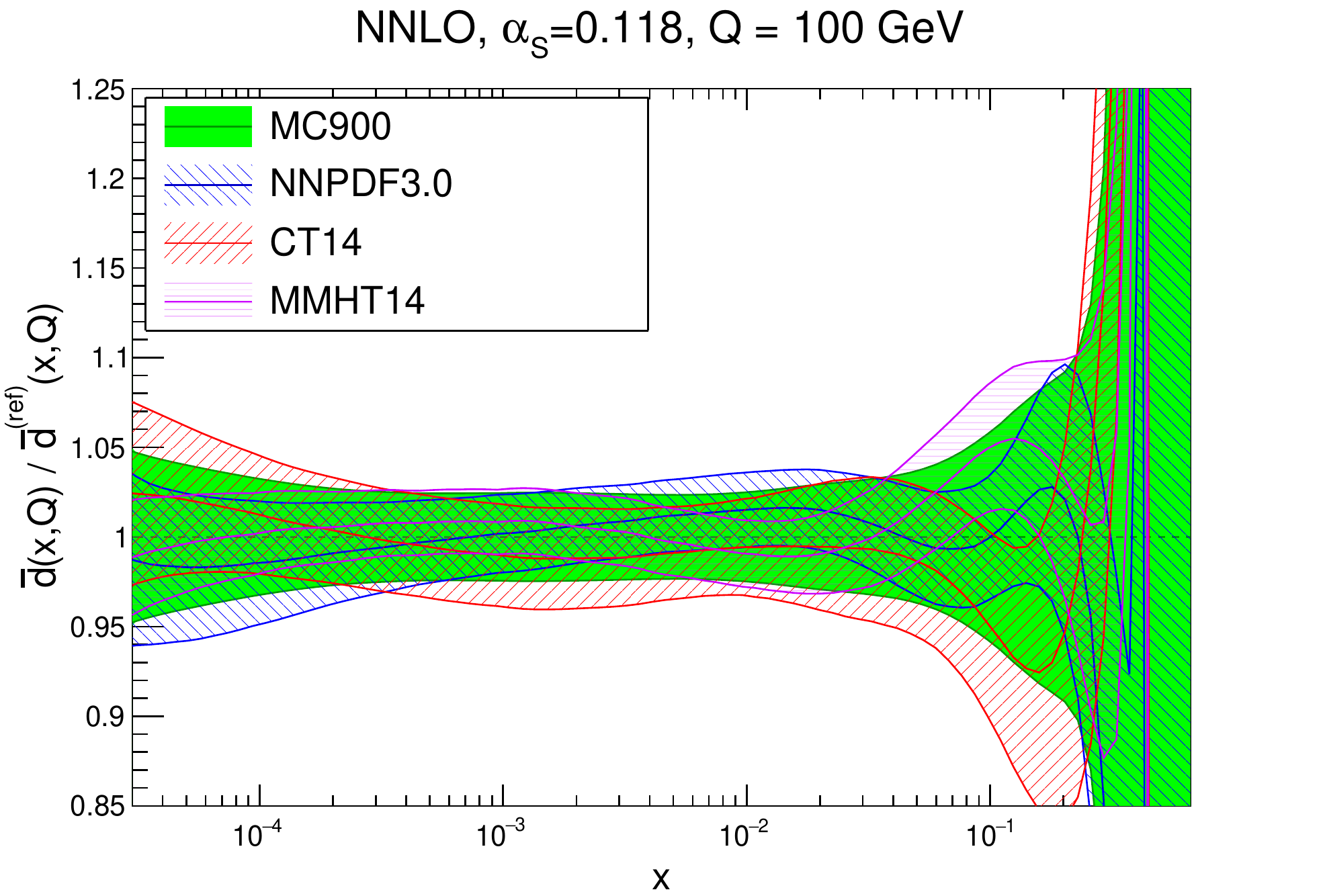}
   \includegraphics[width=.48\textwidth]{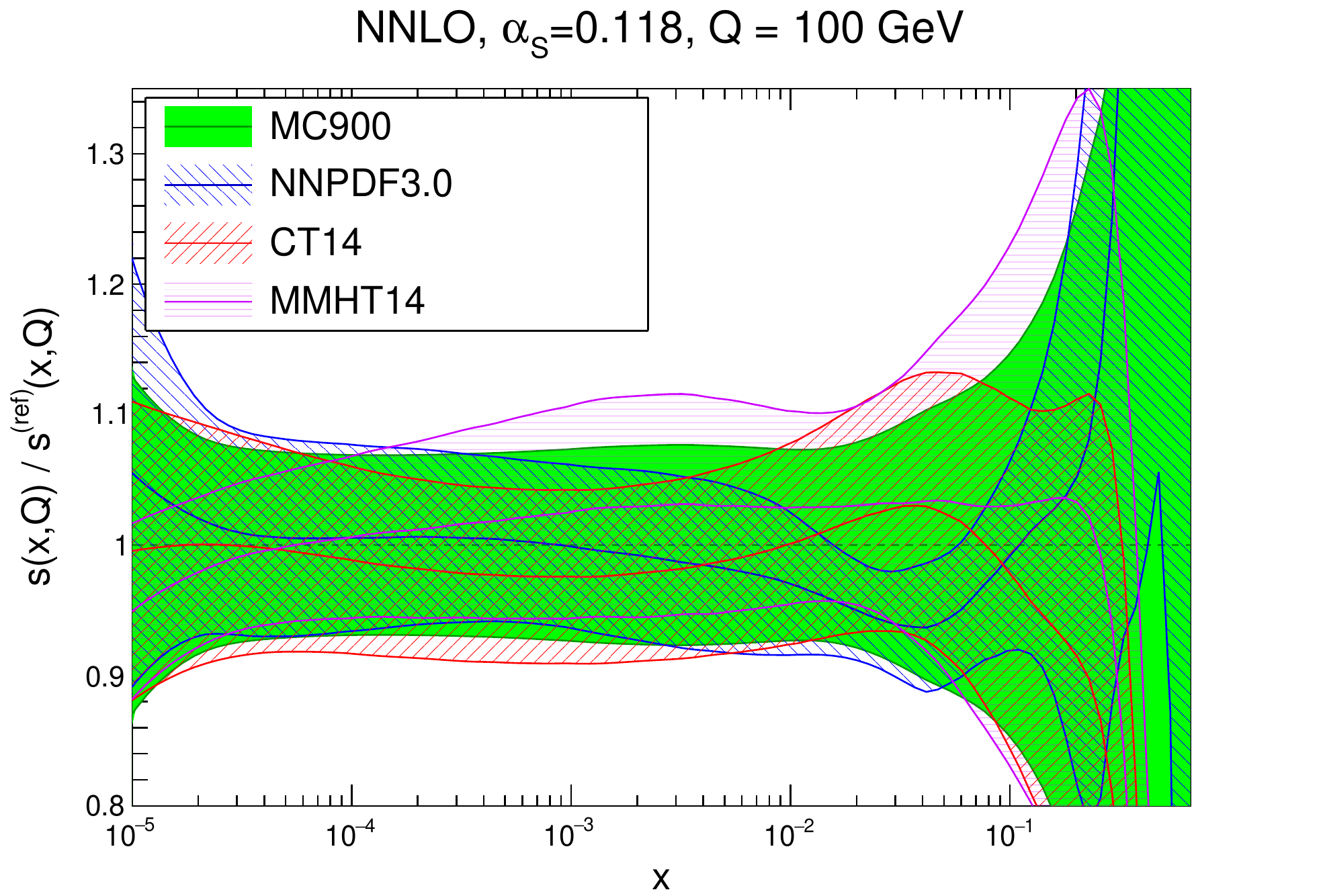}
   \caption{\small Comparison
     of the MC900 PDFs with the  sets that
     enter the combination: CT14, MMHT14 and NNPDF3.0 at
     NNLO.
     We show the gluon and the up, anti-down and strange quarks
     at $Q=100$ GeV.
     Results are normalized to the central value of MC900.
}  
\label{fig:MCPDFcombV2}
\end{figure}

In Fig.~\ref{fig:MCPDFcombV2} we compare the MC900 combined
PDF set with the three individual PDF sets,
CT14, MMHT14 and NNPDF3.0, at NNLO, for
$\alpha_s(m_Z^2)=0.118$ at 
$Q^2=10^4$ GeV$^2$.
Results are normalized to the central value of MC900. Because of the
good consistency of the PDF sets which enter the combination, the
combined uncertainty is  not much larger than that of
individual PDF sets.
A somewhat more significant increase
observed in some cases,
for instance, for the strange
$s(x,Q)$ around $x\simeq 0.05$,
due to the larger differences between the individual
PDF sets.
By construction,
the uncertainty on the statistical combination
remains always smaller than that of the envelope method.

\begin{figure}[t]
  \centering
  \includegraphics[width=.48\textwidth]{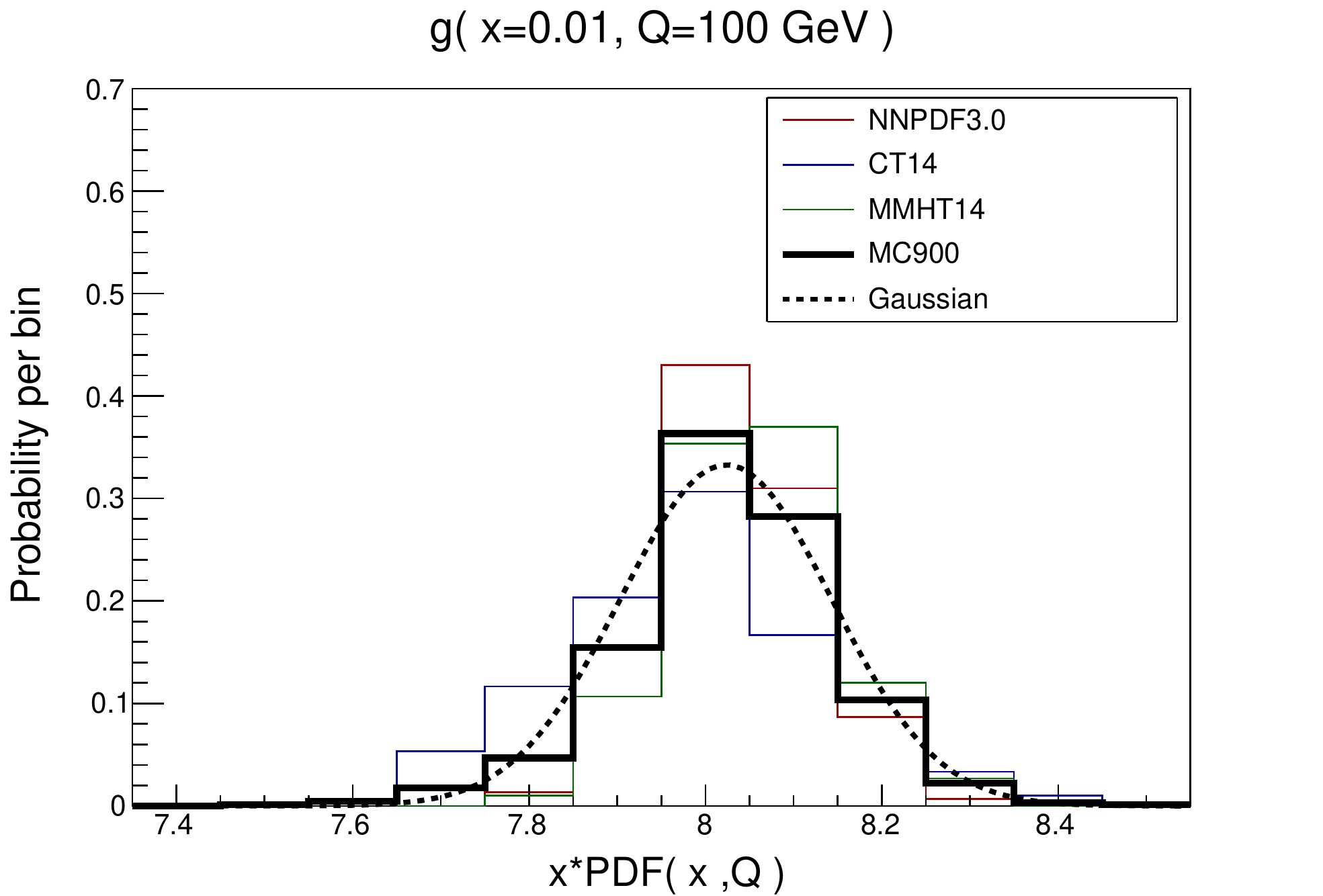}
  \includegraphics[width=.48\textwidth]{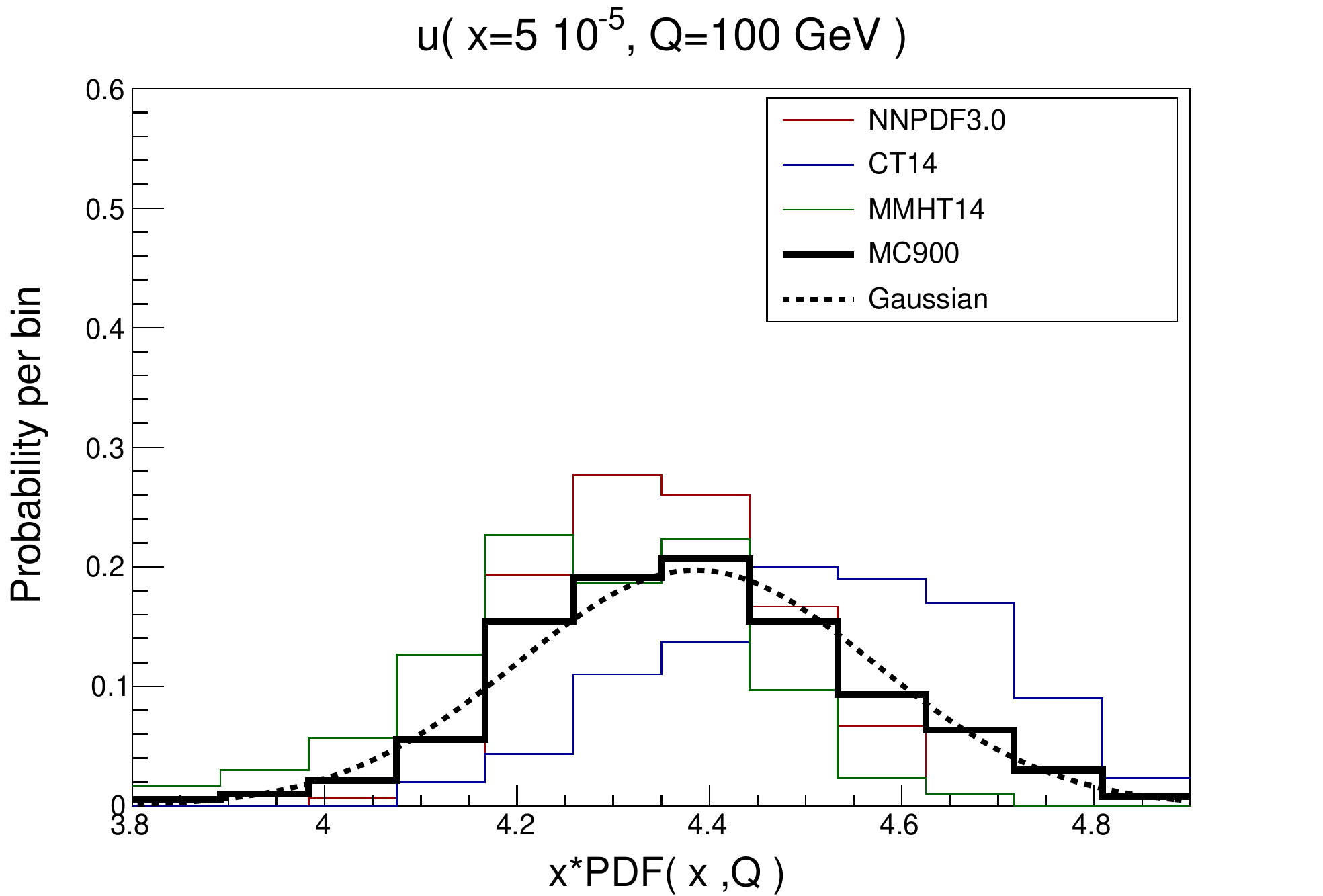}
  \includegraphics[width=.48\textwidth]{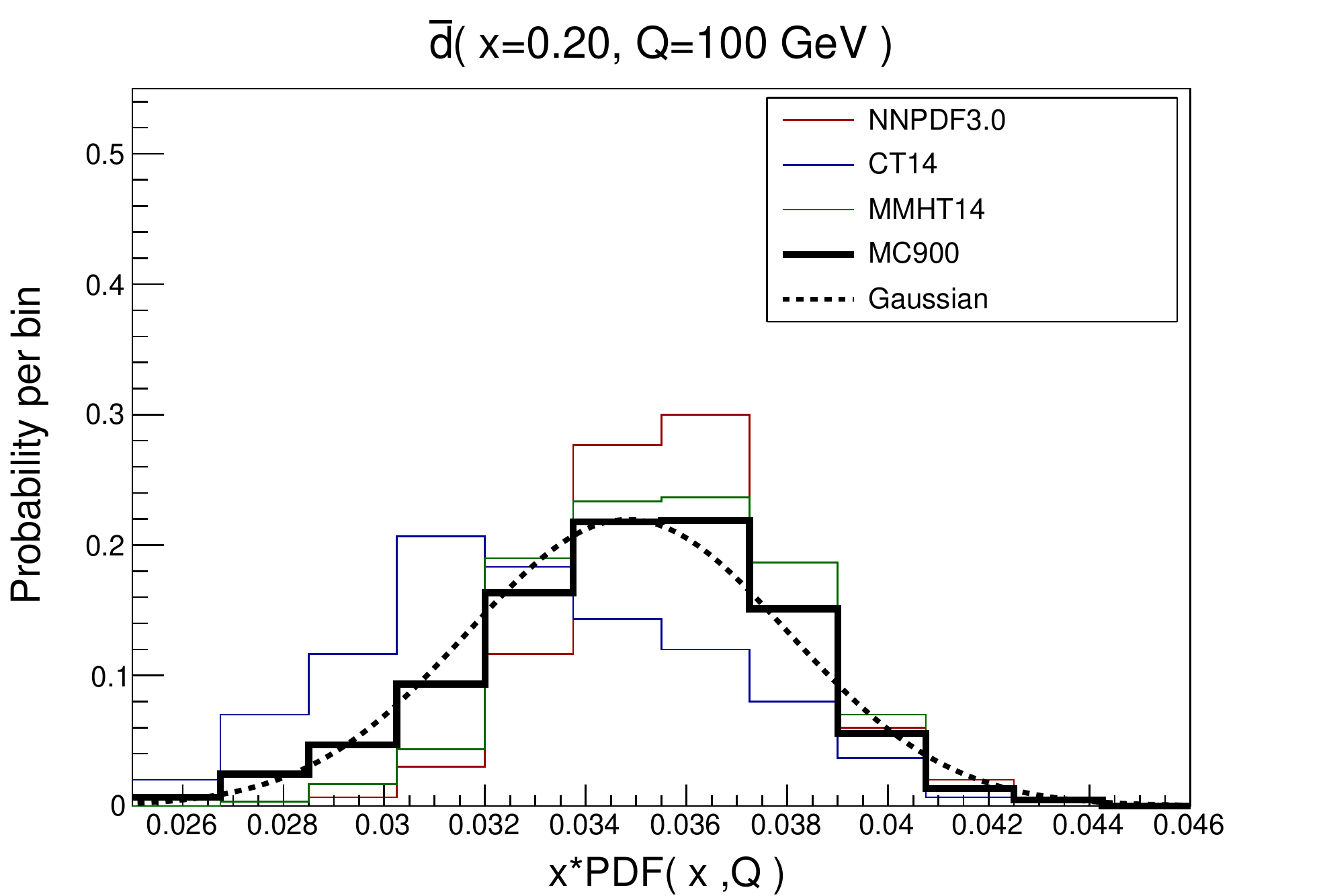}
  \includegraphics[width=.48\textwidth]{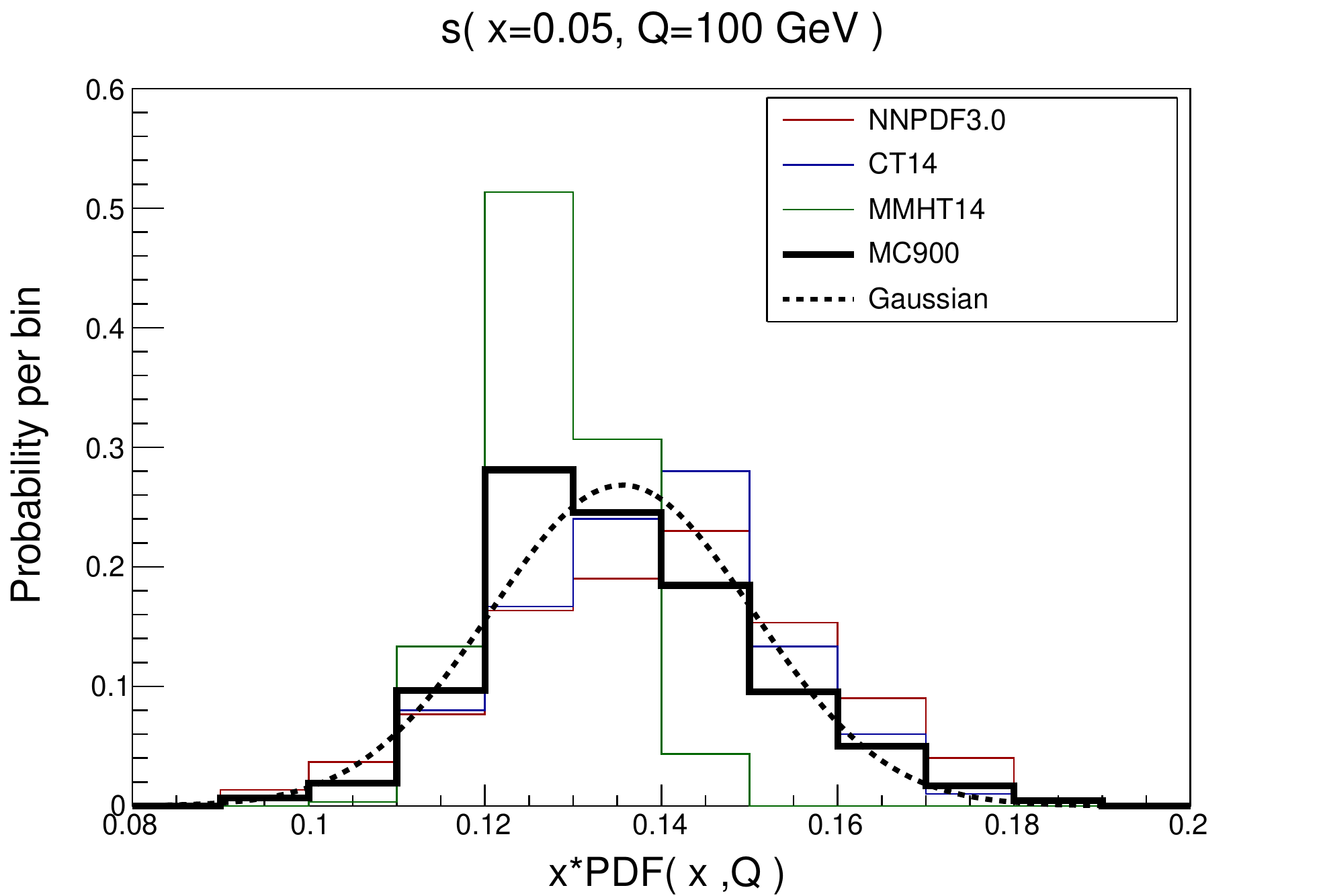}
  \caption{\small Probability
    distribution of MC900, compared to that
    of the individual input sets CT14, MMHT14 and
    NNPDF3.0 NNLO.
    Results are shown for the gluon at $x=0.01$, the up quark
    at $x=5\cdot 10^{-5}$, the
    down antiquark at $x=0.2$ and the strange PDF at $x=0.05$.
    All PDFs have been
    evaluated at $Q=100$ GeV.
    The histograms represent the probability per bin of each
    PDF value, quantified with the number of replicas that fall into
    that bin.
    For comparison, a Gaussian distribution with the same mean and variance
    of the MC900 histogram is plotted. 
}  
\label{fig:cmc3}
\end{figure}

In Fig.~\ref{fig:cmc3} we show the
probability distribution for MC900 compared to that
of the individual input PDF sets.
 Results are shown for the gluon at $x=0.01$, the up quark
    at $x=5\cdot 10^{-5}$, the
    down antiquark at $x=0.2$ and the strange PDF at $x=0.05$.
    The histograms represent the probability per bin of each
    PDF value, quantified by the number of replicas that fall into
    that bin.
    For comparison, a Gaussian distribution with the same mean and variance
    as that of the MC900 histogram is also plotted.
    The combination appears to be generally  Gaussian to good
    approximation, though  some features
    of the combined distribution provide evidence of deviation from Gaussian behaviour,
    for example a non-vanishing skewness in the $\bar{d}(x,Q)$
    and $s(x,Q)$ distributions.

\begin{figure}[t]
  \centering
  \includegraphics[scale=.65]{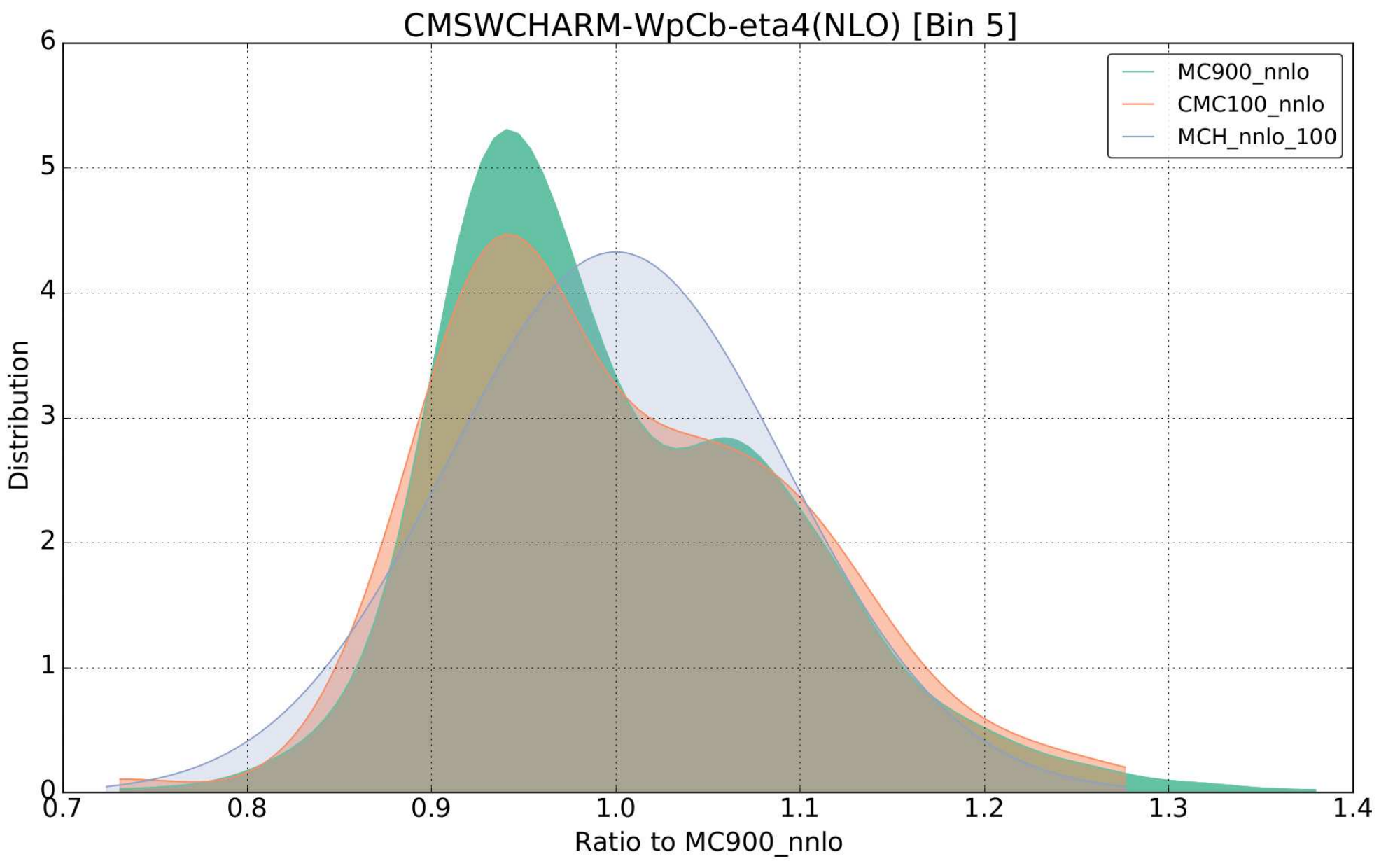}
  \caption{\small The probability distribution
    computed with the Kernel Density
    Estimation method for the MC900 prior and
    the CMC100 and MCH100 reduced sets, for
    the most forward bin of the CMS 
    $W$+charm differential cross-section
    measurement~\cite{Chatrchyan:2013uja}.
}  
\label{fig:nongaussian1}
\end{figure}

\begin{figure}[t]
  \centering
  \includegraphics[scale=.65]{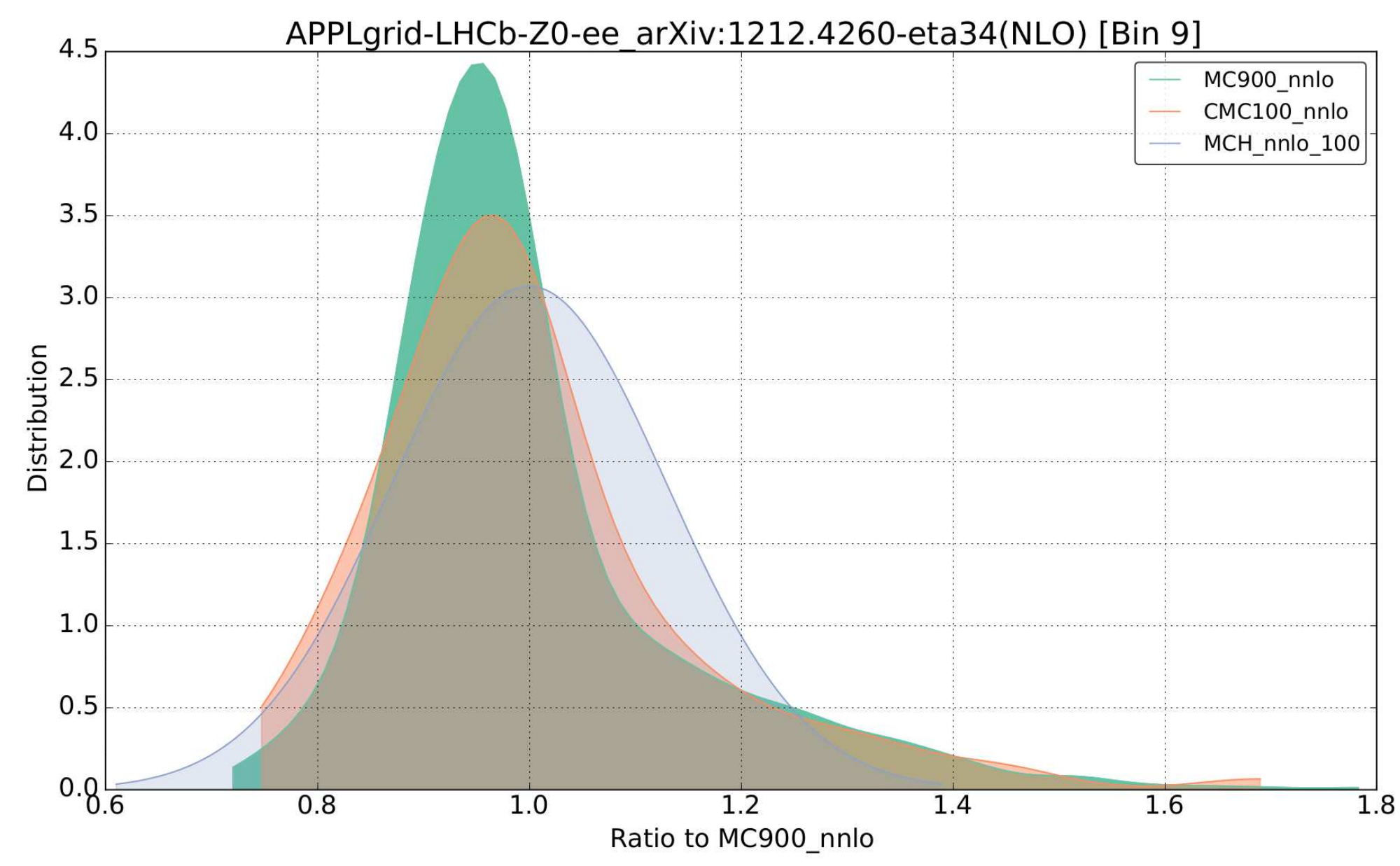}
  \caption{\small Same as Fig.~\ref{fig:nongaussian1}
    for forward Drell-Yan measurement from
LHCb~\cite{Aaij:2012vn}.
}  
\label{fig:nongaussian2}
\end{figure}

The non-Gaussian features of the MC combination can
also be observed for specific
LHC cross-sections, in particular for extreme kinematic regions
or for those processes that involve PDF combinations affected by large uncertainties.
As an illustration, in Figs.~\ref{fig:nongaussian1}
and~\ref{fig:nongaussian2} we show
the probability distribution for MC900 using
the Kernel Density Estimation
(KDE) method~\cite{Carrazza:2015hva} for two specific
observables.
First of all,
the $W$+charm differential cross-section
in the lepton rapidity from the CMS measurement~\cite{Chatrchyan:2013uja}
in the range $\eta_l \in \lc 2.1,2.5\rc$, and then
the forward Drell-Yan process in dileptons from the
LHCb measurement~\cite{Aaij:2012vn} in the range for
$\eta_l \in \lc 4.2,4.5\rc$.
These probability distributions are
clearly non-Gaussian: double-hump structure for $W$+charm;
significant skewness for LHCb forward Drell-Yan.
In
the same plot, results obtained after applying two reduction techniques
are also shown, to be discussed below.

The non-Gaussian features for $W$+charm production
shown in Fig.~\ref{fig:nongaussian1} are related
to the different assumptions on the parameterization
of the strange PDF adopted by the various groups.
For instance, the MMHT14 analysis allows the
$D \to \mu$ branching ratio to be determined from the fit,
increasing the associated uncertainties in the strange
PDF.
Here the usefulness of the present combination method
is clear, allowing to combine in a statistically
consistent way the different assumptions used by each
group (under the assumption that each of the various choices
has equal prior likelihood - for the sets that
enter the combination).

The discussion above refers to the combination
of PDF sets for a common value of $\alpha_s(m_Z^2)=0.118$.
In addition, we need to provide
combined sets for the values $\alpha_s(m_Z^2)=0.1165$
and $\alpha_s(m_Z^2)=0.1195$ in order to be able compute
the combined PDF+$\alpha_s$ uncertainty.
These varying-$\alpha_s$
PDF sets  can be
constructed from a simple average of existing sets from the individual
groups.
If we denote by $q^{\rm pdf4lhc}$ the central value of
combined PDF4LHC set for a specific quark
flavour (or the gluon),
then the sets with different
values of $\alpha_s$ can be constructed as follows:
\bea
q^{\rm pdf4lhc}(\alpha_s=0.1165)&=&\frac{1}{6}\lp
q^{\rm mmht}(\alpha_s=0.116)+q^{\rm mmht}(\alpha_s=0.117)\rp
\nonumber \\ &+& \frac{1}{6}\lp
q^{\rm ct}(\alpha_s=0.116)+q^{\rm ct}(\alpha_s=0.117)\rp\\ &+& \frac{1}{6}\lp
q^{\rm nnpdf}(\alpha_s=0.115)+q^{\rm nnpdf}(\alpha_s=0.118)\rp \, ,
\nonumber \eea
and
\bea
q^{\rm pdf4lhc}(\alpha_s=0.1195)&=&\frac{1}{6}\lp
q^{\rm mmht}(\alpha_s=0.119)+q^{\rm mmht}(\alpha_s=0.120)\rp
\nonumber  \\ &+& \frac{1}{6}\lp
q^{\rm ct}(\alpha_s=0.119)+q^{\rm ct}(\alpha_s=0.120)\rp\\ &+& \frac{1}{6}\lp
q^{\rm nnpdf}(\alpha_s=0.121)+q^{\rm nnpdf}(\alpha_s=0.118)\rp \nonumber \, ,
\eea
with $q^{\rm mmht}$, $q^{\rm ct}$ and  $q^{\rm nnpdf}$ the corresponding
central values of the MMHT14, CT14 and NNPDF3.0 sets for those
specific values of $\alpha_s(m_Z^2)$.
We have verified that other possible ways of constructing these sets
(such as different interpolation options) do not change the result
in any appreciable way.

\subsection{The Monte Carlo reduction method: CMC-PDFs}
\label{sec:cmcpdfs}

Compressed Monte Carlo PDFs (CMC-PDFs)~\cite{Carrazza:2015hva} are
determined by using  a compression algorithm, that, starting from a MC prior
with $N_{\rm rep}$ replicas, determines the set of $\widetilde{N}_{\rm rep}
< N_{\rm rep}$ replicas that most faithfully reproduce the original
probability distribution in terms of central values, variances,
higher moments and corrections.
Therefore with the CMC-PDFs one ends up with a Monte Carlo
representation of the original MC900 combination but based
on a much reduced number of replicas (about
an order of magnitude reduction), with minimal
information loss.

The compression algorithm is based on the minimization of
a figure of merit, 
  \begin{equation}
\label{eq:erf}
      {\rm ERF}=\sum_{k} \frac{1}{N_{k}}
      \sum_{i}\left(\frac{C^{(k)}_{i}-O^{(k)}_{i}}{O^{(k)}_{i}}\right)^{2},
  \end{equation}
  where $k$ runs over the number of statistical estimators
used to quantify the distance between the original
and compressed distributions,
$N_{k}$ is a normalization factor,
$O^{(k)}_{i}$ is the value of the estimator $k$
(for example, the mean or the variance) computed
at the generic point $(x_i,Q_i)$ and $C^{(k)}_{i}$ is
the corresponding value of the same estimator in the compressed set.
The various contributions to Eq.~(\ref{eq:erf}) include
the mean, variance, skewness,
  kurtosis, the Kolmogorov distance and the correlations between
  PDFs.
  The minimization is performed
  using  Genetic Algorithms.
The main advantage of the CMC-PDF method is that
not only the  central value
and variance of the original distribution are reproduced, but also its
higher moments: this is of course crucial when the underlying probability
distribution is non-Gaussian, as illustrated in
Figs.~\ref{fig:nongaussian1}
and~\ref{fig:nongaussian2}.

The quality of the compression has been extensively
validated in Ref.~\cite{Carrazza:2015hva} 
at the level of PDFs and of LHC cross-sections, including
their correlations; the general conclusion is that
$N_{\rm rep}\simeq 100$
replicas are enough  to reproduce
all relevant statistical estimators of the original combined PDF set and
can be reliably used in LHC phenomenology: a comparative quantitative
assessment will be given below.

The choice of estimators $O^{(k)}_{i}$ included in the figure of merit  
Eq.~(\ref{eq:erf}) is
a compromise between the goal of reproducing reasonably
well both the low moments,
and specifically the mean and standard
deviation of the distributions (which fully determine the Gaussian
approximation), and the higher moments of the distribution.
Specifically, if
only the Kolmogorov distance were included, all moments would be
reproduced equally well, while if only the mean and standard deviation
were included, only these would be optimized.
Therefore, in general there is a trade-off in accuracy:
the CMC-PDFs will perform slightly worse in reproducing the central values
and variances of the prior as compared to
a Hessian reduction, but the advantage
of keeping the non-Gaussian features which are missed by the latter case.

The CMC100 PDFs
have been used to construct the default Monte
Carlo representation of the PDF4LHC combined sets
that are made available on {\tt LHAPDF6}:
\begin{center}
  \tt PDF4LHC15\_nlo\_mc \\
  \tt PDF4LHC15\_nnlo\_mc
\end{center}
In addition,
the {\tt compressor} code use to generate the CMC-PDF sets is also publicly available from
\begin{center}
\url{https://github.com/scarrazza/compressor}
\end{center}
together with the corresponding user documentation.
For completeness, Sect.~\ref{sec:formulae} reports the formulae
that should be used to compute the PDF and PDF+$\alpha_s$ uncertainties
whenever the {\tt PDF4LHC15\_mc} sets are used.

\subsection{Hessian reduction methods}

A Hessian representation of PDF uncertainties has certain advantages
and disadvantages in comparison to a Monte Carlo representation.
The
main disadvantage is that a Hessian representation assumes that the
underlying probability distribution is multi-Gaussian (though in
general possibly
asymmetric, i.e. with different upper and lower uncertainties).
The main advantage is that Hessian uncertainties can be treated as
nuisance parameters, and thus on the same footing as other nuisance
parameters: this can be useful in analysis, for example for the determination
of  PDF-induced correlations in
large datasets, and also when using  profiling in order to
 understand the dominant PDF
contributions to a given process.

Two techniques have been suggested  to turn a Monte Carlo
PDF set into a Hessian set: META-PDF and MCH-PDFs.
By carefully tuning the number of Hessian
sets which are used, these techniques also provide a suitable reduction method.
Clearly, when choosing the number of eigenvectors $N_{\rm eig}$ to be included in a
Hessian representation of PDF errors there is a trade-off between speed
and accuracy.
The optimal number of  $N_{\rm eig}$ thus depends then on the
specific application, {\it i.e.},  whether speed or accuracy is the most
important consideration. We will now present each of the two methods, and
then discuss the choice of the number of PDF error members.

\subsubsection{META-PDFs \label{sec:metapdfs}}

The meta-analysis of Ref.~\cite{Gao:2013bia}
was the first method proposed for reducing the number
of Monte-Carlo replicas in the combined PDF ensemble.
The META analysis starts from the 
Monte Carlo representation of the input PDF set.
Then, each MC replica
is re-fitted using a flexible ``meta-parametrization'',
so that all input replicas are cast into a common parametrized
form.
The probability distribution is examined as a function of
parameters $a_{i}$ of the meta-parametrizations. The least and
best constrained combinations of $a_{i}$ are found by diagonalization
of the covariance matrix on the PDF parameter space. 

From this information, one can construct
Hessian error PDFs that reproduce intervals of given confidence along
each poorly constrained direction, centered on the average PDF set
of the input ensemble.
In contrast, error PDFs corresponding to displacements
along well-constrained directions contribute little to the total PDF
uncertainty and can be discarded.
In the end, one obtains a central
PDF, corresponding to an average of the input replicas; as well as
an ensemble consisting of a relatively small number $N_{\rm eig}$ of Hessian
eigenvectors, reproducing the principal components of the original covariance
matrix.

Introduction of meta-parametrizations thus leads to a fairly intuitive
method for combination, which reduces to averaging of discrete parameters
and diagonalization of a covariance matrix in a quasi-Gaussian approximation.
While the choice of the functional form for the meta-parametrizations
is not unique, it does not bias the reduced META-PDFs in practice,
which has been verified by trying various parametrization forms and
fitting procedures.

In the current study, we construct a META-PDF combination with one central
and $N_{\rm eig}=30$ symmetric eigenvectors
according to the following procedure.
We start from the same prior MC900, see Sect.~\ref{sec:prior}
as done in the 
other reduction methods.
Each PDF replica
$f_{\alpha}^{(k)}(x,Q_{0})$ is then re-fitted, at a given
input scale $Q_0$, using the form 
\begin{equation}
\Phi_{\alpha}^{(k)}(x;\{a\})=f_{\alpha}^{(0)}(x,Q_{0})\left(1+\sum_{i=1,...}a_{\{\alpha,i\}}^{(k)}b_{i}(x)\right),
\end{equation}
where $f_{\alpha}^{(0)}(x,Q_{0})$ is the average PDF set for flavor
$\alpha$. The basis functions are \begin{equation}
b_{i}(x)=\left\{ \ln(x),\ln(1-x),\mathcal{B}_{14,i-2}(x)\right\} ,
\end{equation}
 for $i=1,2,...$ and include Bernstein polynomials of degree $n=14,$
\begin{equation}
\mathcal{B}_{n,i}(x)=\left(\begin{array}{c}
n\\
i\end{array}\right)x^{i}(1-x)^{n-i}.
\end{equation}
 The parameters $a_{\{\alpha,i\}}^{(k)}\equiv a_{l}^{(k)}$ are zero
 for the central PDF set, but vary for each of the Monte-Carlo replicas.
 They
are chosen to minimize a metric function $E$ on a grid $\{x_{n}\}$
of momentum fraction values in the interval $x=[3\cdot10^{-5},0.9]$
at the initial scale $Q_{0}$: \begin{equation}
E\left[f_{\alpha}^{(k)}(x,Q_{0}),\ \Phi_{\alpha}^{(k)}(x;\{a\})\right]=\sum_{x{\rm \, grid}}\left[\frac{\ln f_{\alpha}^{(k)}(x_{n},Q_{0})-\ln\Phi_{\alpha}^{(k)}(x_{n};\{a\})}{\delta(\ln f_{\alpha}(x_{n},Q_{0}))}\right]^{2},\end{equation}
 where $\delta(\ln f(x,Q_{0}))\equiv\delta f(x,Q_{0})/f(x,Q_{0})$,
and $\delta f(x,Q_{0})$ is the symmetric PDF uncertainty of $f(x,Q_{0})$. 

The value of $Q_{0}$ where the MC replicas are parametrized is
taken to be $Q_0=8$ GeV, well above the bottom-quark
mass $m_{b}$, where all input heavy-quark schemes lead to close predictions.
Thus we fit $9\times16=144$ parameters for 9 independent PDF flavors
at $Q_{0}$, and setting $\bar{c}(x,Q_{0})=c(x,Q_{0})$, $\bar{b}(x,Q_{0})=b(x,Q_{0})$
for simplicity.
Small differences between sea quarks and antiquarks
induced purely by NNLO evolution effects are negligible here, given
that $Q_{0}$ is still relatively low.

The covariance matrix in the space of
PDF parameters
is computed according to \begin{equation}
\textrm{cov}(a_{l},a_{m})=\frac{N_{\rm rep}}{N_{\textrm{rep}}-1}\sum_{k=1}^{N_{\textrm{rep}}}a_{l}^{(k)}a_{m}^{(k)},\end{equation}
 and diagonalised by an orthogonal transformation $O$:
\begin{equation}
O_{ll'}\cdot{\rm cov}(a_{l'},\, a_{m'})\cdot O_{m'm}^{T}=\lambda_{l}\delta_{lm}.\end{equation}
 The eigenvalues $\lambda_{l}$ of the covariance matrix are positive-definite.
The eigenvectors are associated with the parameters $a_{l}^{'}\equiv O_{lm}a_{m}$.
We can then reasonably interpret $\sqrt{\lambda_{l}}$ to be the width
of the effective Gaussian distribution describing $a_{i}'$, and neglecting
any asymmetry of this distribution. 

To construct the META ensemble we select the largest, or
principal, eigenvalues $\lambda_l$, and discard the 
eigenvectors associated with well-constrained $a_{i}'$.
For the discarded directions, the respective $a_{i}'$ are set to zero, their
average value; the number $N_{\rm eig}$ of eigenvectors for the final META-PDFs
thus reduces below the original number of 144.
Using the META representation,  a standard deviation
on a quantity $X$ can be computed using
the usual Hessian symmetric master formula,
\begin{equation}
\delta X=\sqrt{\sum_{k=1}^{N_{\rm eig}}\left(X^{(k)}-X^{(0)}\right)^{2}} \, ,
\end{equation}
where $X^{(0)}$ and $X^{(k)}$ are the predictions for the central
and $k-$th eigenvector of the META ensemble.
This is a symmetrized
master formula, which provides an estimate of the 68\% CL uncertainty
at all $Q$ values, providing the DGLAP evolution of all input PDFs
is numerically consistent.%
\footnote{Agreement between the codes for numerical DGLAP evolution at NNLO
  has been confirmed by a benchmarking exercise in the appendix of~\cite{Gao:2013bia}. An already small difference between NNLO evolution of MSTW'08 grid files and HOPPET was further reduced in the MMHT14 release.
At NLO, the NNPDF3.0 sets evolve according to a truncated solution
of the DGLAP equation, which deviates marginally from the evolution
of the CT14 and MMHT14 sets. We apply a correction to the LHAPDF
grid files of the META-NLO ensemble after the combination to compensate
for this weak effect. %
}
In this approach, common DGLAP evolution settings are used for
all the META eigenvectors using the {\tt HOPPET} program~\cite{Salam:2008qg}.

The Hessian META uncertainty provides a lower estimate for the
MC900 uncertainty.
For example, the META-PDF set with $N_{\rm eig}=100$ symmetric
eigenvectors (denoted by META100 in the following) exhibits
PDF uncertainties  closer
to the MC900 uncertainty than the META30 uncertainty (the
META-PDF set with $N_{\rm eig}=30$ symmetric
eigenvectors), but requires
the evaluation of a larger number of eigenvectors.
We find that the META30 combination efficiently
captures information about the lowest two (Gaussian) moments of the
original MC900 distribution, which are the most robust and predictable.
The number $N_{\rm eig}\simeq25-30$ corresponds to about the minimal number of parameters 
that are needed to describe PDF degrees of freedom 
in at least three independent $x$ regions (small, intermediate, and large $x$) 
for the nine physical flavors.

In a sense, META30 provides a ``minimal" estimate of the PDF
uncertainty, so that ``the true uncertainty must be at least
as much with good confidence".
META100, on the other hand, provides a more complete estimate for the
given prior, by including subleading contributions that are also more prone
to variations.

The META30 PDFs are parametrized at a starting scale of $Q_0=8$ GeV,
above which one can neglect differences in the treatment
of heavy-quark flavors in the CT14, MMHT2014, and NNPDF3.0 PDF sets.
This is sufficient for computing hard-scattering cross sections for typical LHC
observables using the $N_f=5$ massless approximation.
However, in those calculations that match the finite-order cross sections with
initial-state parton showers, knowing the PDFs
at scales below 8 GeV may also be necessary in order to
calculate the Sudakov form factors.
For these specific applications, the META30 PDFs have been extended
down to $Q_{\rm min}=1.4$ GeV by backward DGLAP evolution,
and the  stability of this low-$Q$ extension was validated using several
tests.
The resulting uncertainties at low $Q$
are in good agreement with those from the the MCH-100 PDFs.
In the publicly provided
LHAPDF grids for the {\tt PDF4LHC15\_30 set},
the whole range of $Q \geq 1.4$ GeV is covered;
the transition
across the fitting scale of 8 GeV is invisible to the user.
Only above 8 GeV the treatment of heavy flavors is fully consistent;
below this scale the combined PDFs are suitable for
specific calculations, such as those involving
parton showering generators. 

\subsubsection{MCH-PDFs}
\label{sec:mc2h}

The  {\tt mc2Hessian} algorithm for construction of
a Hessian representation of a Monte
Carlo PDF set uses the replicas themselves as a linear
expansion basis.
This idea was realized in two different ways in
Ref.~\cite{Carrazza:2015aoa}.
The first method first identifies the region where the Gaussian approximation
is valid, and uses Genetic Algorithms to identify the optimal expansion basis.
The second method uses Singular Value Decomposition (SVD) to represent the
covariance matrix on a basis of replicas.

This second method is the
preferred one if the goal is achieving an optimal reproduction of the
covariance matrix, and it is the method which is reviewed here.
In this method, the multi-Gaussian distribution of replicas is viewed
as a distribution of deviations of PDFs from their central value. If
$N_{\rm pdf}$ independent PDFs $f_{\alpha}^{(k)}(x_{i},Q)$,
$\alpha=1,\dots,N_{\rm pdf}$, at scale $Q$  are sampled on a
fine enough grid of $N_x$ points in $x_{i}$, the deviations 
from
the central PDF $f_{\alpha}^{(0)}(x_{i},Q)$
can be collected in a $N_x N_{\rm pdf}$-dimensional vector
$X_{lk}(Q)=f_{\alpha}^{(k)}(x_{i},Q)-f_{\alpha}^{(0)}(x_{i},Q)$, and the
multi-Gaussian is fully determined by the corresponding $N_xN_{\rm
  pdf}\times N_x N_{\rm
  pdf}$ covariance matrix. This, in turn, is represented as a linear
combination of the starting replicas   $f_{\alpha}^{(k)}(x_{i},Q)$,
$k=1,\dots,N_{\rm rep}$.

This is done in the following way.
The set of vectors   $X$ of deviations from the central value, for
each replica,
\begin{equation}
X_{lk}(Q)=f_{\alpha}^{(k)}(x_{i},Q)-f_{\alpha}^{(0)}(x_{i},Q)\,,\label{eq:Xmat}
\end{equation}
where $\alpha$ labels PDFs, $i$ points in the $x$ grid, $l\equiv N_{x}(\alpha-1)+i$
runs over all $N_{x}N_{\textrm{pdf}}$ grid points, $k$ runs over
all the $N_{\rm rep}$   MC replicas, and $f_{\alpha}^{(0)}(x_{i},Q)$, can be viewed as a
$N_{x}N_{\textrm{pdf}}\times N_{\rm rep}$ rectangular matrix.
The $N_{x}N_{\textrm{pdf}} \times N_{x}N_{\textrm{pdf}}$ covariance
matrix can then be written as
\begin{equation}
  \label{eq:covmat}
\textrm{cov}_{ll'}(Q) =
\frac{1}{N_{\textrm{rep}} - 1}\sum_k^{N_{\textrm{rep}}}X_{lk}X^t_{kl'}
\ ,
\end{equation}
and therefore,
\begin{equation}
\textrm{cov}(Q) =
\frac{1}{N_{\textrm{rep}} - 1}XX^t \ .
\end{equation}

A representation of the covariance matrix in terms of replicas is
found by Singular Value Decomposition (SVD) of the matrix $X$.
Namely, we can write $X$  as: 
\begin{equation}
X=USV^t\ .  
\end{equation}
Assuming that  $N_{\rm pdf}N_x<N_{\rm rep}$,  $U$ is
an orthogonal matrix of dimensions $N_{\rm pdf}N_x\times N_{\rm rep}$
which contains the orthogonal eigenvectors of the covariance matrix
with nonzero eigenvalues; $S$ is a  diagonal matrix of real positive
elements, constructed out of the singular values of $X$, {\it i.e.},
the
square roots of the nonzero eigenvalues of the covariance matrix
Eq.~(\ref{eq:covmat}),
multiplied by the normalization constant
$(N_{\rm rep} -1)^\frac{1}{2}$; and $V$ is an orthogonal $N_{\rm rep} \times N_{\rm
rep}$ matrix of coefficients.

Because
\begin{equation}
XX^t = US^2U^t = (US)(US)^t\ ,
\end{equation}
the matrix 
$Z= US$  
has the property that
\begin{equation}\label{eq:ztox}
ZZ^t= XX^t .
\end{equation}
But also,
\begin{equation}\label{eq:zlin}
Z= XV
\end{equation}
and thus $Z$ provides the sought-for representation of the covariance
 matrix as a linear combination of Monte Carlo replicas.

Note that by `linear combination of replicas' here we mean that
each eigenvector is a combination of the original PDF replicas
$f_{\alpha}^{(k)}(x_{i},Q)$, i.e.,
with coefficients which depend on the replica index $k$, but do not
depend on either the PDF index $\alpha$ or the value of $x$; nor, because
of linearity of QCD evolution, on the scale $Q$.
The method therefore
allows for combination of PDFs that evolve in slightly different ways,
with results that do not depend on the scale at which the SVD is performed.

A reduction of the number of eigenvectors $N_{\rm eig}$
can now be performed using Principal
Component Analysis (PCA).
Since we observe
that in practice many of the eigenvectors  lead to a very
small contribution to the covariance matrix, we can select a
smaller set of  $N_{\rm eig}<N^{(0)}_{\rm eig}$ eigenvectors which
still provides a good approximation to the covariance matrix by
selecting the $N_{\textrm eig}$ eigenvectors with largest
eigenvalues.

Denoting with  $u$,
$s$, and $v$  the $N_{\rm pdf}N_x\times N_{\rm eig}$, $N_{\rm eig}\times
N_{\rm eig}$ and $N_{\rm eig} \times N_{\rm rep}$ reduced matrices
computed using these eigenvalues,
for a given
value of $N_{\rm eig}$, using $v$ instead of $V$ in
Eq.~(\ref{eq:zlin}) 
minimizes the difference between the original and reduced  covariance
matrix
\begin{equation}
\label{eq:deltaind}
\Delta\equiv \left\Vert US^2U^t - us^2u^t  \right\Vert \, .
\end{equation}
This method is guaranteed by construction to reproduce the covariance matrix of the
initial prior set, that is, both PDF uncertainties and PDF correlations with
an arbitrary level of precision, upon including more eigenvectors.
It
is also numerically stable (in the sense that the SVD yields well
conditioned matrices), and the same methodology can be applied
no matter which input PDF sets are used in the Monte Carlo combination.

In practice, we   
add eigenvectors until the size of the differences between
the new Hessian and the original MC representations becomes
comparable to the accuracy of the Gaussian approximation in that specific region.
This can be
characterized by the difference between the $1\sigma$ and  $68\%$ CL
intervals of the distribution of results when computing
LHC cross-sections prior Monte Carlo set MC900.
The $1\sigma$ interval
will overestimate the effect of the outlier replicas in comparison
with the $68\%$ CL intervals.
Note that insisting that the method reproduces the  $1\sigma$
band exactly would actually lead to bigger differences
between the prior and the Hessian distribution, as measured for
example by the Kolmogorov distance between both.
Therefore we  add eigenvectors until the Hessian prediction
for the $1\sigma$ ends up between the $68\%$ CL and the $1\sigma$
intervals of the
prior MC900.

As will be discussed in the comparison section below, we find that
in the {\tt mc2hessian} algorithm provides
a satisfactory description of both the data
regions and the extrapolation regions can be realized with $N_{\rm eig}=100$
symmetric eigenvectors, for all relevant of observables.
Decreasing this number does not seem advisable, since then one needs to provide
{\it ad-hoc}
assumptions to increase the correlation lengths of the points in the
small and large-$x$ regions.

\subsubsection{Choice of Hessian sets}
\label{sec:hchoice}

Extensive benchmarking of the two available Hessian methods has been
performed, both at the level of parton distributions, luminosities,
and LHC cross-sections.
The complete set of comparison plots can be found at the PDF4LHC
website:
\begin{center}
\url{http://www.hep.ucl.ac.uk/pdf4lhc/mc2h-gallery/website}
\end{center}
It turns out that for a low number of eigenvectors, of the order of $N_{\rm eig}\simeq 30$,
META-PDFs has a somewhat superior performance, leading to a generally satisfactory
representation of the PDF covariance matrix except in outlying regions, 
where the META-PDF uncertainties are
smaller than those of the MC900 prior
because of its minimal nature (cf. Sect.~\ref{sec:metapdfs}).
On the other hand, for a larger number of eigenvectors, $N_{\rm eig}\simeq 100$, 
the MCH-PDFs yield an extremely accurate representation of the
covariance matrix with a somewhat improved
performance as compared to  the META-PDF method. For instance,
as will be shown below, the entries of the PDF correlation matrix
can be reproduced at the per-mile level.

For these reasons, in the following the Hessian PDF4LHC15 combined
sets with $N_{\rm eig}=30$ eigenvectors
\begin{center}
\tt PDF4LHC15\_nlo\_30, \\
\tt  PDF4LHC15\_nnlo\_30. 
\par\end{center}
are constructed from the META30 sets,
while
those with $N_{\rm eig}=100$ eigenvectors
\begin{center}
  \tt PDF4LHC15\_nlo\_100 \\
  \tt PDF4LHC15\_nnlo\_100
\end{center}
are constructed from the MCH100 sets.

The \texttt{MP4LHC} code used to generate the META sets is publicly
available from the {\tt HepForge} website
\begin{center}
\url{https://metapdf.hepforge.org/} .
\par\end{center}
The
{\tt mc2Hessian} code use to generate the MC2H sets is also 
publicly available from:
\begin{center}
\url{https://github.com/scarrazza/mc2Hessian} .
\end{center}
together with the corresponding user documentation.
For completeness, Sect.~\ref{sec:formulae} reports the formulae
that should be used to compute the PDF and PDF+$\alpha_s$ uncertainties
whenever any of the two Hessian sets, {\tt PDF4LHC15\_100} and
 {\tt PDF4LHC15\_30},  are used.

\subsection{Comparisons and benchmarking}
\label{sec:comparisonsandbenchmarking}

Now we compare the three combined
PDF sets resulting from
applying different reduction strategies to the MC900 prior:
the Monte Carlo set {\tt PDF4LHC15\_mc} with $N_{\rm rep}=100$
replicas, and the two Hessian sets {\tt PDF4LHC15\_100} and
{\tt PDF4LHC15\_30}, with $N_{\rm eig}=100$ and 30
symmetric eigenvectors respectively.
We will only show here some representative comparison plots
between the three NNLO sets,
with the complete set of plots of
PDFs, luminosities and LHC cross-sections (as well as the corresponding
comparisons at NLO)
are available from the PDF4LHC website
\begin{center}
\url{http://www.hep.ucl.ac.uk/pdf4lhc/mc2h-gallery/website}
\end{center}

\begin{figure}[t]
  \centering
  \includegraphics[width=.48\textwidth]{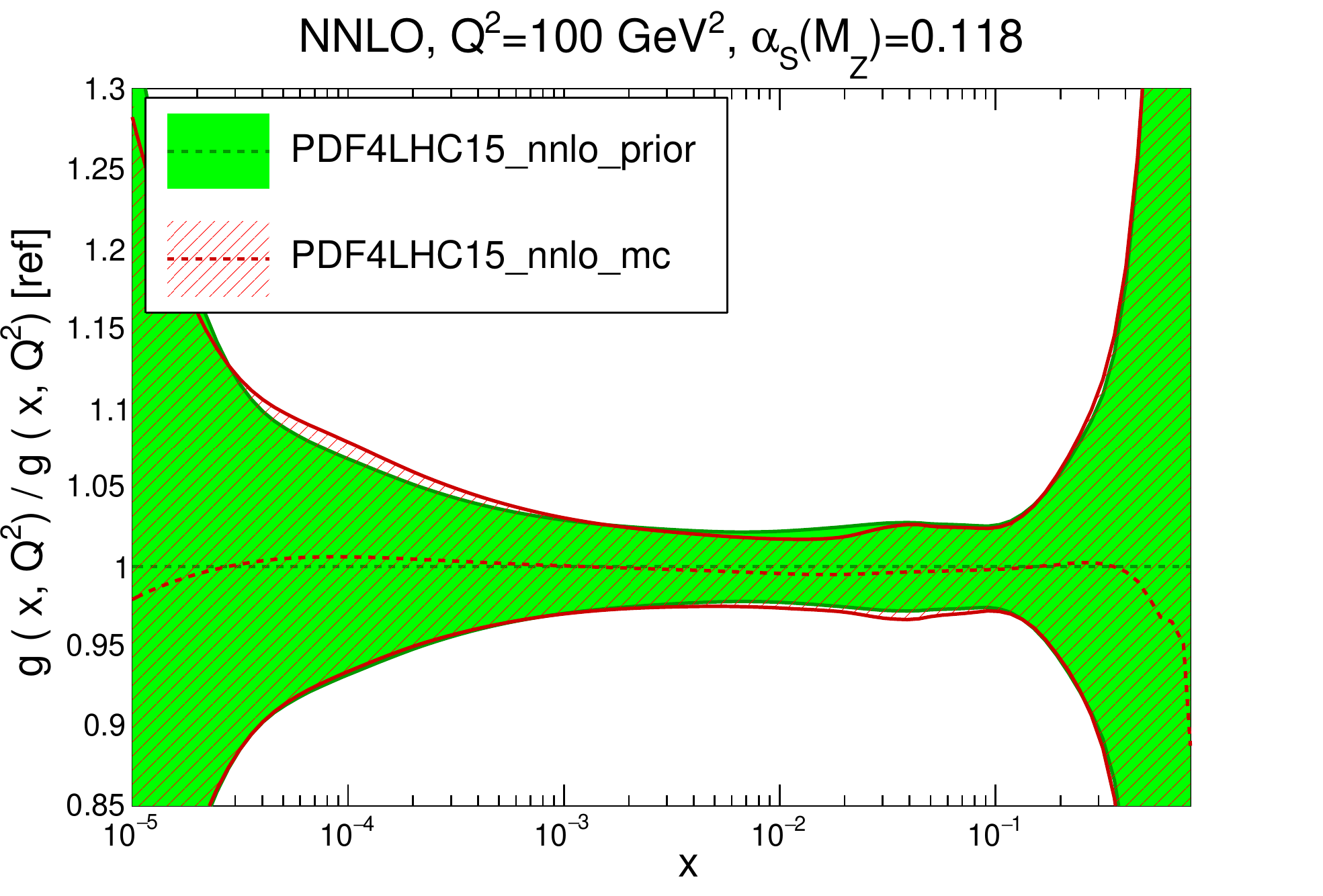}
  \includegraphics[width=.48\textwidth]{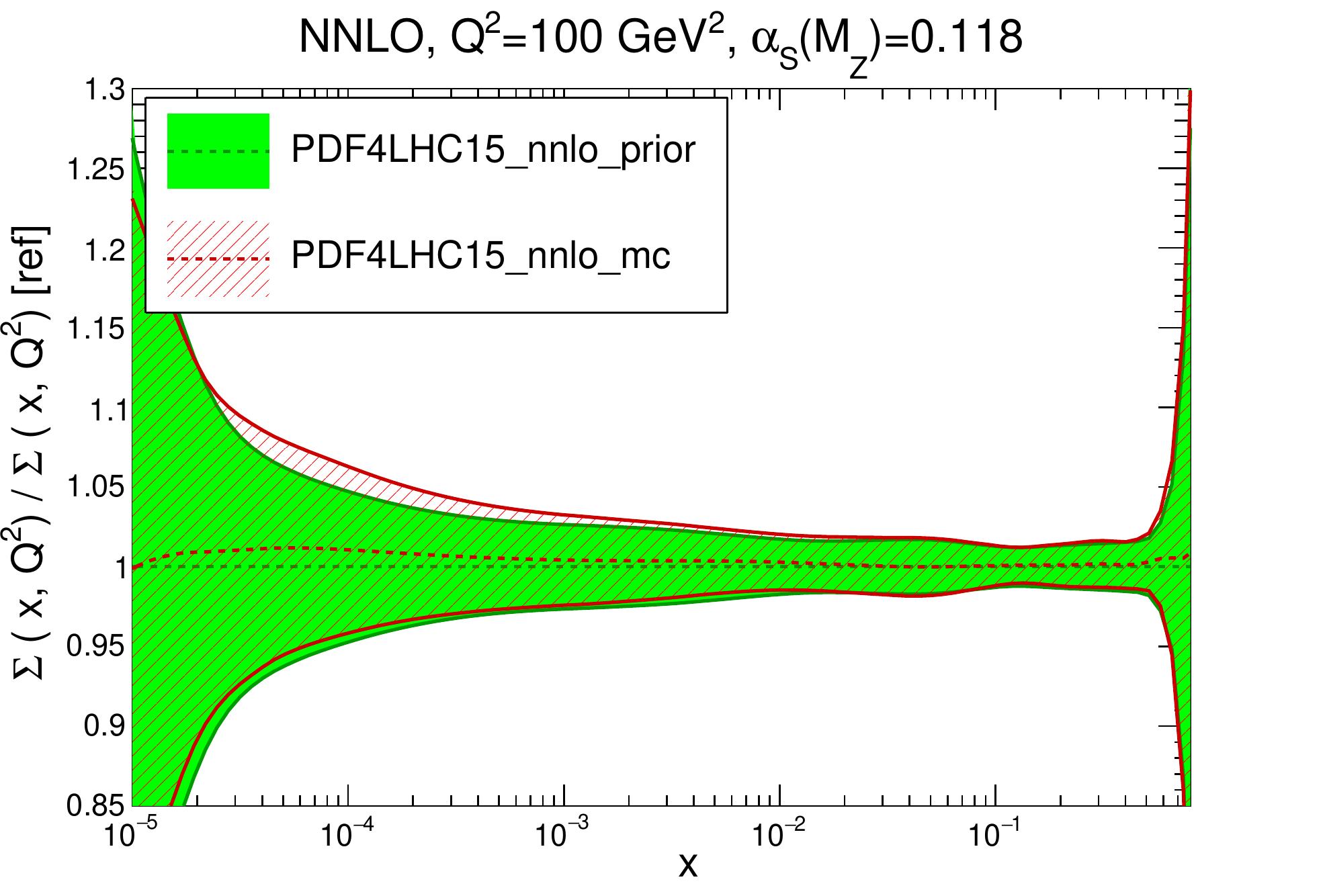}
  \includegraphics[width=.48\textwidth]{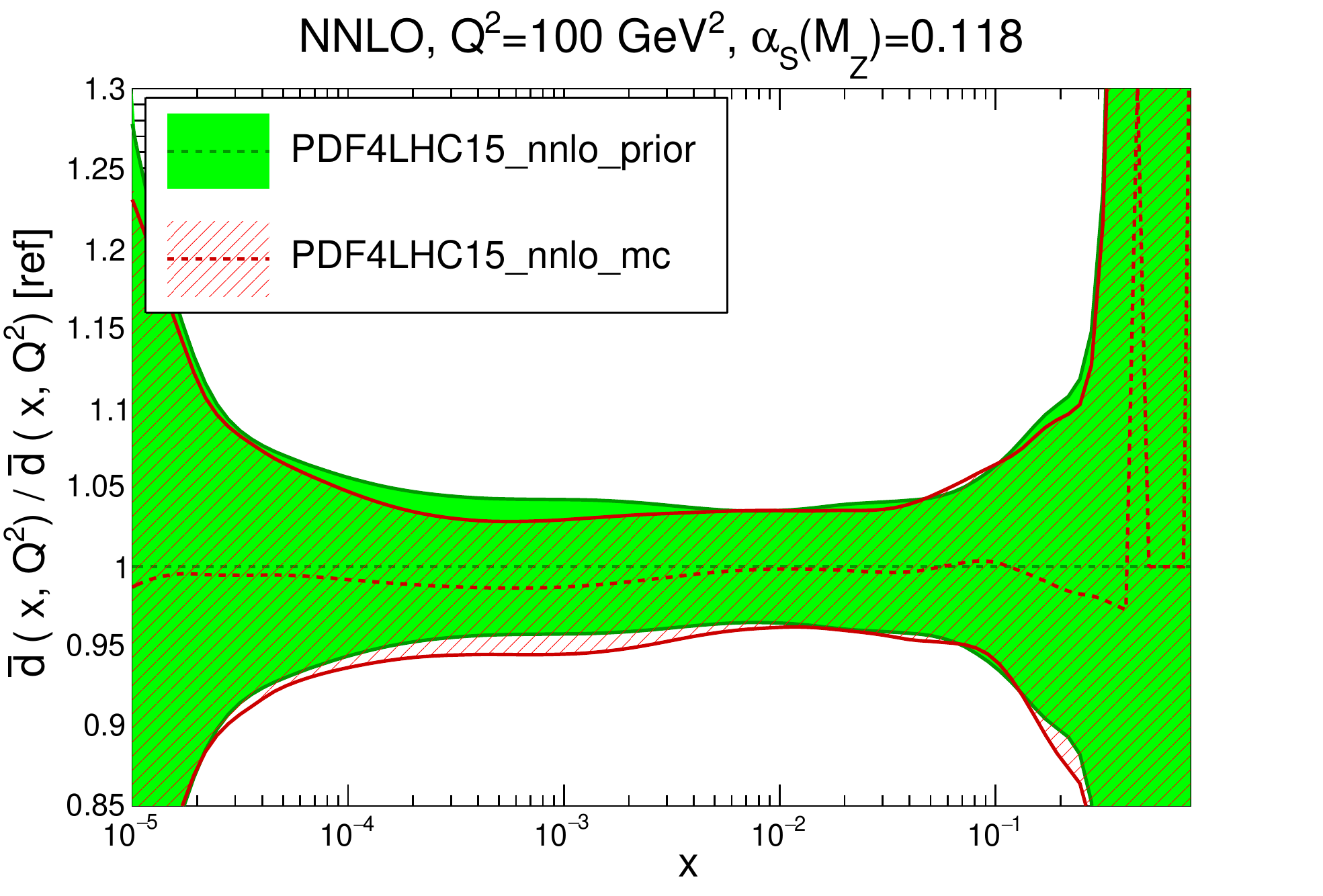}
  \includegraphics[width=.48\textwidth]{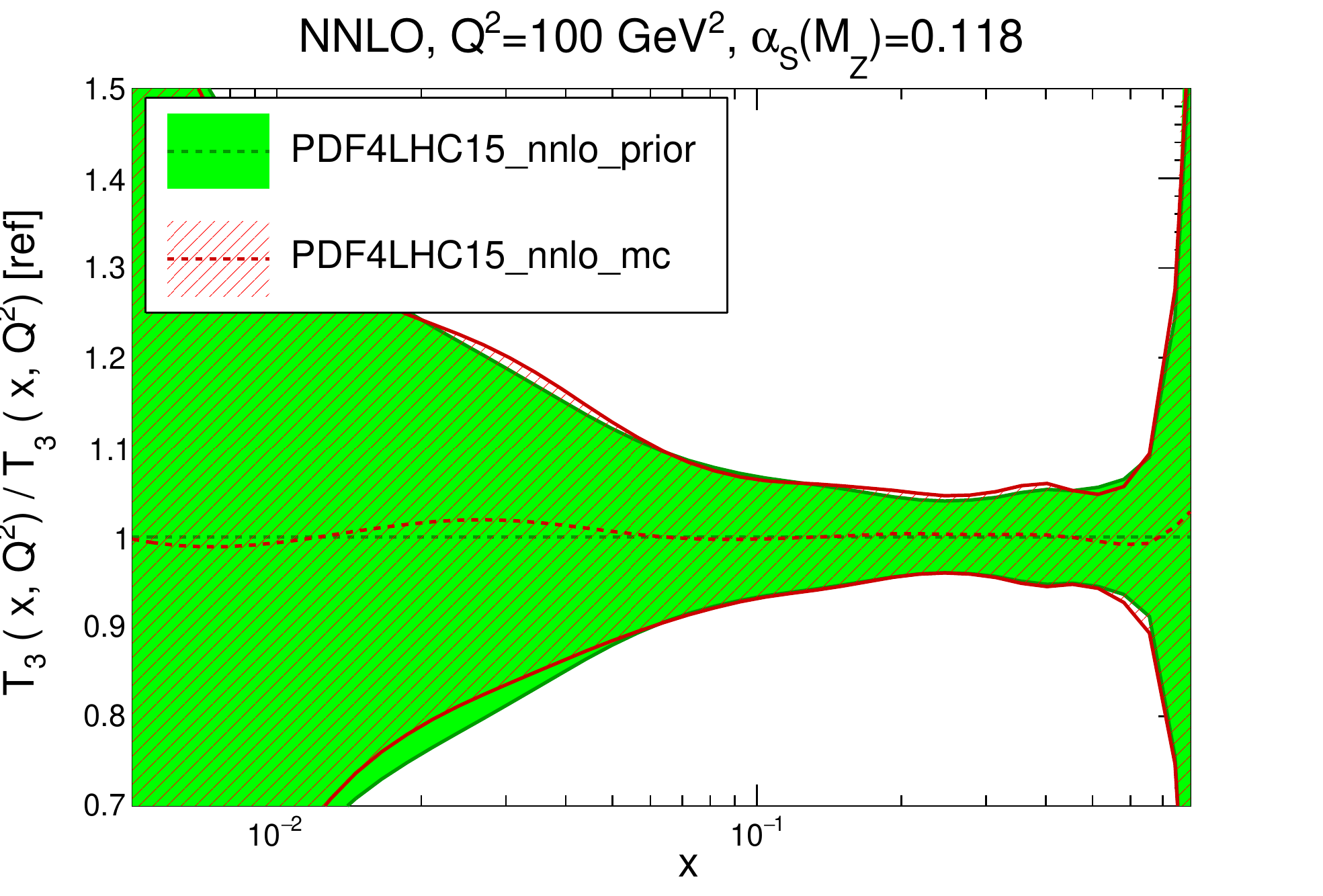}
  \caption{\small
    Comparison of PDFs at a scale $Q^2=10^2$ GeV$^2$ from
    the starting prior 900 replica set  {\tt
      PDF4LHC15\_nnlo\_prior} and the MC compressed 
    set {\tt PDF4LHC15\_nnlo\_mc}.
    We show results for the gluon, total quark singlet,
    anti-up quark and isoscalar triplet,
    normalized to the central value of {\tt PDF4LHC15\_nnlo\_prior}.
}  
\label{fig:xpdf-mc900_vs_cmc100}
\end{figure}

We start the comparisons of  the prior with the various reduced
sets at the level of PDFs, which
we perform at a representative LHC scale of $Q^2=10^2$ GeV$^2$.
In Fig.~\ref{fig:xpdf-mc900_vs_cmc100} we compare the prior   
and the MC compressed
    set {\tt PDF4LHC15\_nnlo\_mc}.
    We show results for the gluon, total quark singlet,
    anti-up quark and the  isoscalar triplet,
    normalized to the central value of {\tt PDF4LHC15\_nnlo\_prior}.
    As can be seen, there is good agreement in all the range in $x$
    both for central values and for PDF
    uncertainties.

\begin{figure}[t]
  \centering
  \includegraphics[width=.48\textwidth]{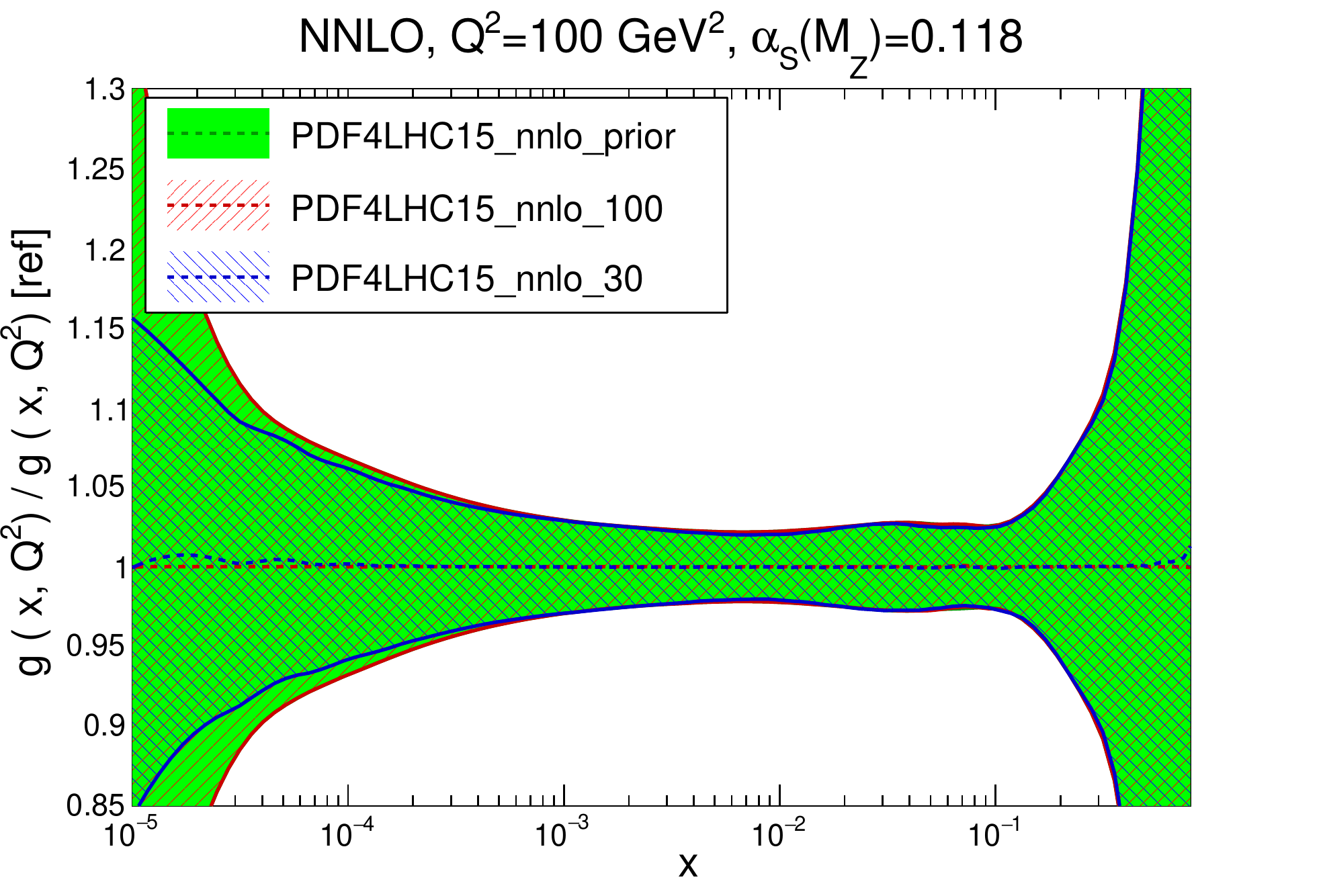}
  \includegraphics[width=.48\textwidth]{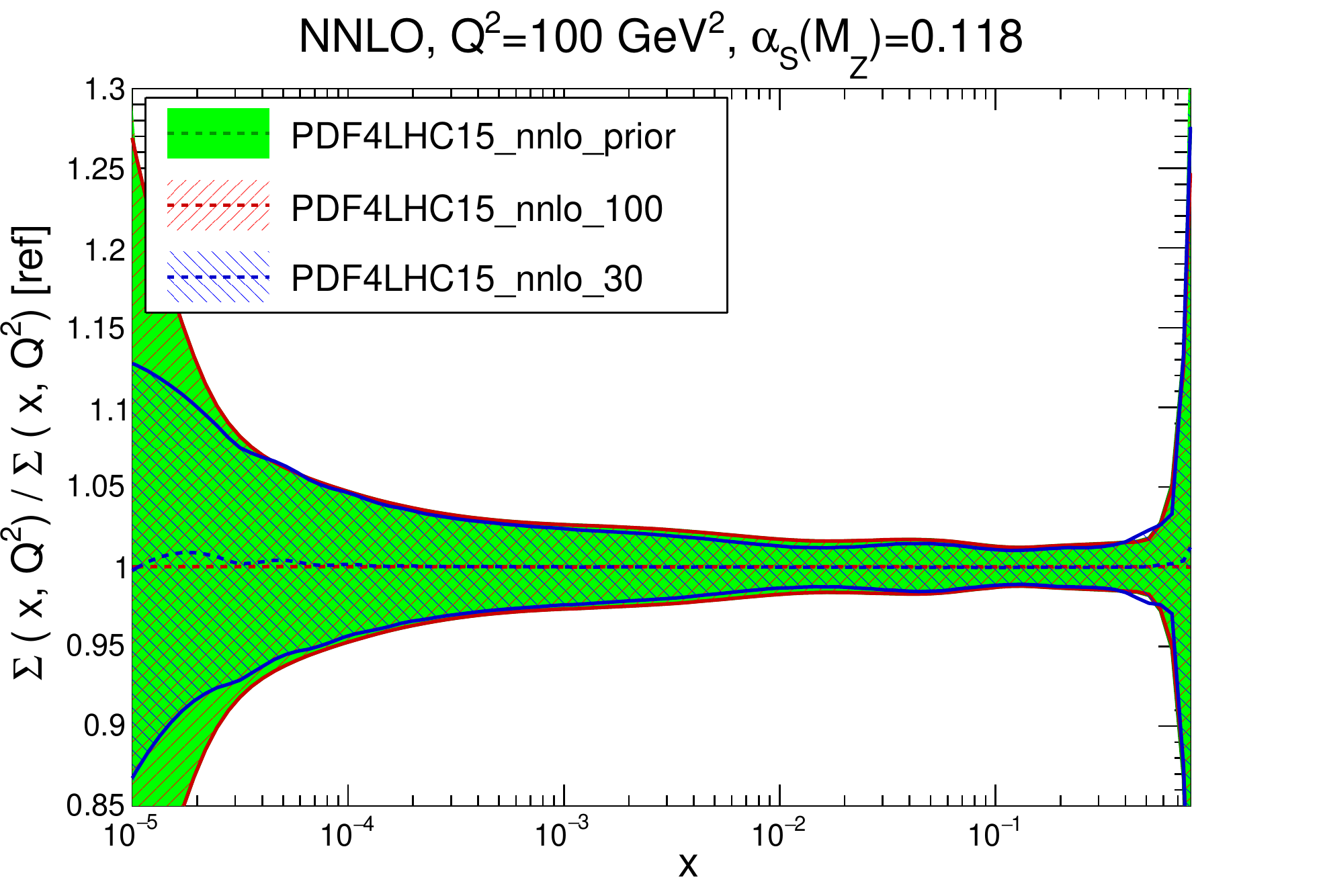}
  \includegraphics[width=.48\textwidth]{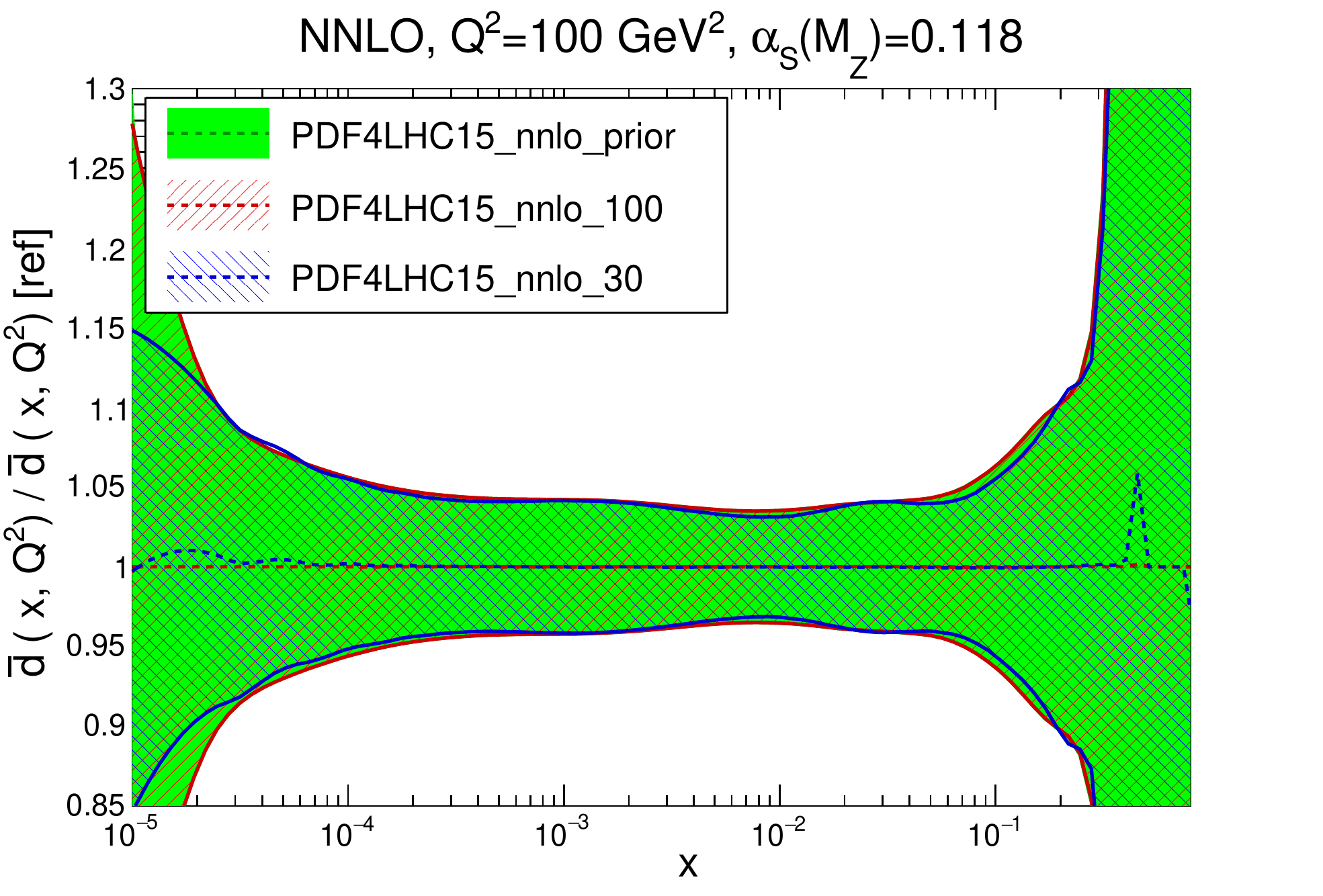}
  \includegraphics[width=.48\textwidth]{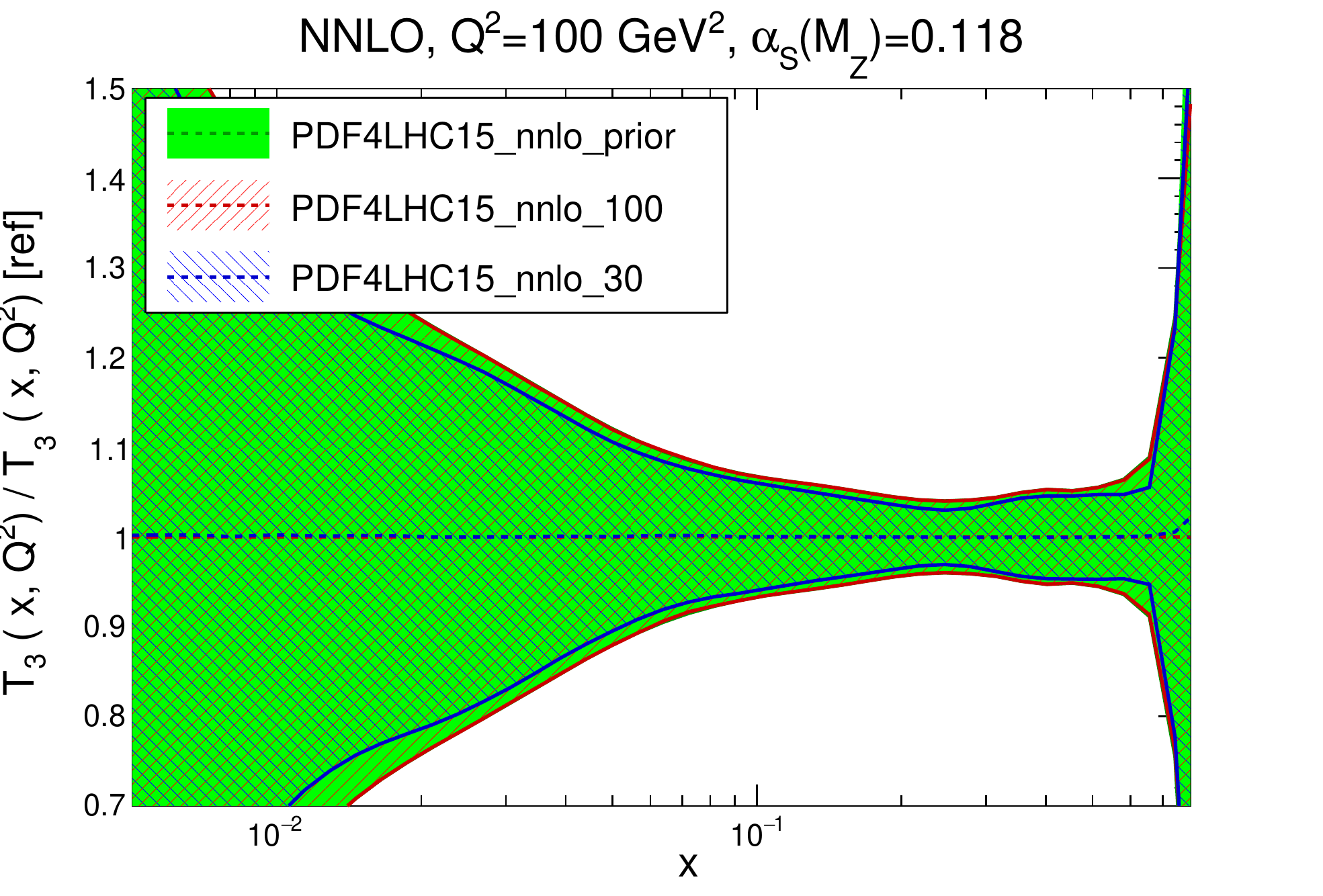}
  \caption{\small
    Same as Fig.~\ref{fig:xpdf-mc900_vs_cmc100} but now
    comparing the prior to the two Hessian sets,
     {\tt  PDF4LHC15\_nnlo\_100} and   {\tt  PDF4LHC15\_nnlo\_30}.
}  
\label{fig:xpdf-mc900_vs_mch100_vs_meta30}
\end{figure}

The corresponding comparison between the prior MC900
and the two reduced Hessian sets,
{\tt PDF4LHC15\_nnlo\_100} and {\tt  PDF4LHC15\_nnlo\_30}, is shown
in Fig.~\ref{fig:xpdf-mc900_vs_mch100_vs_meta30}.
In the case of the set with $N_{\rm eig}=100$ eigenvectors,
the agreement is good for all PDF combinations in the complete range
of $x$.
For the case of the set with $N_{\rm eig}=30$ eigenvectors,
the agreement is also good in the PDF $x$ range corresponding to
precision physics measurements,  but renders a slightly smaller uncertainty
 in the extrapolation regions at small- and large-$x$.
Also for PDF combinations that are not known very well, 
like  the isoscalar triplet, again the uncertainty of 
$N_{\rm eig}=30$ sets can be below that of the prior in the poorly
constrained regions.  
%

\begin{figure}[t]
  \centering
  \includegraphics[width=.48\textwidth]{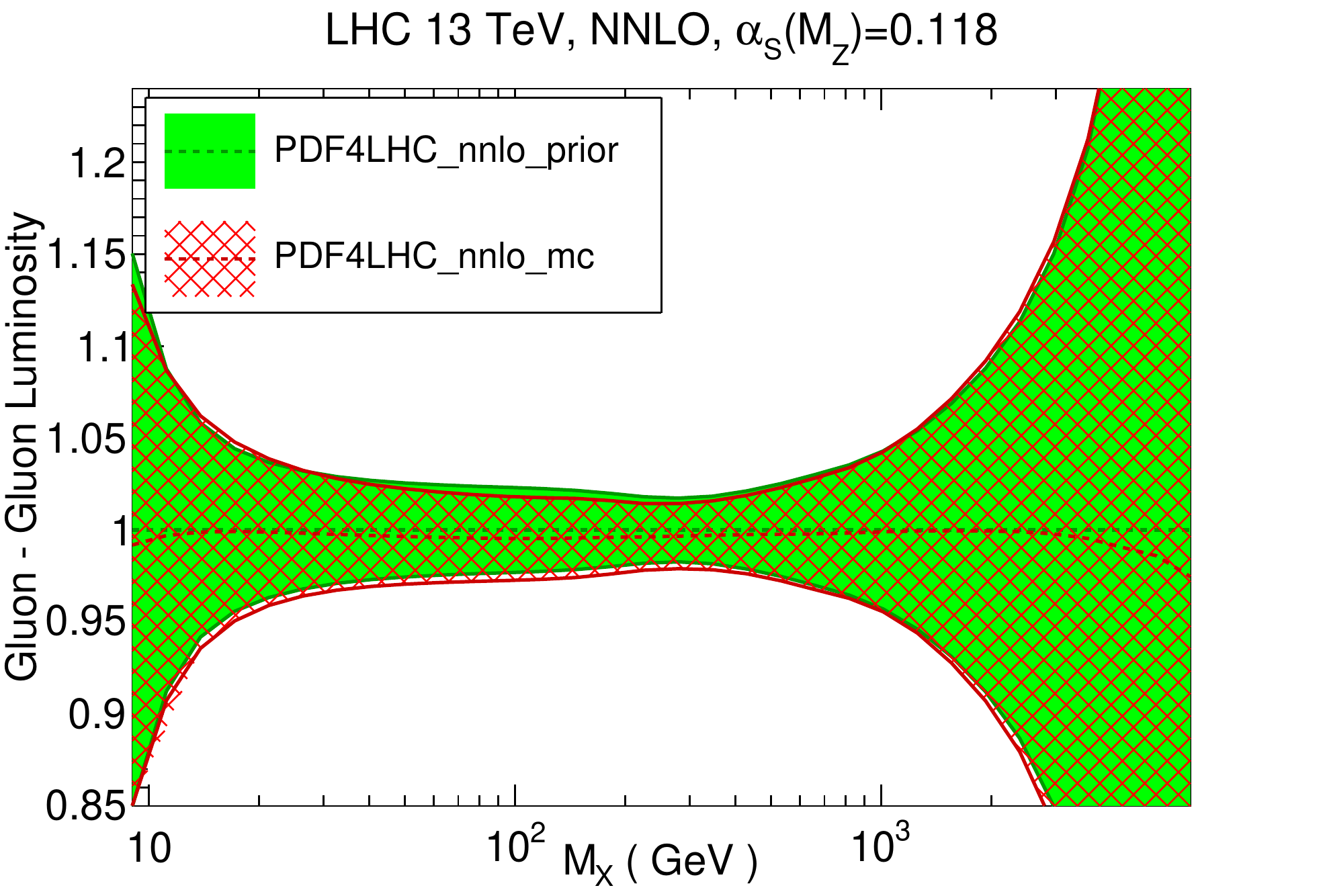}
  \includegraphics[width=.48\textwidth]{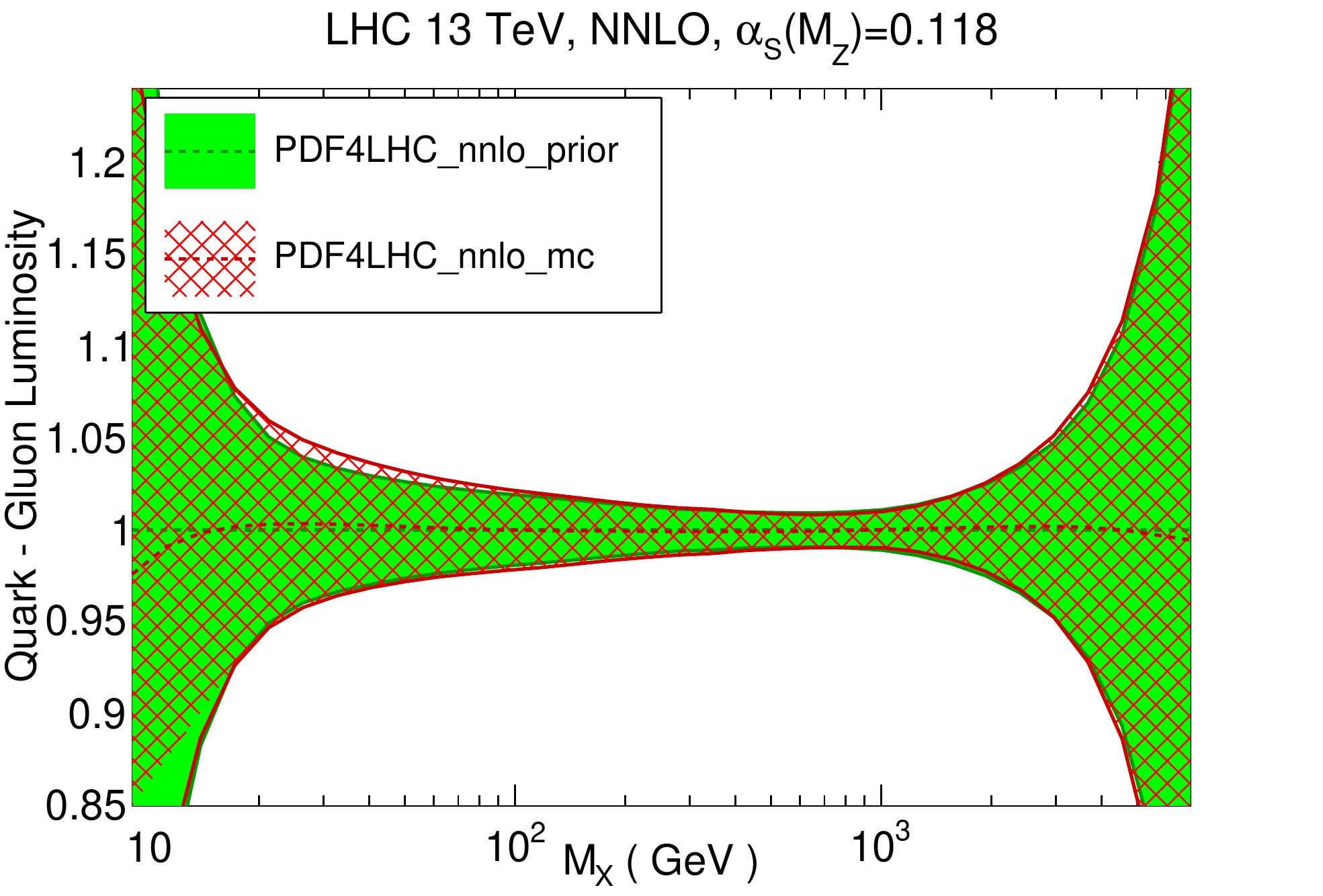}
   \includegraphics[width=.48\textwidth]{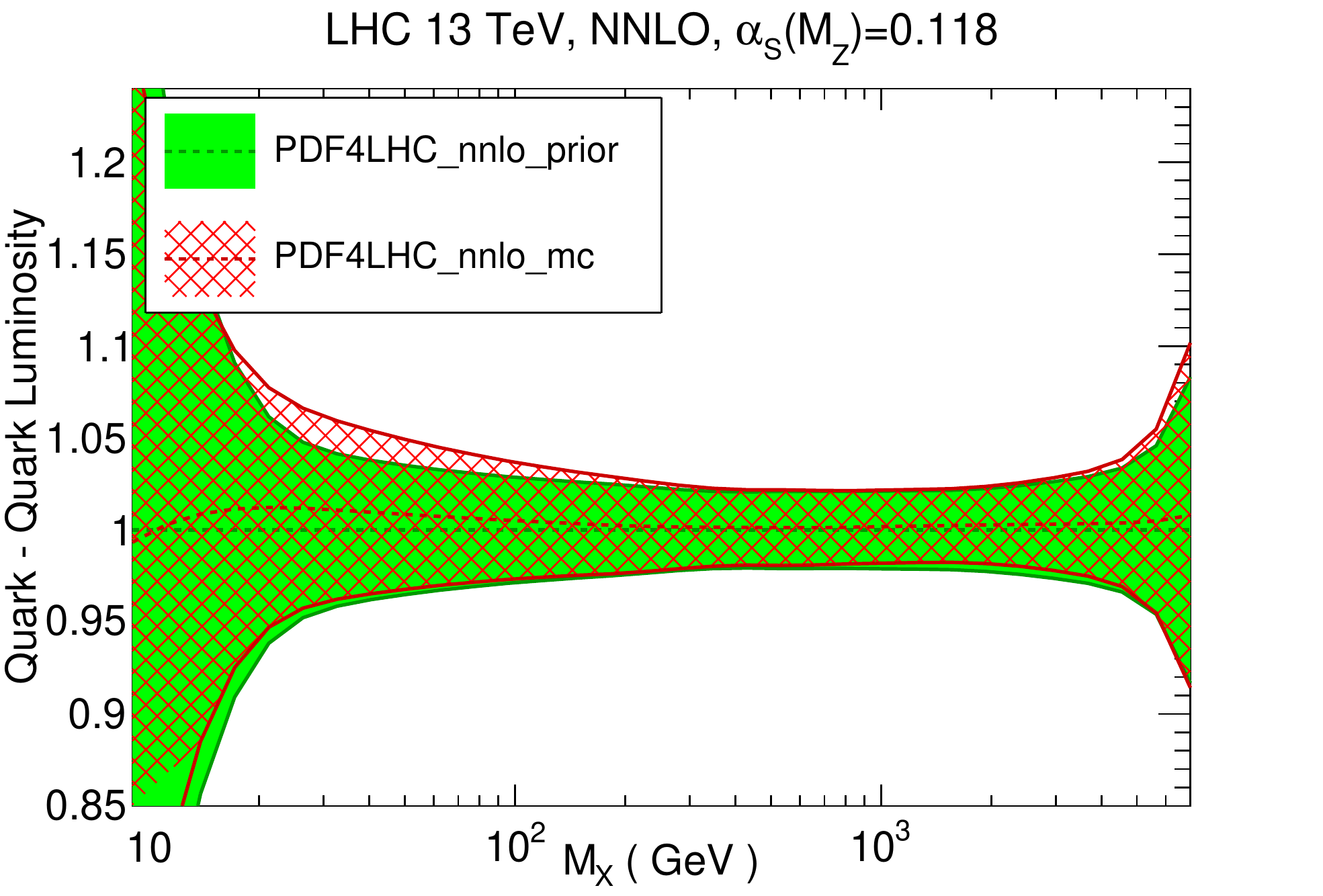}
   \includegraphics[width=.48\textwidth]{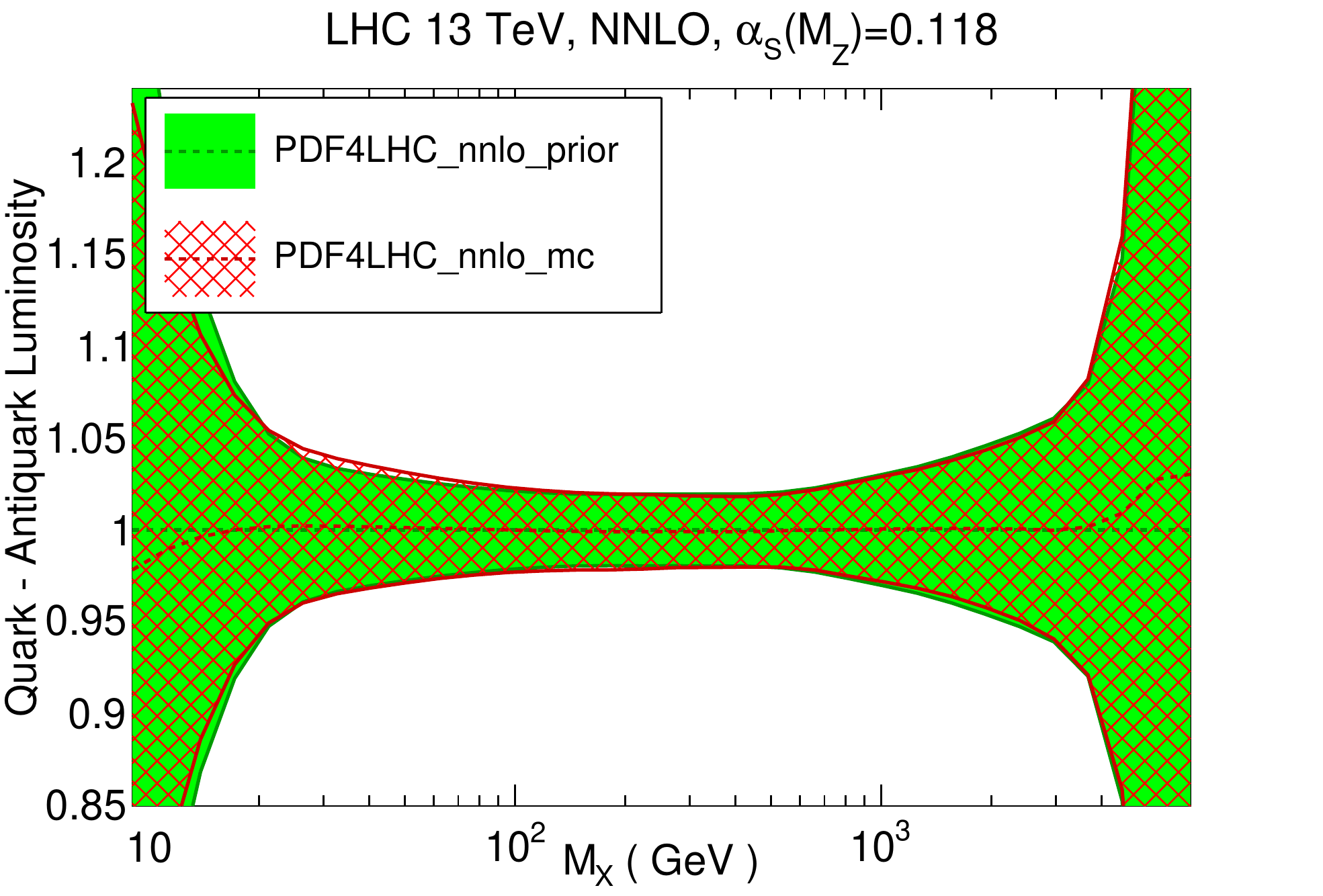}
   \caption{\small Comparison of parton luminosities at the LHC 13 TeV
     computed using the prior set {\tt PDF4LHC15\_nnlo\_prior} with
     $N_{\rep}=900$, and its compressed Monte Carlo representation,
     {\tt PDF4LHC15\_nnlo\_mc}, for $\alpha_s(m_Z^2)=0.118$.
     We show the $gg$, $qg$, $qq$ and $q\bar{q}$ luminosities
     as a function of the invariant mass of the final state
     $M_X$, normalized to the central value of
     {\tt PDF4LHC15\_nnlo\_prior}.
}  
\label{fig:lumi_mc900_vs_cmc100}
\end{figure}

Now we turn to a comparison of parton luminosities. In Fig.~\ref{fig:lumi_mc900_vs_cmc100} we compare
the parton luminosities at the LHC 13 TeV computed with
     the prior set {\tt PDF4LHC15\_nnlo\_prior}, with
     $N_{\rep}=900$, and its compressed Monte Carlo representation,
       {\tt PDF4LHC15\_nnlo\_mc}, for $\alpha_s(m_Z^2)=0.118$.
     We show the $gg$, $qg$, $qq$ and $q\bar{q}$ luminosities
     as a function of the invariant mass of the final state
     $M_X$, normalized to the central value of
       {\tt PDF4LHC15\_nnlo\_prior}.
     We see reasonable agreement in all cases.

\begin{figure}[t]
  \centering
  \includegraphics[width=.48\textwidth]{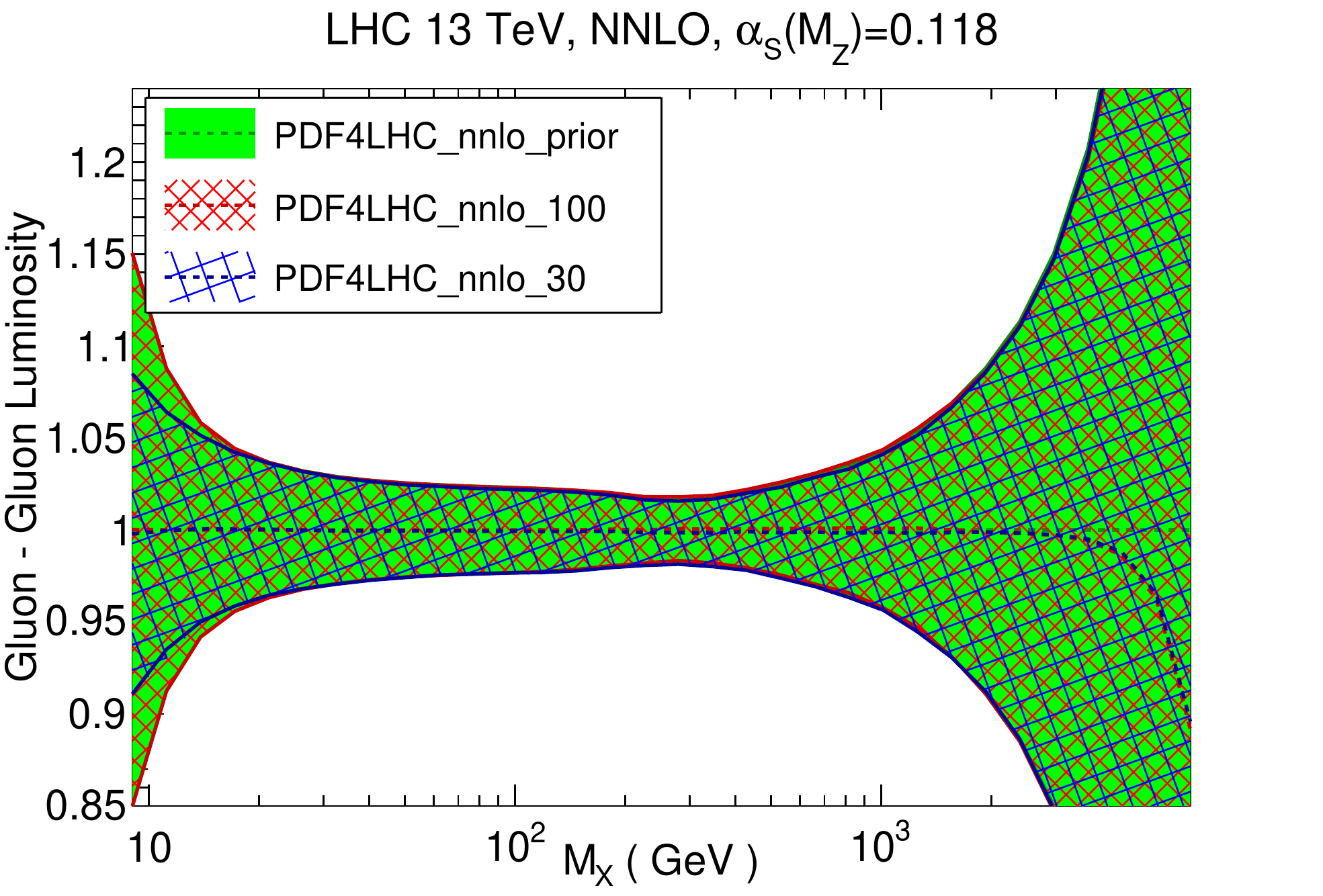}
  \includegraphics[width=.48\textwidth]{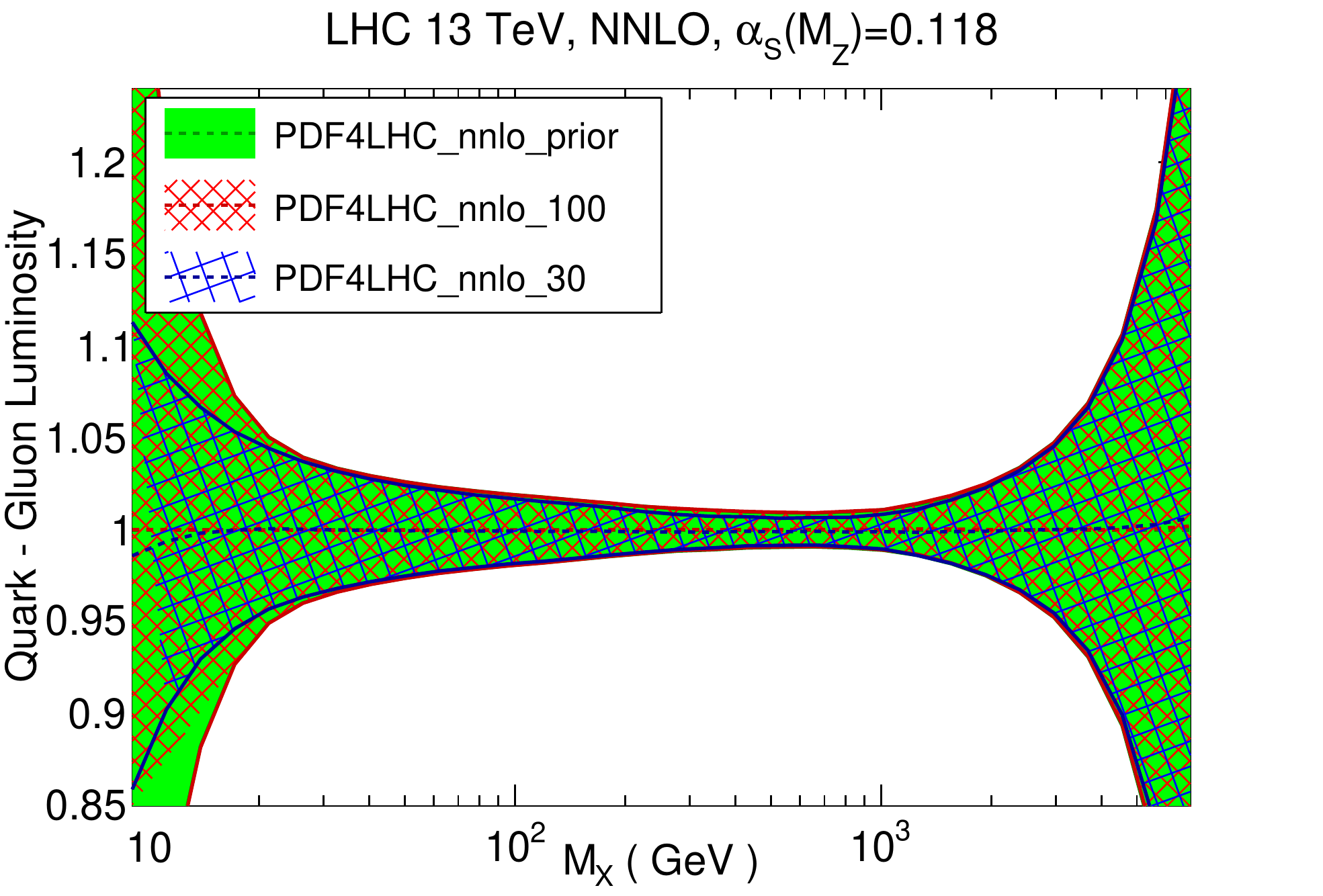}
   \includegraphics[width=.48\textwidth]{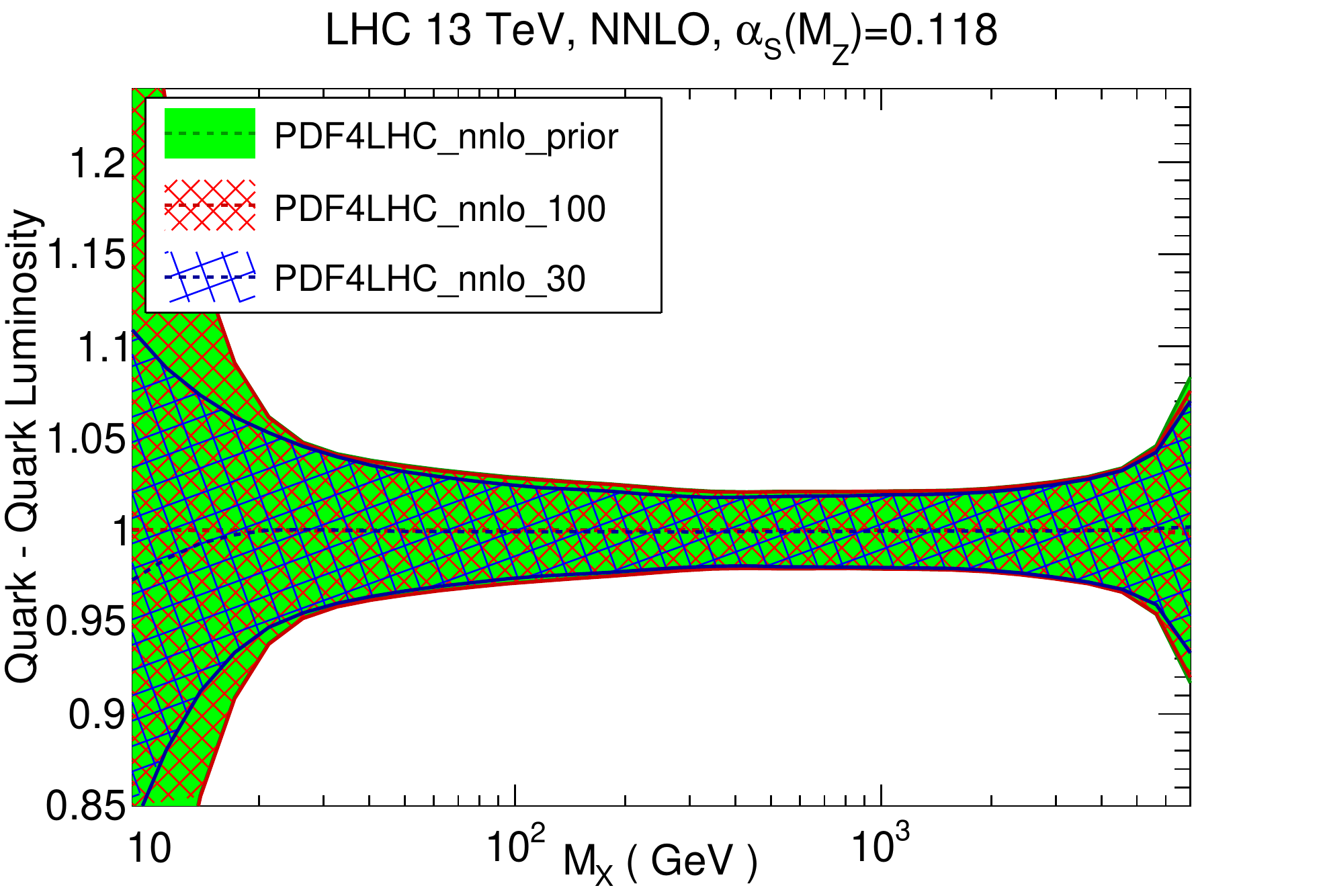}
   \includegraphics[width=.48\textwidth]{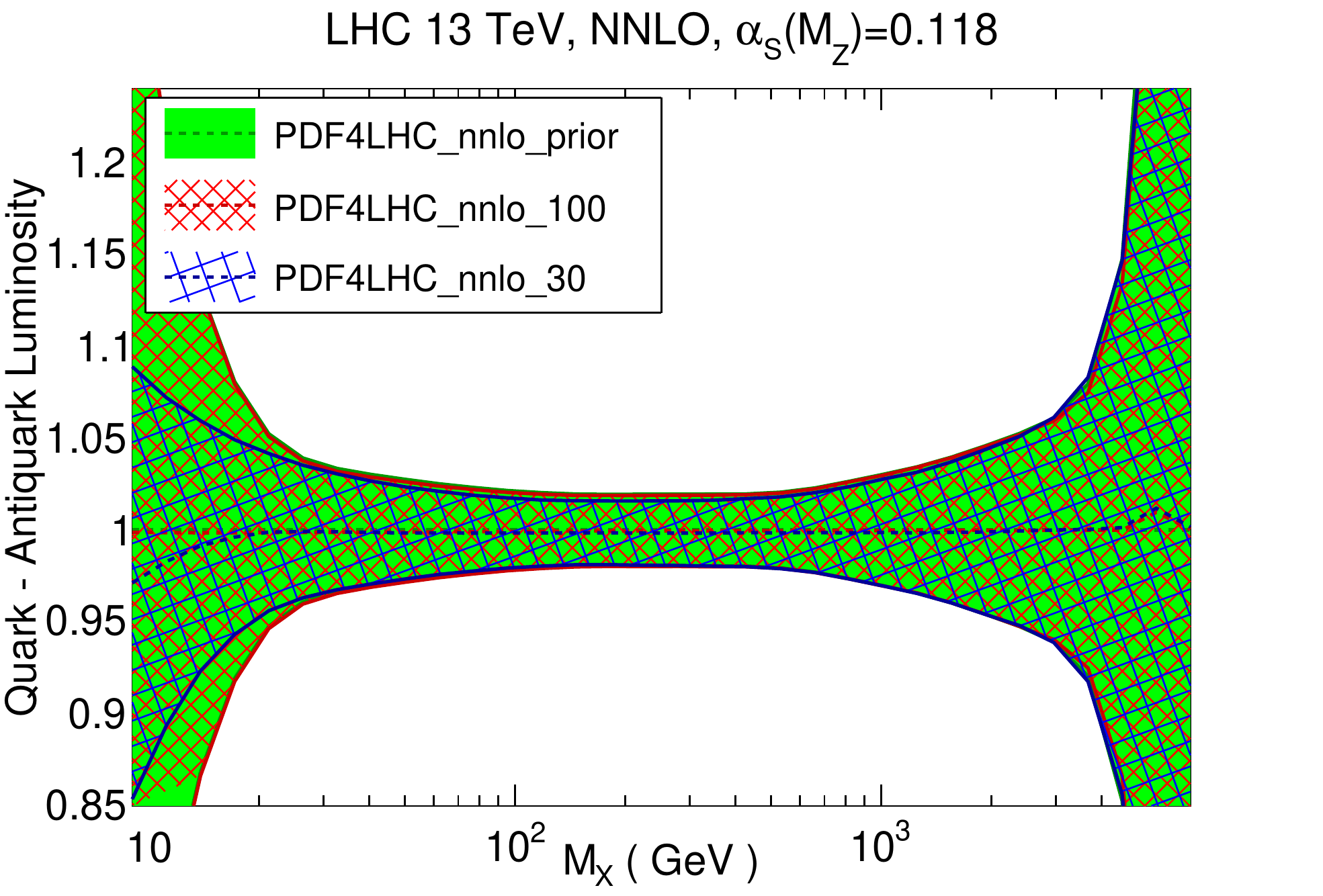}
   \caption{\small Same as Fig.~\ref{fig:lumi_mc900_vs_cmc100}
     now comparing the  parton luminosities from
the prior set {\tt PDF4LHC15\_nnlo\_prior}
     and the two Hessian sets,
        {\tt PDF4LHC15\_nnlo\_100} and  {\tt  PDF4LHC15\_nnlo\_30}.
}  
\label{fig:lumi_mc900_vs_mch100_vs_meta30}
\end{figure}

The corresponding comparison of PDF luminosities
in the case
of the reduced Hessian sets
is shown in Fig.~\ref{fig:lumi_mc900_vs_mch100_vs_meta30},
where we compare the prior set {\tt PDF4LHC15\_nnlo\_prior}
and the two Hessian sets, {\tt  PDF4LHC15\_nnlo\_100} and {\tt  PDF4LHC15\_nnlo\_30}.
Both the $N_{\rm eig}=100$ set and the $N_{\rm eig}=30$ provide a good 
representation of the 900 replica prior in the region of
$M_X$ relevant for precision physics while the the $N_{\rm eig}=30$ performs slightly worse
in the low mass region, sensitive to the small-$x$ PDFs.

A more detailed quantification of the differences in the three reduction
methods can be obtained by computing the percentage difference of
variances between the prior and its various reduced representations.
Recall that in the Hessian approach, central values are reproduced
automatically, as  is the deviation from the central set which is
parametrized by the error sets; in the Monte Carlo approach in
principle, there can be a small shift in the mean when reducing the
sample size, though this is of the order of the standard deviation of
the mean (i.e. 10 time smaller than the standard deviation, for a
100-replica sample), further reduced by the compression method. We
have explicitly 
checked that indeed in all cases the accuracy with which central
values are reproduced is that of the {\tt LHAPDF6} interpolation with which
final grids are delivered.

\begin{figure}[t]
  \centering
  \includegraphics[width=.48\textwidth]{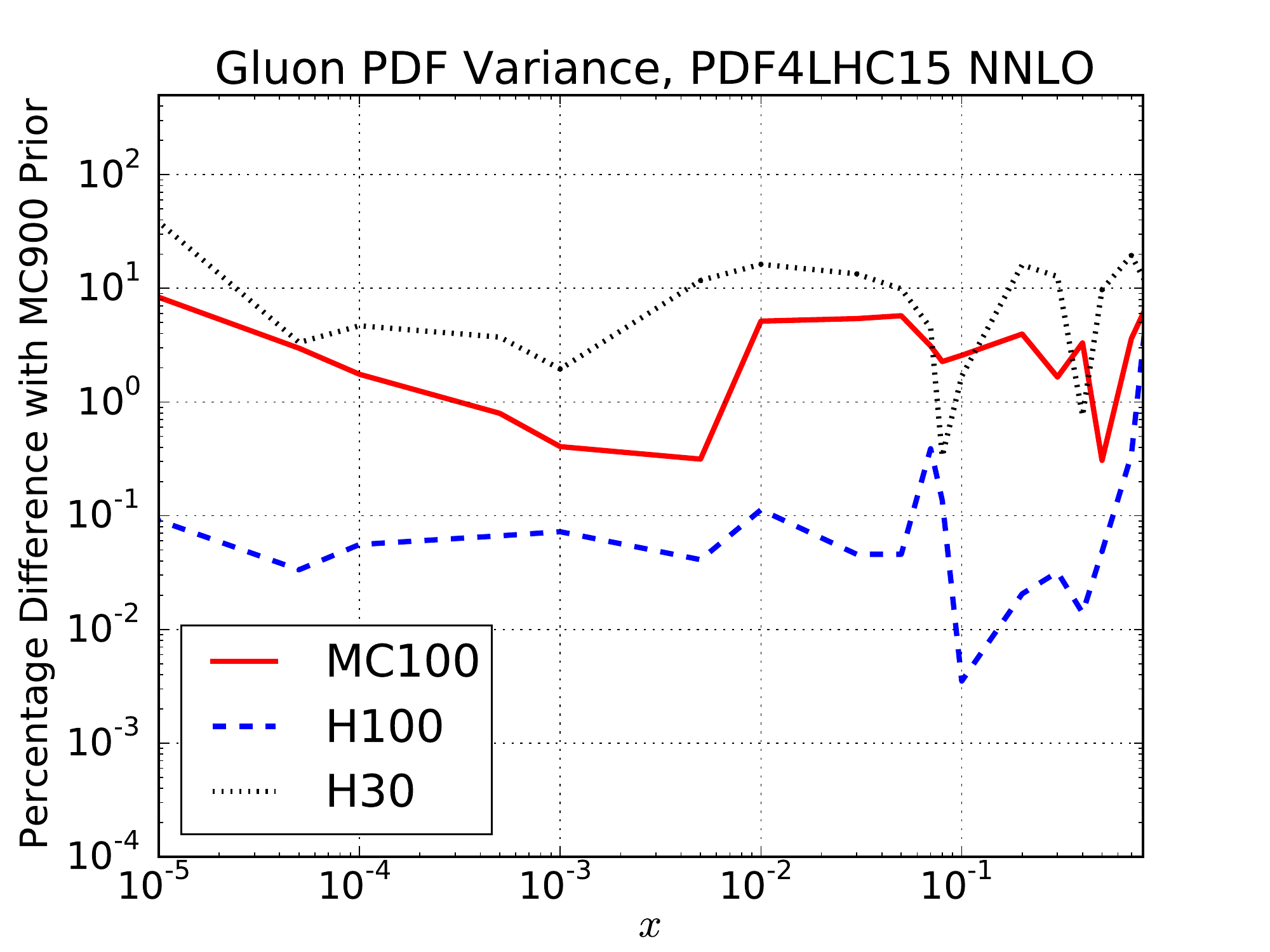}
  \includegraphics[width=.48\textwidth]{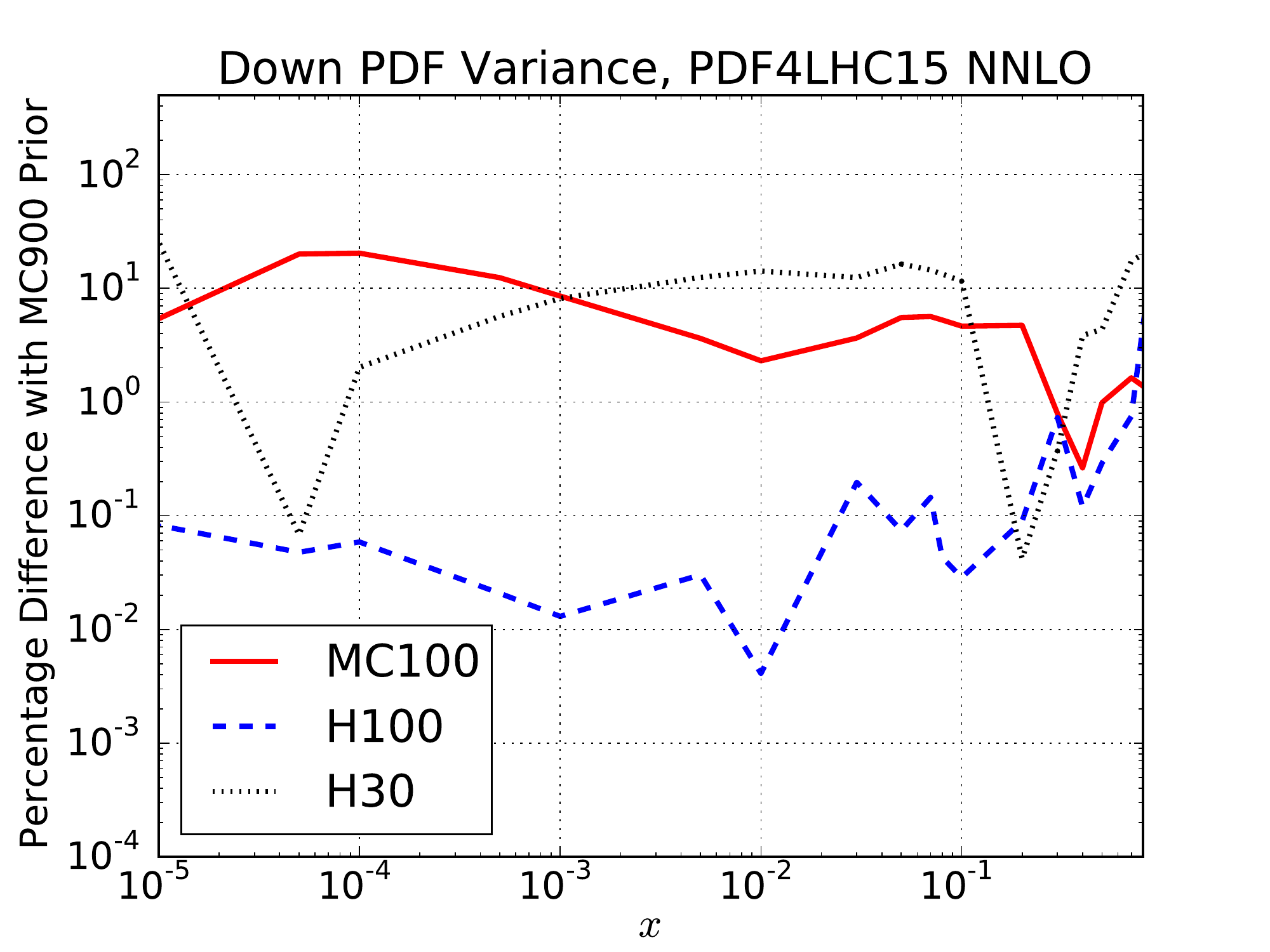}
   \includegraphics[width=.48\textwidth]{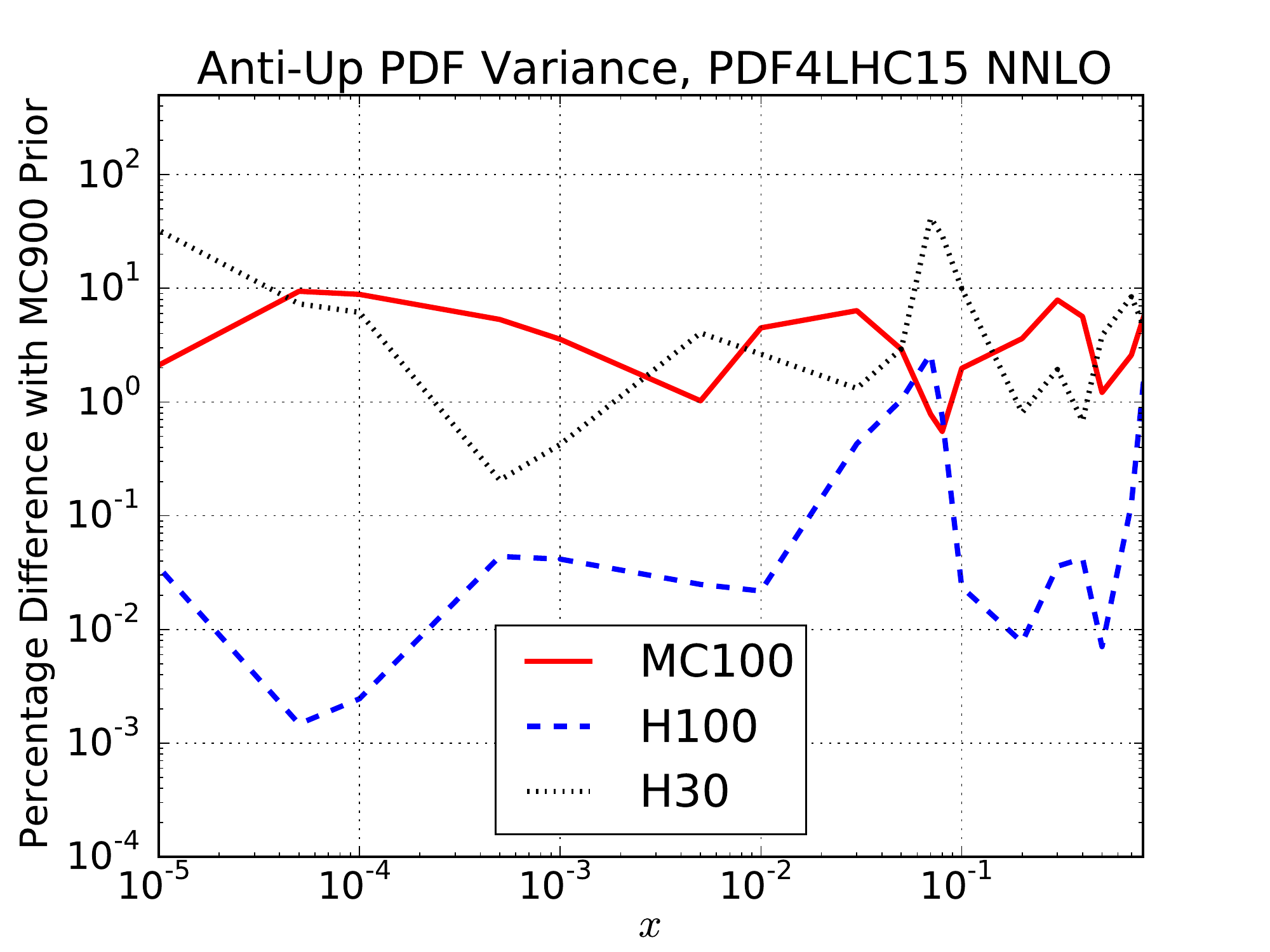}
   \includegraphics[width=.48\textwidth]{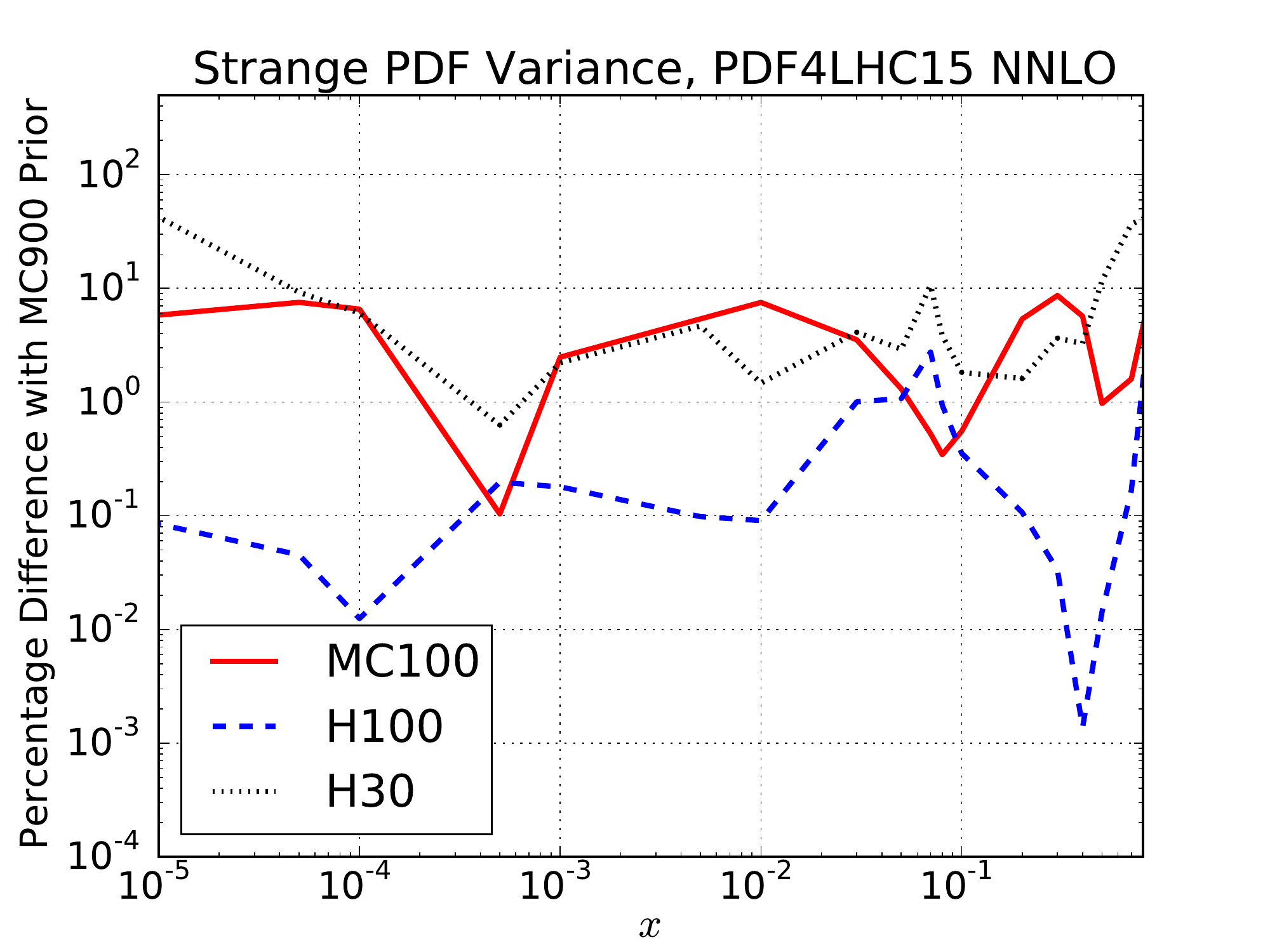}
   \caption{\small Percentage difference between 
     the variances on the gluon, down, anti-up and strange
     distributions computed using the prior MC900 set, and each of the
     three reduced sets (see text).
}  
\label{fig:comp_sig_4}
\end{figure}

Results for the percentage differences in the variance (that is, in
the PDF uncertainties) between the prior
and each of the three reduced sets are shown in
Fig.~\ref{fig:comp_sig_4}.
We see that Hessian set with 100 eigenvectors 
reproduces in all cases the variances of the prior
with precision at worst of 1\%, typically much better than that.
The Monte Carlo set and the Hessian with 30 eigenvectors  
lead to  similar performances in terms of reproducing
the variances of the prior, with differences that can be up to
10\%.

\begin{figure}[t]
  \centering
  \includegraphics[scale=.63]{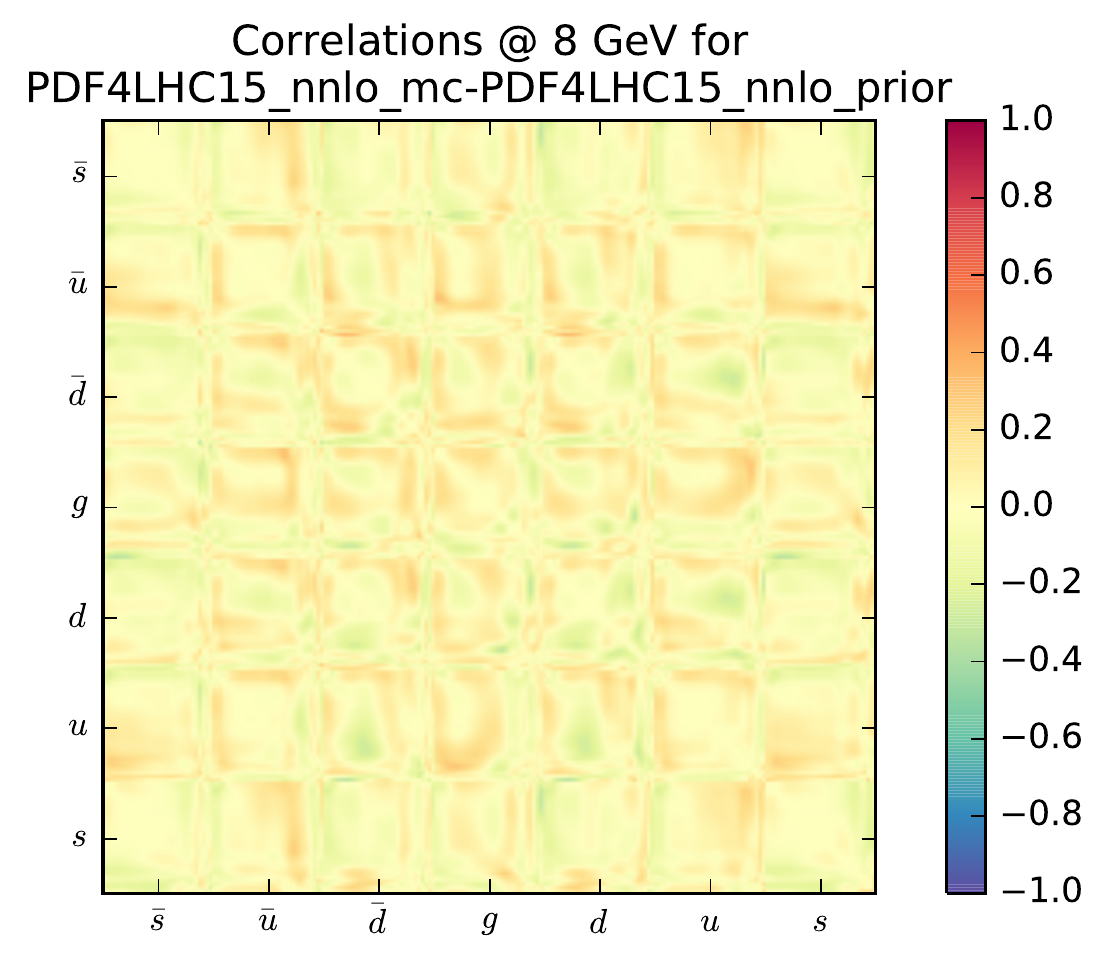}
  \includegraphics[scale=.63]{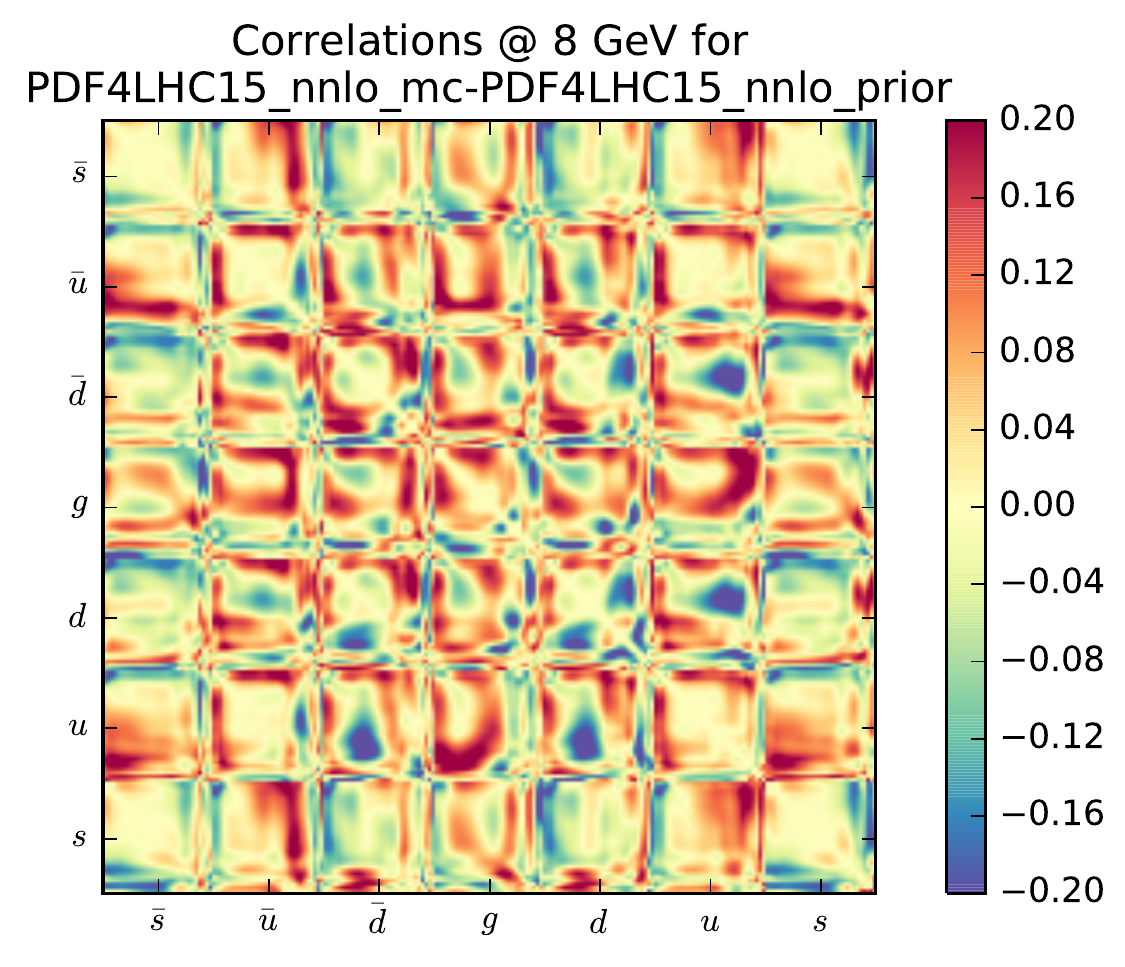}
  \caption{\small
    Left plot: difference between the PDF correlation coefficients
    computed with the prior {\tt PDF4LHC15\_nnlo\_prior}
    and its compressed Monte Carlo representation,
    {\tt PDF4LHC15\_nnlo\_mc}.
    The plot shows the $7\times 7$ matrix for all possible
    comparisons of light flavor PDFs, with in each case $x$ ranging 
   between $x_{\rm min}=10^{-5}$
    and $x_{\rm max}=0.9$ (on a logarithmic scale) and fixed
    $Q=8$~GeV.
    Right plot: same, but with the scale for the difference magnified
    to only cover the range from $-0.2$ to $0.2$.
}  
\label{fig:cmc100nlocorr}
\end{figure}

We then turn to a comparison of PDF correlations.
In Fig.~\ref{fig:cmc100nlocorr} we show the
 difference between the PDF correlation coefficients
    computed with the prior set {\tt PDF4LHC15\_nnlo\_prior}
    and its compressed Monte Carlo representation,
     {\tt PDF4LHC15\_nnlo\_mc}.
    The plot shows the $7\times 7$ matrix of all possible correlations
    of light flavor PDFs (three quarks and antiquarks
    and the gluon), sampled on logarithmically spaced $x$ values
   between $x_{\rm min}=10^{-5}$
    and $x_{\rm max}=0.9$, and with
    $Q=8$~GeV.
    In the left plot of Fig.~\ref{fig:cmc100nlocorr} the range
    of difference shown is between -1 and 1, while
    in the right plot we zoom in to a range of difference between -0.2 and 0.2.
    We find good agreement in general, with differences never
    exceeding 0.2 in modulus.
    Note that the correlation coefficients can easily vary
    by 0.1-0.2 or more
    among the input PDF sets.

\begin{figure}[t]
  \centering
  \includegraphics[scale=.63]{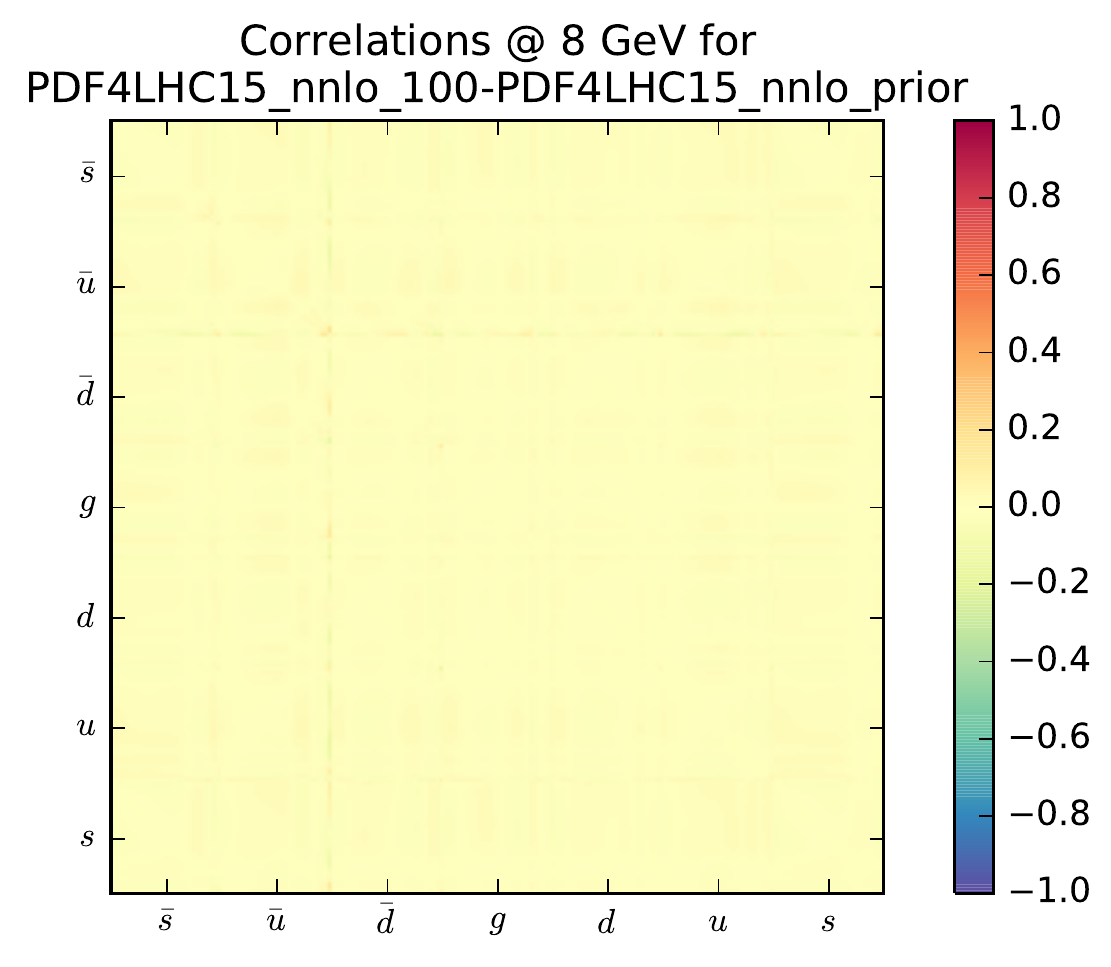}
  \includegraphics[scale=.63]{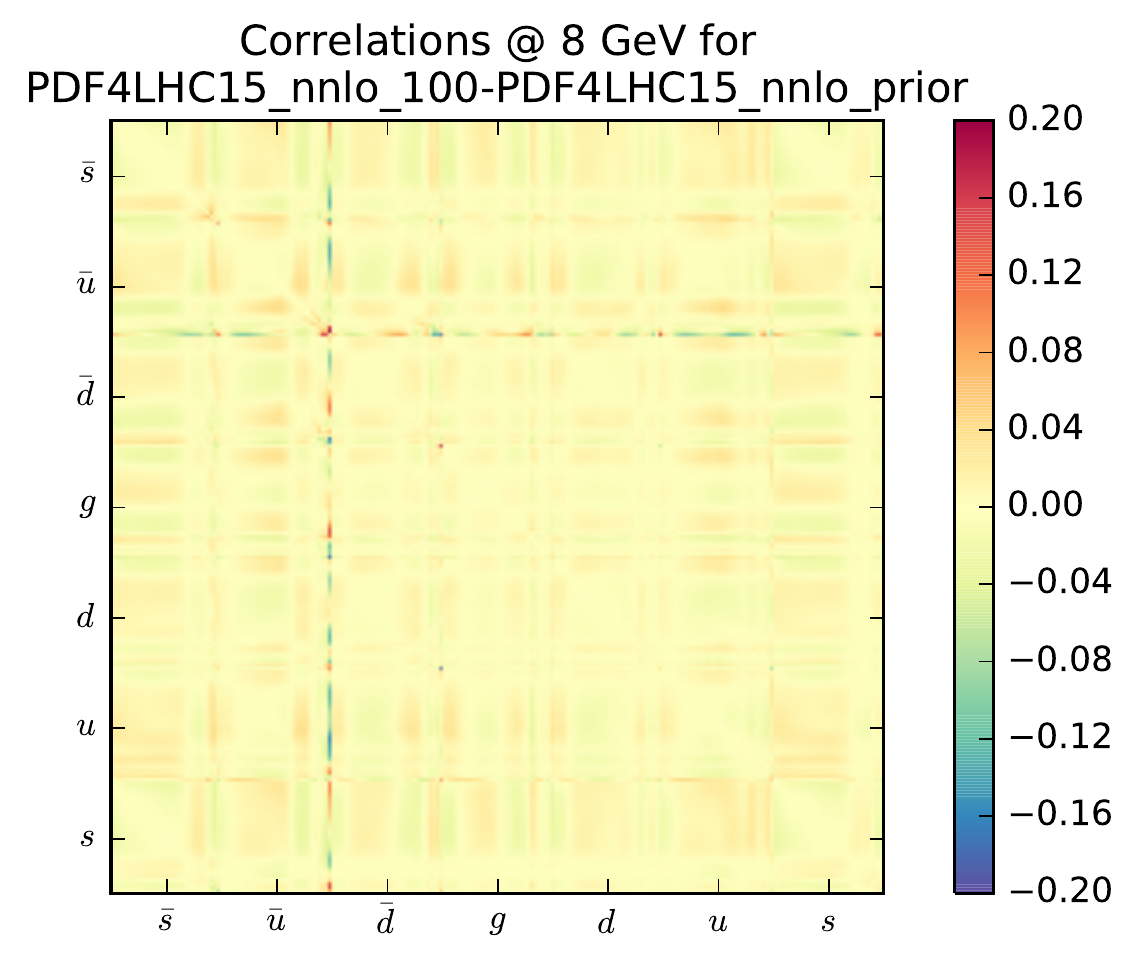}
  \includegraphics[scale=.63]{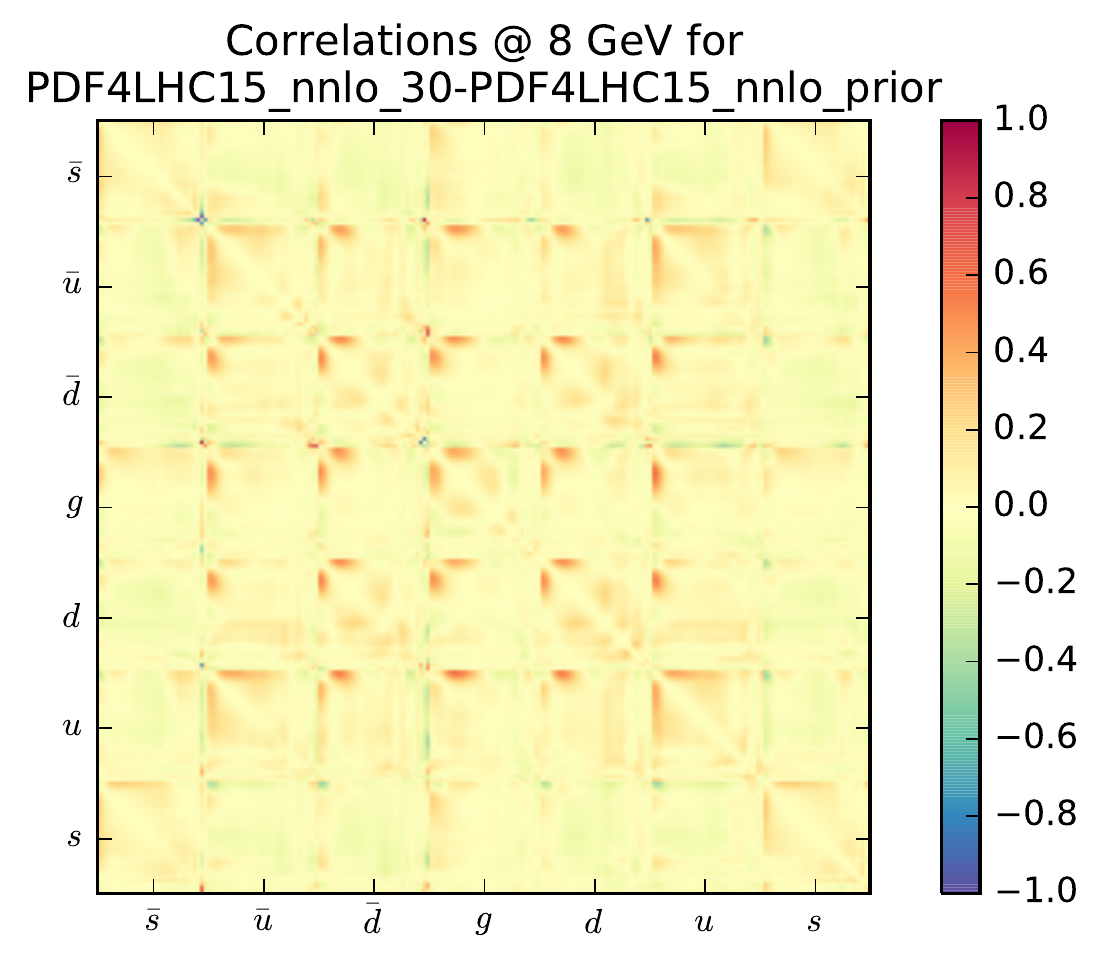}
  \includegraphics[scale=.63]{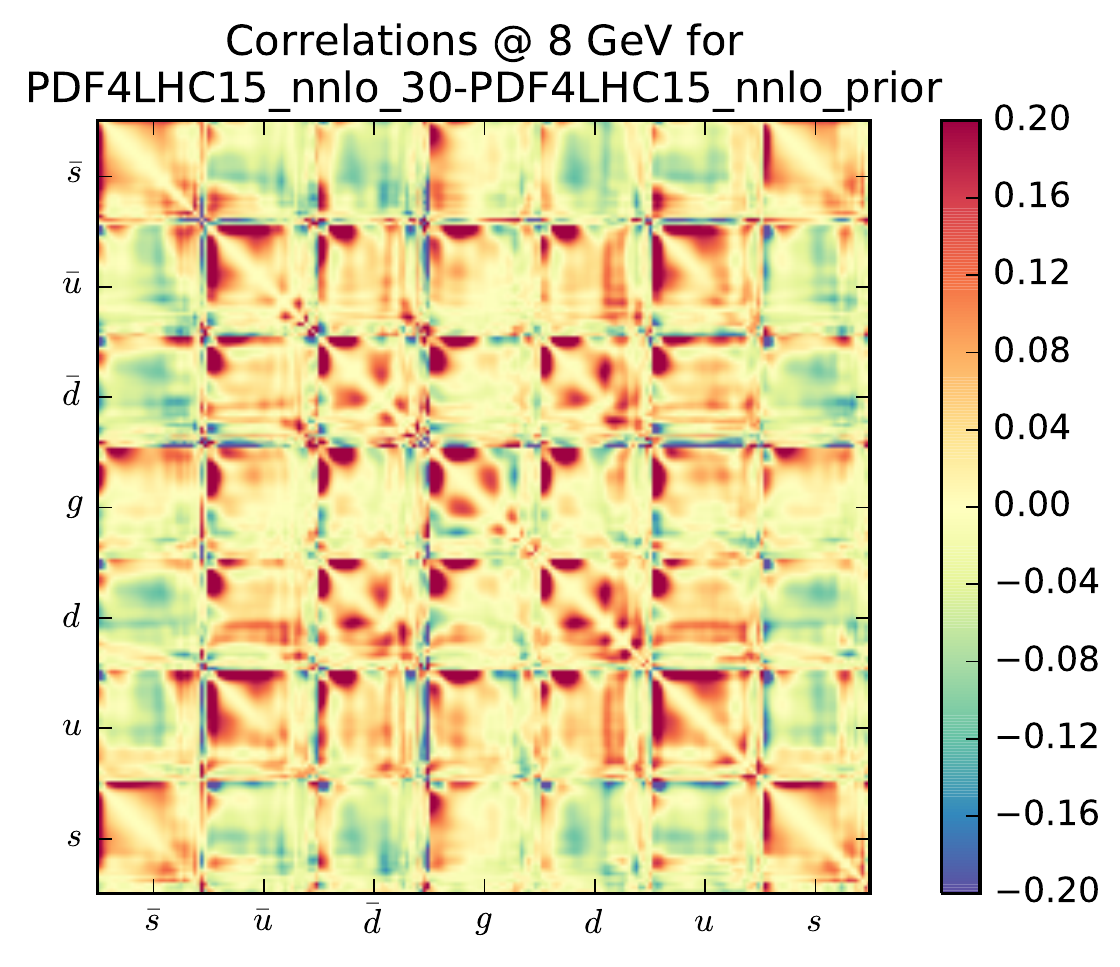}
  \caption{\small
    Same as Fig.~\ref{fig:cmc100nlocorr} now comparing the
    the prior to
    the Hessian sets,  {\tt  PDF4LHC15\_nnlo\_100} (upper plots)
    and  {\tt  PDF4LHC15\_nnlo\_30} (lower plots).
}  
\label{fig:cmc100nlocorr2}
\end{figure}

The corresponding comparison of PDF correlations for the two Hessian
sets,  {\tt  PDF4LHC15\_nnlo\_100} (upper plots)
and  {\tt  PDF4LHC15\_nnlo\_30} (lower plots), is shown
in Fig.~\ref{fig:cmc100nlocorr2}.
For the case of the $N_{\rm eig}=100$ set the correlations are essentially
identical to those of the prior, with differences below 0.01.
For the $N_{\rm eig}=30$ set instead, typical differences in the
correlation coefficients are somewhat larger, comparable to those
found when using 
the MC set {\tt PDF4LHC15\_nnlo\_mc}
in Fig.~\ref{fig:cmc100nlocorr}. 

\subsection{Implications for LHC processes}

We  now consider
the comparison at the level of LHC cross-sections and distributions.
All comparisons reported in this document have been  performed using
NLO matrix elements, for which fast computational tools are available,
even when using  NNLO PDFs: the comparison is
performed for PDF validation purposes only.
The NLO calculations that have been used for the
validation presented in this report have been produced used {\tt APPLgrid}~\cite{Carli:2010rw}
interfaced
to {\tt NLOjet++}~\cite{Nagy:2003tz} and {\tt MCFM}~\cite{Campbell:2002tg},
and with {\tt aMCfast}~\cite{amcfast}
interfaced to {\tt Madgraph5\_aMC@NLO}~\cite{Alwall:2014hca}.
In all cases we use the default theory settings, including the scale
choices, of the respective codes.
Theory calculations have been performed at 7 TeV for these processes
for
which data is already available and that have been used
for PDF fits, see for example~\cite{Ball:2014uwa}; in addition,
a number of dedicated grids for
13 TeV processes have also been generated.
In the former case the binning follows that to the corresponding
experimental measurements, which we indicate in the list below.

In this report we will
 only show a representative subset of processes.
The complete
list of processes and kinematic
distributions (available from the PDF4LHC15 webpage) is described now.
At 7 TeV, the processes where experimental LHC measurements are available,
and for which {\tt APPLgrid} grids matching the experimental binning have been produced,
are the following:
\begin{itemize}
\item Drell-Yan rapidity distributions in the LHCb forward region~\cite{Aaij:2012vn,Aaij:2012mda}.
\item Invariant mass distribution from high-mass Drell-Yan (DY)~\cite{Aad:2013iua}.
\item Rapidity distributions of $W$ and $Z$ production~\cite{Chatrchyan:2013mza,Aad:2011dm}.
\item $p_T$ distribution of inclusive $W$ production~\cite{Aad:2011fp}.
\item Double differential DY distributions in dilepton mass and rapidity~\cite{CMSDY}.
\item Lepton rapidity distributions in $W$+charm production~\cite{Chatrchyan:2013uja}.
\item Inclusive jet production in the central and forward regions~\cite{Aad:2011fc,Aad:2013lpa}.
\end{itemize}
In addition to these grids, at 13 TeV we have generated specifically for this benchmark
exercises a number of new fast NLO grids for differential distributions in
Higgs, $t\bar{t}$ and vector boson production using
{\tt Madgraph5\_aMC@NLO} interfaced to {\tt aMCfast},
namely:
\begin{itemize}
\item Rapidity and $p_T$ distributions in inclusive $gg\to h$ production,
  as well as total cross-sections for $hZ$, $hW$ and $ht\bar{t}$ production.
\item Rapidity, $p_T$ and $m_{t\bar{t}}$ distributions in top-quark pair production.
\item Missing $E_T$, lepton $p_T$ and rapidity,
  and transverse mass distributions in inclusive $W$  and $Z$ production.
\end{itemize}

First of all, let us go back to Figs.~\ref{fig:nongaussian1} and~\ref{fig:nongaussian2}, which
illustrated two cases of LHC cross-sections where departures from the Gaussian regime
were particularly striking.
In these figures, the
probability distributions computed from the MC900 prior, is compared to those
computed using the CMC100 ({\tt PDF4LHC15\_mc}) and MCH100
({\tt PDF4LHC15\_100}) reduced sets.
It is clear that while the Monte Carlo {\tt PDF4LHC15\_mc} set is able to
capture these non-Gaussian features, this is not the case for the
Hessian set {\tt PDF4LHC15\_100}.

\begin{figure}
  \centering
  \includegraphics[scale=.38]{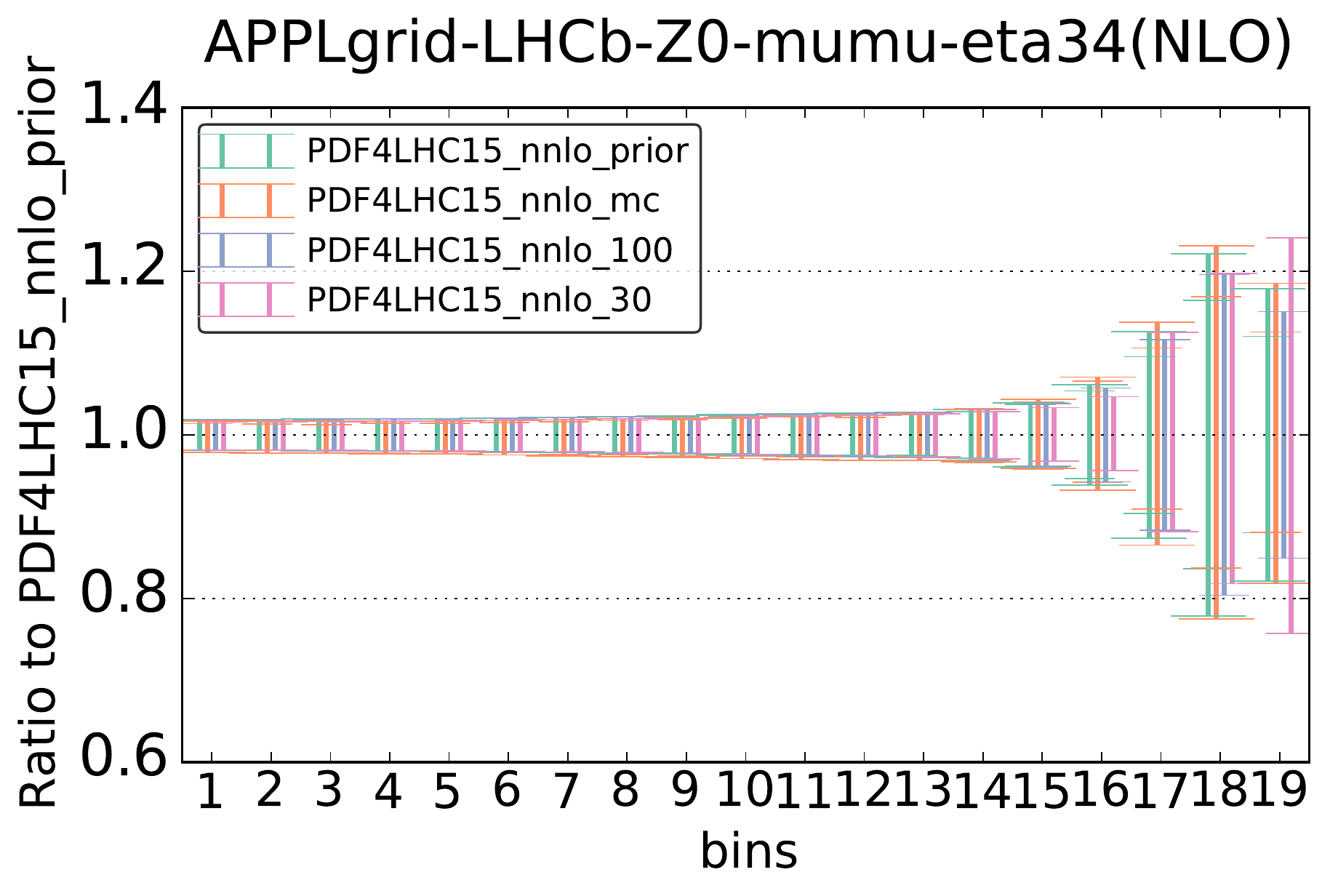}
  \includegraphics[scale=.38]{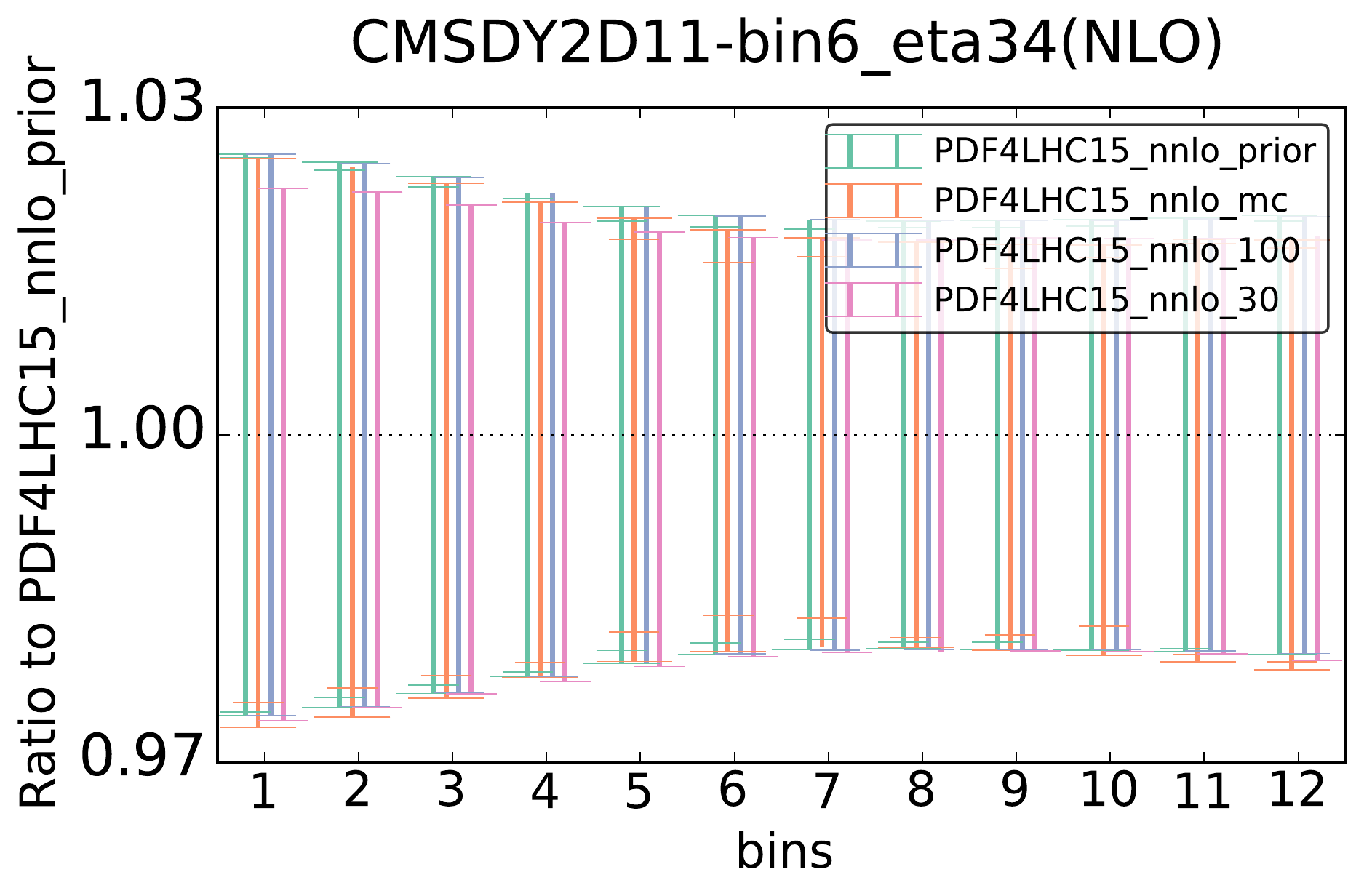}
  \includegraphics[scale=.38]{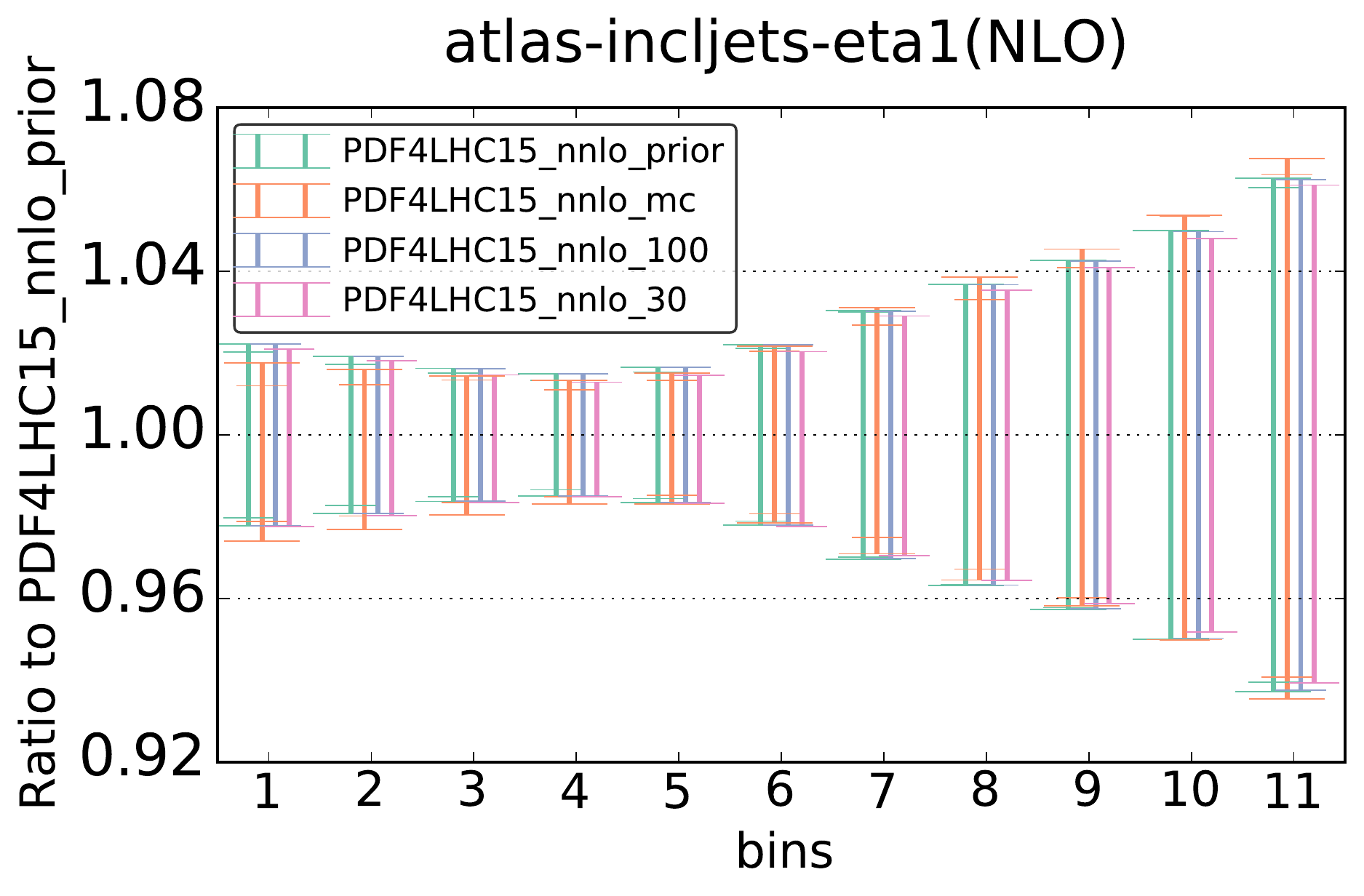}
  \includegraphics[scale=.38]{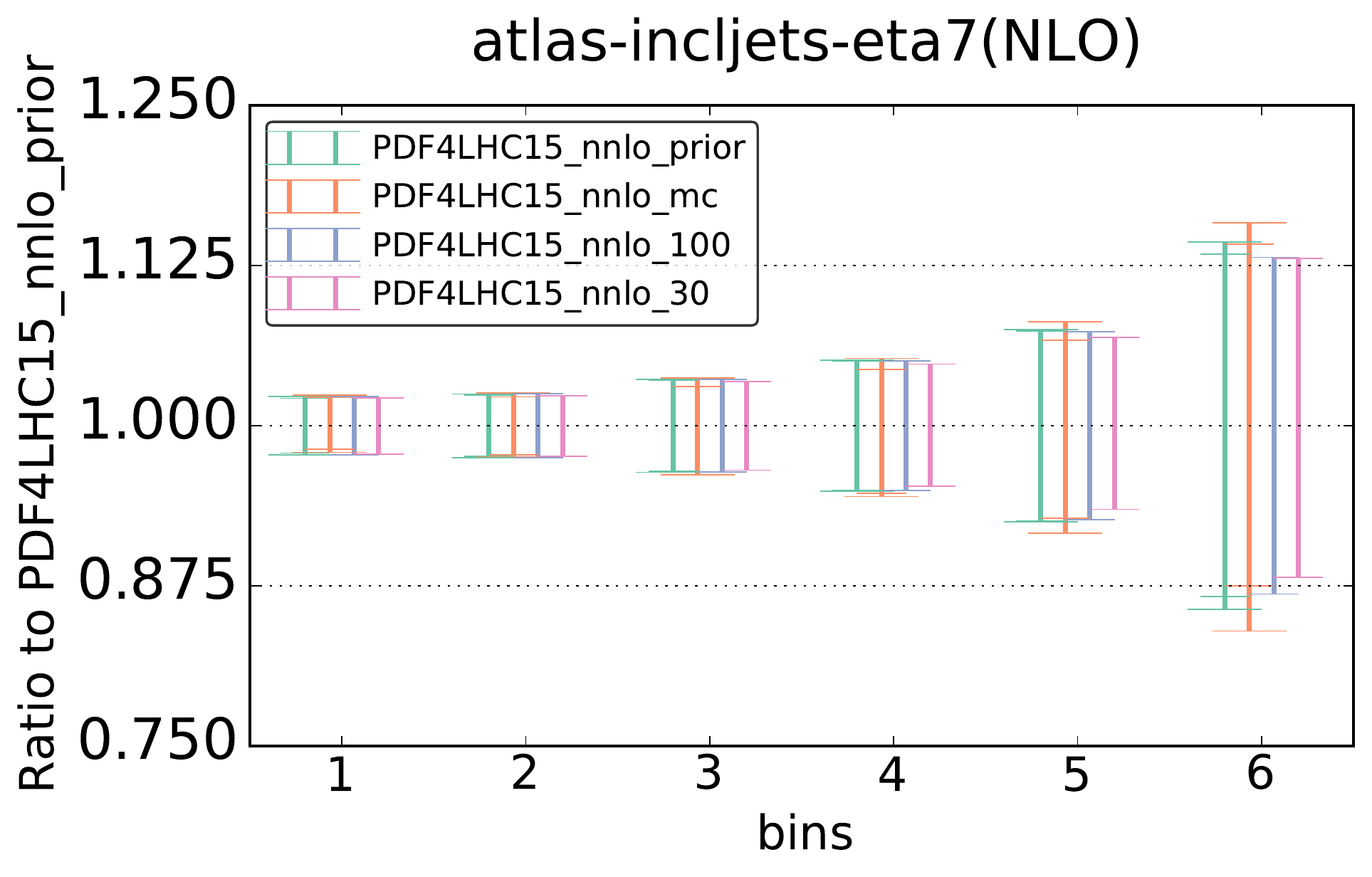}
  \includegraphics[scale=.38]{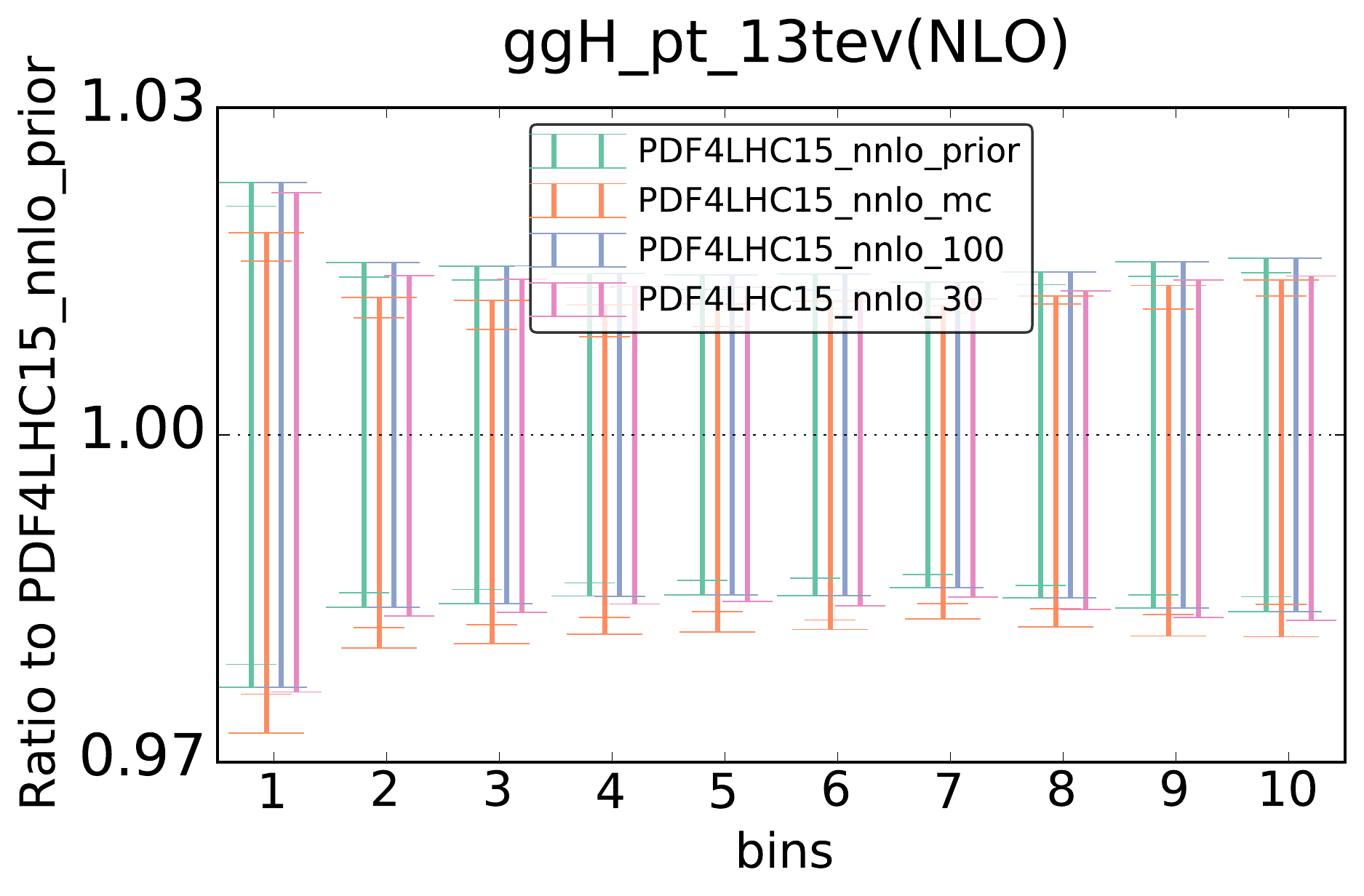}
  \includegraphics[scale=.38]{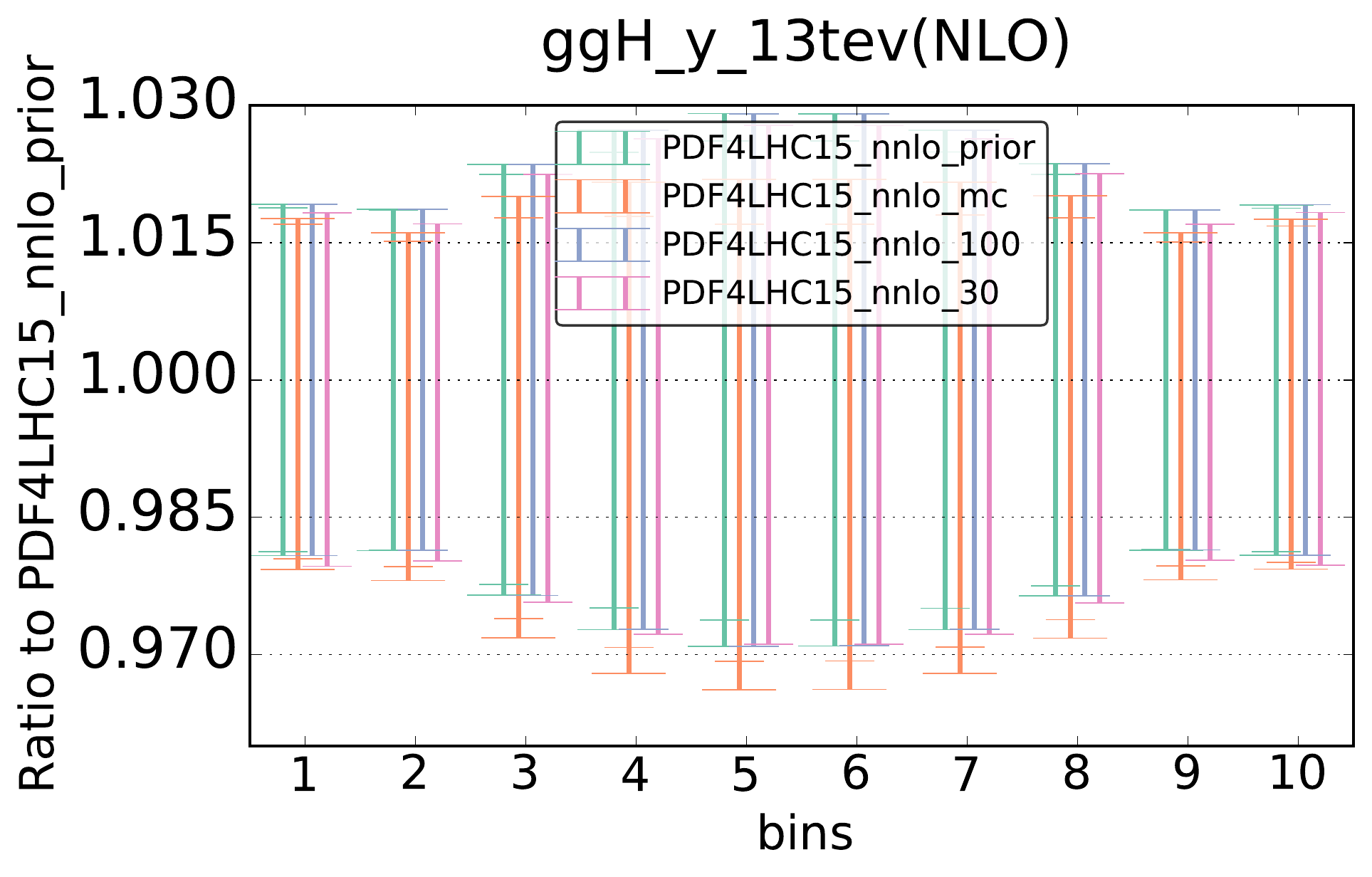}
  \includegraphics[scale=.38]{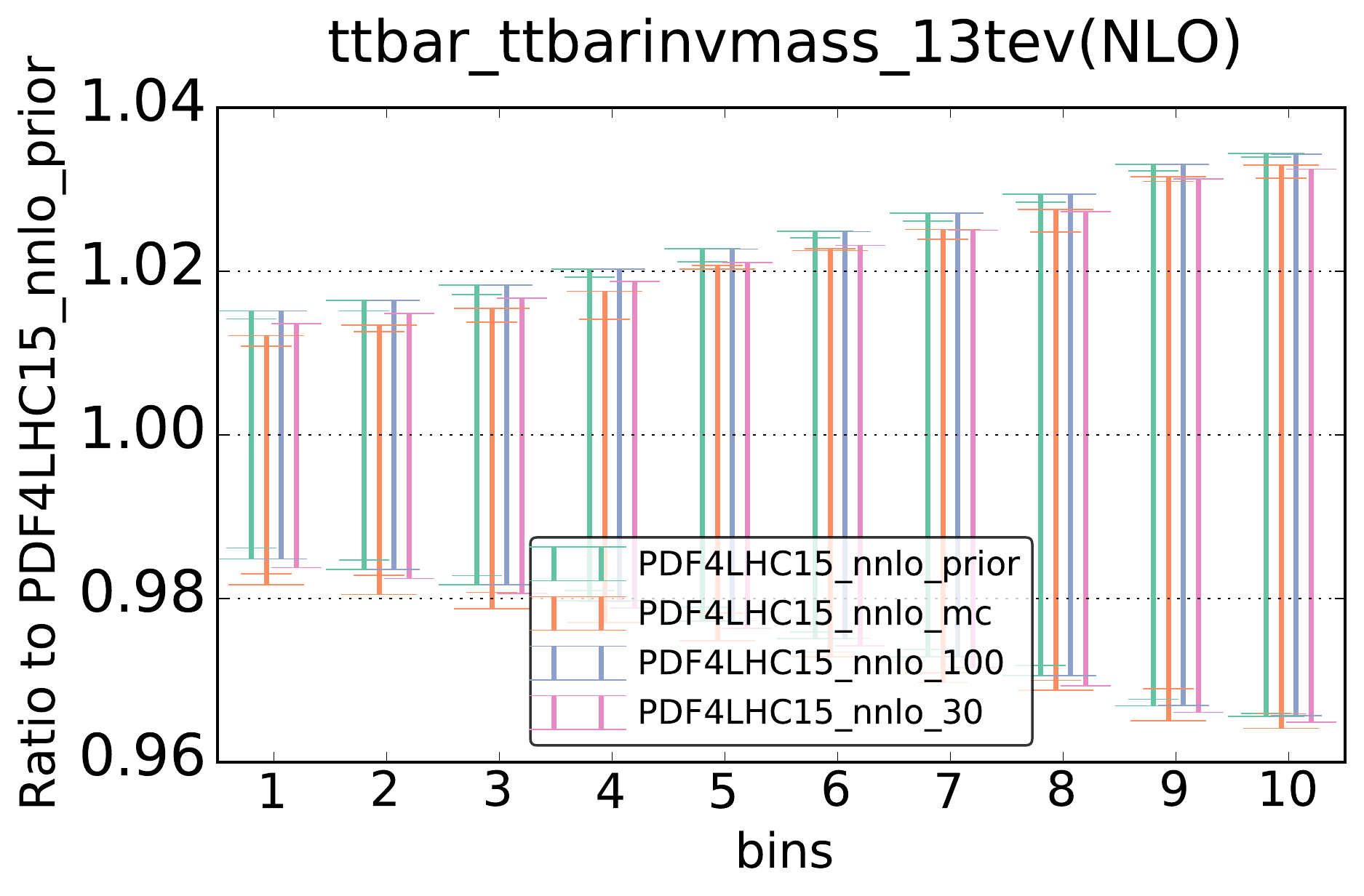}
   \includegraphics[scale=.36]{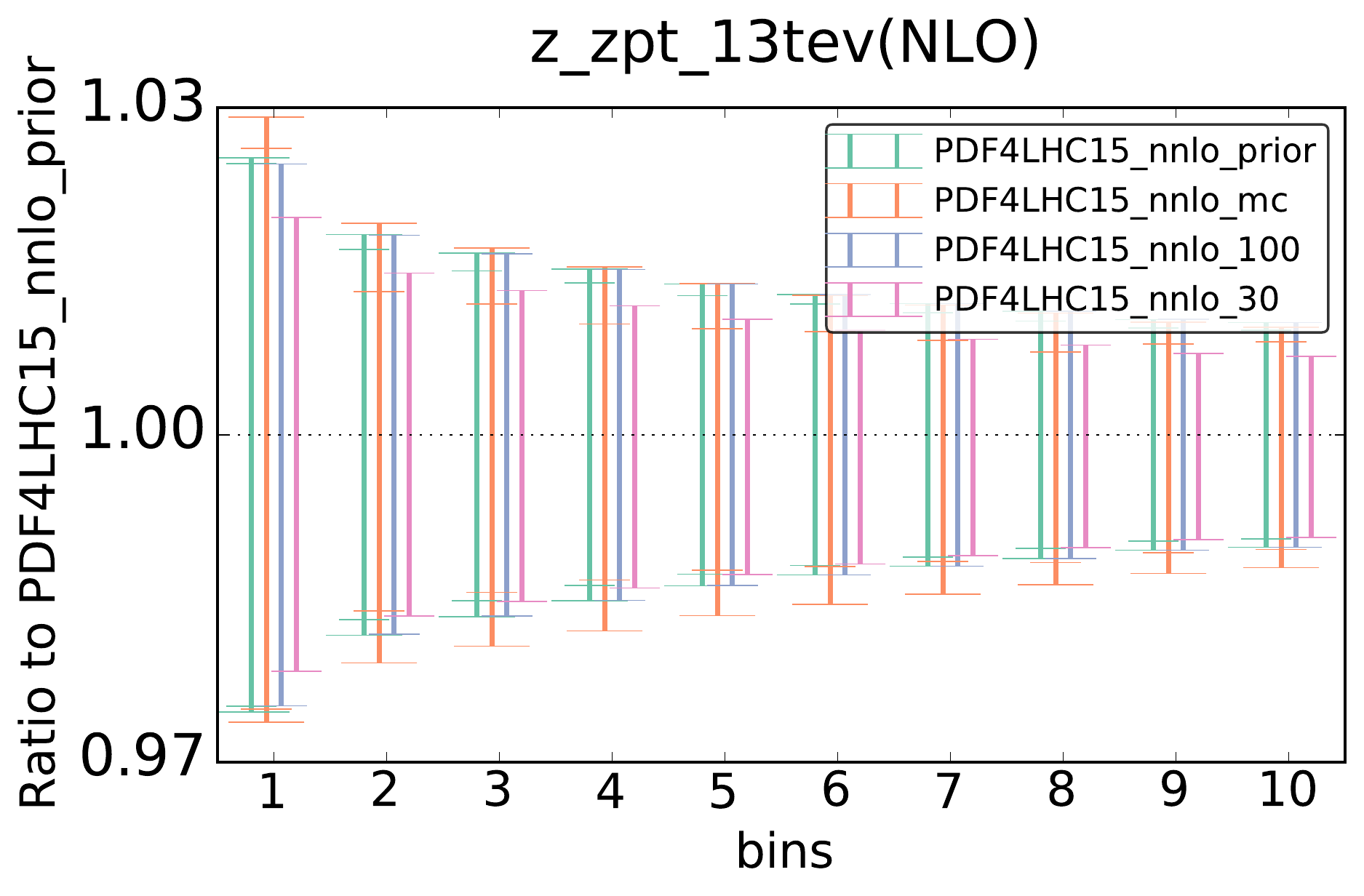}
  \caption{\small
    Comparison of differential distributions for several LHC
    processes, computed using 
   the prior and the three reduced sets.
    From top to bottom and from left to right we show the forward DY rapidity
    distributions at LHCb, the CMS DY double-differential
    distributions, the ATLAS 2010 inclusive jets in the forward
    and central regions (all these at 7 TeV),
    the $p_T$ and rapidity distributions of $gg\to h$ production,
    the $m_{t\bar{t}}$ distribution in $t\bar{t}$ production
    and the $Z$ $p_T$ distribution (all these at 13 TeV).
    See text for more details.
}  
\label{LHCxsec}
\end{figure}

Then, in Fig.~\ref{LHCxsec} we show the comparison between the three
PDF4LHC15 combined sets for a representative selection of
LHC cross-sections.
In particular we show  from top to bottom and from left to right  
the forward DY
 rapidity
    distributions at LHCb, the CMS DY double-differential
    distributions, the ATLAS 2010 inclusive jets in the central
    and forward regions (all these at 7 TeV) and then
    the $p_T$ and rapidity distributions of for Higgs production in gluon fusion,
    the $m_{t\bar{t}}$ distribution in $t\bar{t}$ production
    and the $Z$ $p_T$ distribution (all these at 13 TeV).
We compare the prior with the three reduced sets, the MC and the two Hessian sets.
As can be seen, the agreement is in general good in all cases.

\newpage

An important application of the PDF4LHC15 combined sets
is the computation of  the correlation coefficients
between  LHC processes, such as for instance
signal and background processes in Higgs production.
Within the Higgs Cross Section Working Group (HXSWG),
the correlation coefficients between some of these
processes were computed using the original PDF4LHC recommendation:
see Table 10 of Ref.~\cite{Dittmaier:2012vm}.
We have recomputed these correlation coefficients using MC900,
and compared the results with the three reduction methods.
The processes included, at the LHC 13 TeV, are
vector boson production, both $W$ and $Z$,
$t\bar{t}$ production, and then Higgs production in gluon
fusion $ggh$, in associated production $hZ$ and $hW$, and
in association with a top quark pair, $ht\bar{t}$.

\begin{figure}[t]
  \centering
  \includegraphics[scale=.63]{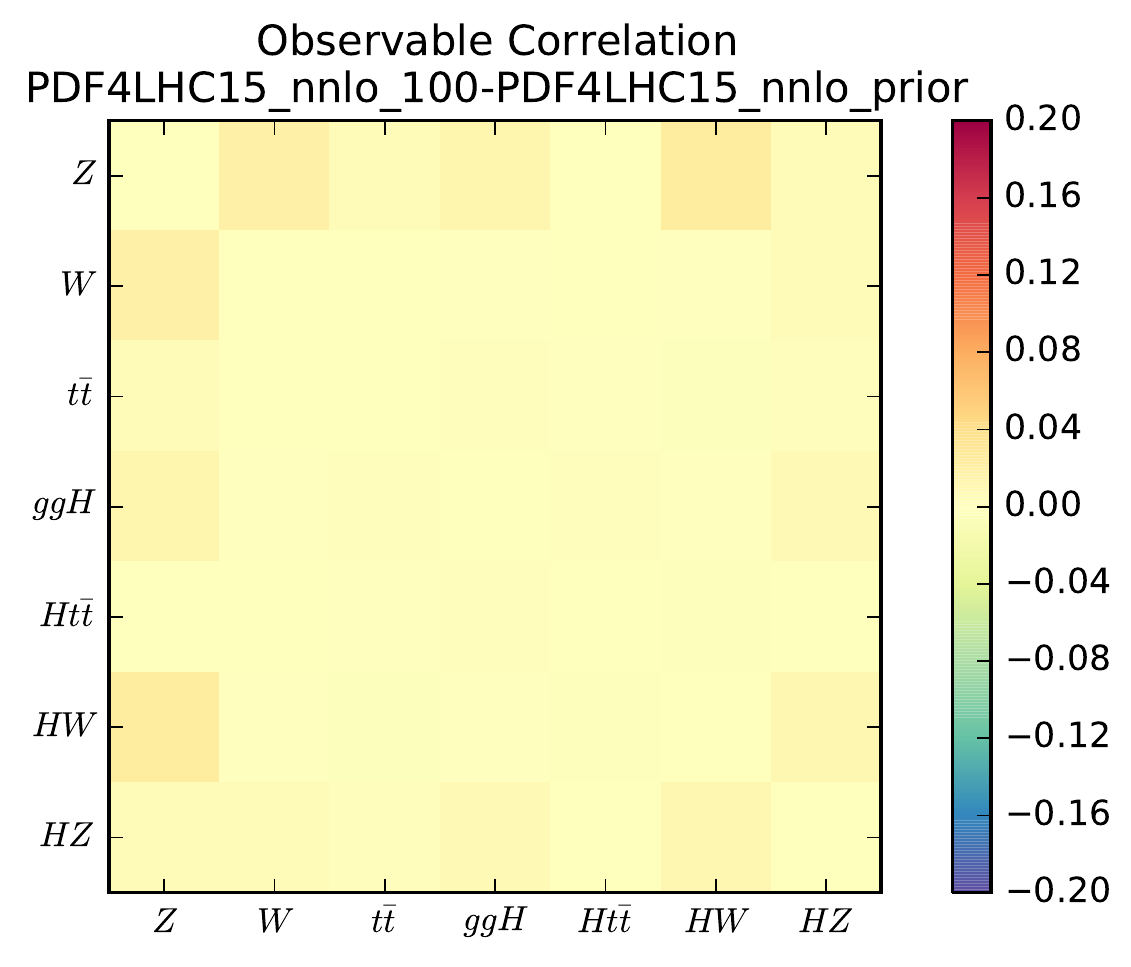}
  \includegraphics[scale=.63]{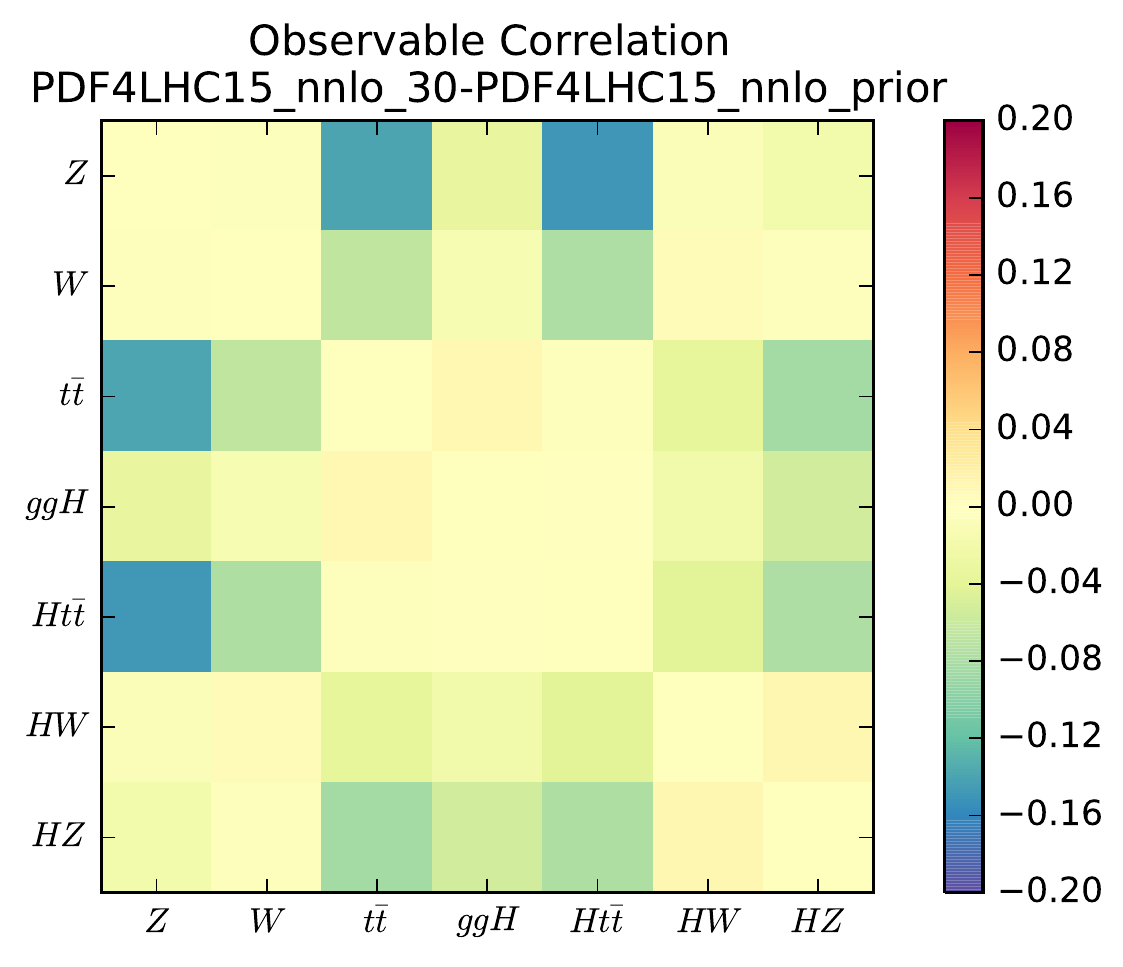}
  \includegraphics[scale=.63]{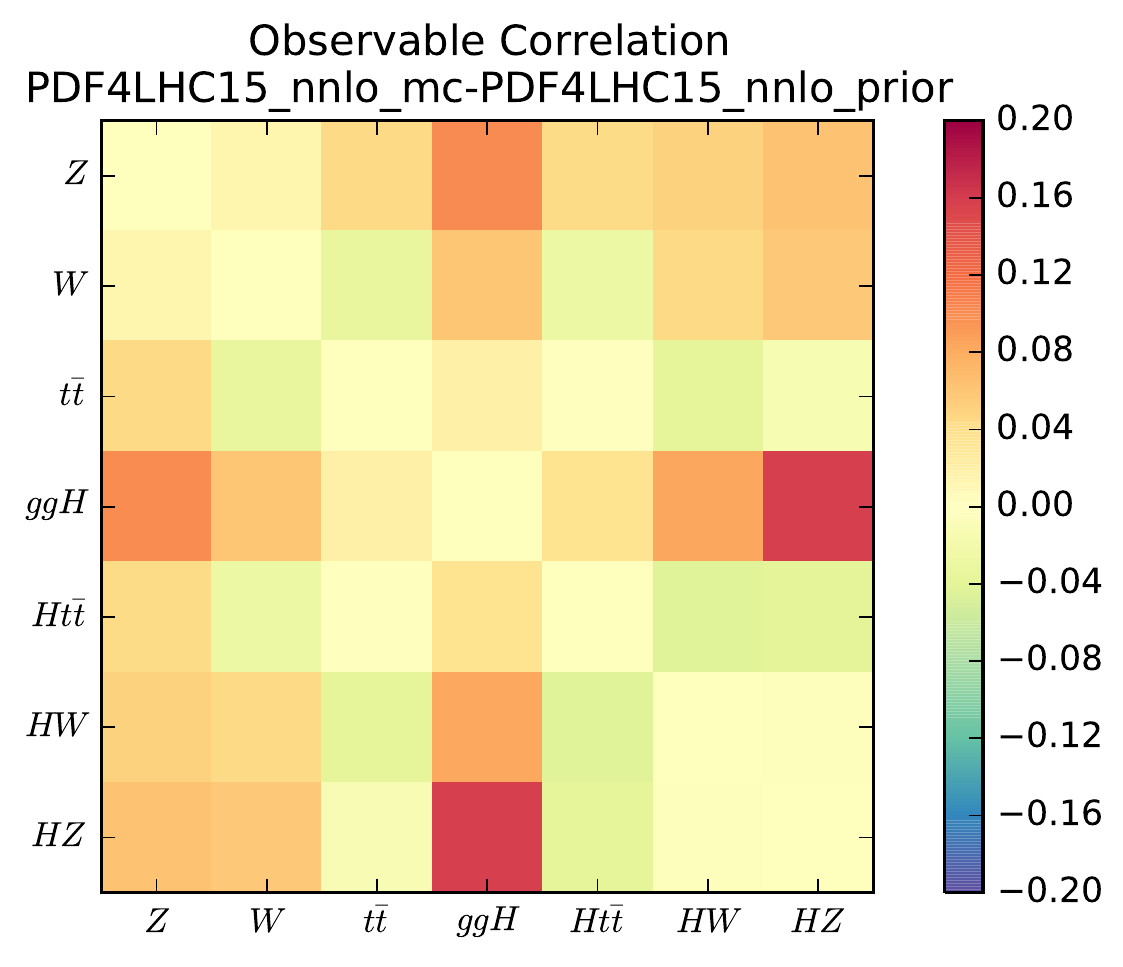}
   \caption{\small
     The difference between the correlation coefficients
     for a number of signal and background processes relevant
     for Higgs production at the LHC 13 TeV, see
     text for more details.
     We compare the correlation coefficients of the
     prior set
      {\tt PDF4LHC15\_nnlo\_prior}
     with those of the two Hessian sets,
     {\tt  PDF4LHC15\_nnlo\_100}
     (upper left plot) and  {\tt  PDF4LHC15\_nnlo\_30} (upper right plot),
     as well as with the {\tt  PDF4LHC15\_nnlo\_mc} set
     (lower plot).
}  
\label{fig:corr_obs_mch30}
\end{figure}

In Fig.~\ref{fig:corr_obs_mch30} we show 
the difference between the correlation coefficients computed using the
     prior set
      {\tt PDF4LHC15\_nnlo\_prior}
and the two Hessian sets,
     {\tt  PDF4LHC15\_nnlo\_100}
     (upper left plot) and  {\tt  PDF4LHC15\_nnlo\_30} (upper right plot)
     as well as with the MC set, {\tt  PDF4LHC15\_nnlo\_mc} (lower plot).
     As we can see, with the $N_{\rm eig}=100$ set the correlation coefficients
     are always reproduced within a few percent, while
     for the $N_{\rm eig}=30$  and the MC sets
     somewhat larger differences are found,
     a few instances
     up to 0.2.

In addition, we have tabulated the
correlation coefficients
for various LHC total cross-sections computed using the prior
and the various reduced sets.
First of all, in Table~\ref{tab:correlation1}
we show the value of the correlation coefficient between the
$Z$ production cross-sections and the $W$, $t\bar{t}$, $ggh$,
$ht\bar{t}$, $hW$ and $hZ$ production cross-sections.
We compare the PDF4LHC15 prior with the Monte Carlo and the two Hessian reduced sets,
both at NLO and at NNLO.

The information from this table is the consistent with that shown 
in Fig.~\ref{fig:corr_obs_mch30}, namely
the Hessian set with 100 eigenvectors reproduces always the correlation
coefficient of the prior, with an accuracy better than 1\%.
Using
the smaller Hessian set and the Monte Carlo set the correlation
coefficients are reproduced not as perfectly, but still sufficiently well
for most phenomenological applications.
It should be kept in mind that the correlations 
themselves from the prior set have sizable uncertainties, in the sense that the 
individual correlations from  the 3 different PDF families may differ significantly
from each other.
Thus, the correlations given in the tables provide an average 
behavior over the 3 PDF sets, and probably only the first digit in the correlation 
coefficient should be considered significant. 
The same conclusion can be derived from a variety of other pairs
of LHC cross-sections, collected in
Tables~\ref{tab:correlation2} to~~\ref{tab:correlation4}.

\begin{table}[t]
\begin{centering}
\begin{tabular}{l|c|c|c|c|c|c}
\hline 
\multirow{2}{*}{PDF Set} & \multicolumn{6}{c}{Correlation coefficient}\tabularnewline
 & $Z,W$ & $Z,t\bar{t}$ & $Z,ggh$ & $Z,ht\bar{t}$ & $Z,hW$ & $Z,hZ$\tabularnewline
\hline 
PDF4LHC15\_nlo\_prior & 0.90 & -0.60 & 0.22 & -0.64 & 0.55 & 0.74\tabularnewline
PDF4LHC15\_nlo\_mc & 0.92 & -0.49 & 0.41 & -0.58 & 0.61 & 0.77\tabularnewline
PDF4LHC15\_nlo\_100 & 0.92 & -0.60 & 0.23 & -0.64 & 0.57 & 0.75\tabularnewline
PDF4LHC15\_nlo\_30 & 0.90 & -0.68 & 0.16 & -0.71 & 0.55 & 0.76\tabularnewline
\hline 
\hline 
PDF4LHC15\_nnlo\_prior & 0.89 & -0.49 & 0.08 & -0.46 & 0.56 & 0.74\tabularnewline
PDF4LHC15\_nnlo\_mc & 0.90 & -0.44 & 0.18 & -0.42 & 0.62 & 0.80\tabularnewline
PDF4LHC15\_nnlo\_100 & 0.91 & -0.48 & 0.09 & -0.46 & 0.59 & 0.74\tabularnewline
PDF4LHC15\_nnlo\_30 & 0.88 & -0.63 & 0.04 & -0.61 & 0.56 & 0.72\tabularnewline
\hline 
\end{tabular}
\par\end{centering}
\caption{\small \label{tab:correlation1} Correlation coefficient between the
$Z$ production cross-sections and the $W$, $t\bar{t}$, $ggh$,
$ht\bar{t}$, $hW$ and $hZ$ production cross-sections.
The PDF4LHC15 prior is compared to  the Monte Carlo and the two
Hessian  reduced sets,
both at NLO and at NNLO.
}
\end{table}

\begin{table}[t]
\begin{centering}
\begin{tabular}{l|c|c|c|c|c|c}
\hline 
\multirow{2}{*}{PDF Set} & \multicolumn{5}{c}{Correlation coefficient} & \tabularnewline
 & $W,t\bar{t}$ & $W,ggh$ & $W,ht\bar{t}$ & $W,hW$ & $W,hZ$ & $t\bar{t},ggh$\tabularnewline
\hline 
{ PDF4LHC15\_nlo\_prior} & -0.46 & 0.32 & -0.51 & 0.77 & 0.78 & 0.27\tabularnewline
{ PDF4LHC15\_nlo\_mc }& -0.35 & 0.49 & -0.46 & 0.81 & 0.80 & 0.27\tabularnewline
{ PDF4LHC15\_nlo\_100} & -0.47 & 0.32 & -0.52 & 0.77 & 0.79 & 0.27\tabularnewline
{ PDF4LHC15\_nlo\_30} & -0.52 & 0.28 & -0.56 & 0.79 & 0.81 & 0.32\tabularnewline
\hline 
\hline 
{ PDF4LHC15\_nnlo\_prior} & -0.40 & 0.20 & -0.40 & 0.76 & 0.77 & 0.30\tabularnewline
{ PDF4LHC15\_nnlo\_mc }& -0.44 & 0.26 & -0.42 & 0.81 & 0.82 & 0.32\tabularnewline
{ PDF4LHC15\_nnlo\_100} & -0.40 & 0.20 & -0.40 & 0.76 & 0.77 & 0.30\tabularnewline
{ PDF4LHC15\_nnlo\_30} & -0.47 & 0.19 & -0.47 & 0.77 & 0.76 & 0.31\tabularnewline
\hline 
\end{tabular}
\par\end{centering}
\caption{\small \label{tab:correlation2} Same as Table~\ref{tab:correlation1}
  for the correlation coefficient of additional pairs
of LHC inclusive cross-sections.}
\end{table}

\begin{table}[t]
\begin{centering}
\begin{tabular}{l|c|c|c|c|c|c}
\hline 
\multirow{2}{*}{PDF Set} & \multicolumn{5}{c}{Correlation coefficient} & \tabularnewline
 & $t\bar{t},Ht\bar{t}$ & $t\bar{t},hW$ & $t\bar{t},hZ$ & $ggh,ht\bar{t}$ & $ggh,hW$ & $ggh,hZ$\tabularnewline
\hline 
PDF4LHC15\_nlo\_prior & 0.93 & -0.22 & -0.50 & -0.02 & 0.15 & 0.08\tabularnewline
PDF4LHC15\_nlo\_mc & 0.92 & -0.14 & -0.41 & -0.04 & 0.33 & 0.27\tabularnewline
PDF4LHC15\_nlo\_100 & 0.93 & -0.22 & -0.48 & -0.03 & 0.15 & 0.08\tabularnewline
PDF4LHC15\_nlo\_30 & 0.93 & -0.25 & -0.54 & 0.02 & 0.11 & -0.01\tabularnewline
\hline 
\hline 
PDF4LHC15\_nnlo\_prior & 0.87 & -0.23 & -0.34 & -0.13 & -0.01 & -0.17\tabularnewline
PDF4LHC15\_nnlo\_mc & 0.87 & -0.27 & -0.35 & -0.10 & 0.07 & -0.01\tabularnewline
PDF4LHC15\_nnlo\_100 & 0.87 & -0.24 & -0.34 & -0.13 & -0.02 & -0.17\tabularnewline
PDF4LHC15\_nnlo\_30 & 0.87 & -0.27 & -0.43 & -0.13 & -0.04 & -0.23\tabularnewline
\hline 
\end{tabular}
\par\end{centering}
\caption{\small \label{tab:correlation3} Same as Table~\ref{tab:correlation1}
  for the correlation coefficient of additional pairs
of LHC inclusive cross-sections.}
\end{table}

\begin{table}[t]
\begin{centering}
\begin{tabular}{l|c|c|c}
\hline 
\multirow{2}{*}{PDF Set} & \multicolumn{3}{c}{Correlation coefficient}\tabularnewline
 & $Ht\bar{t},HW$ & $Ht\bar{t},HZ$ & $HW,HZ$\tabularnewline
\hline 
PDF4LHC15\_nlo\_prior & -0.18 & -0.43 & 0.88\tabularnewline
PDF4LHC15\_nlo\_mc & -0.15 & -0.41 & 0.87\tabularnewline
PDF4LHC15\_nlo\_100 & -0.18 & -0.42 & 0.89\tabularnewline
PDF4LHC15\_nlo\_30 & -0.19 & -0.46 & 0.88\tabularnewline
\hline 
\hline 
PDF4LHC15\_nnlo\_prior & -0.13 & -0.17 & 0.90\tabularnewline
PDF4LHC15\_nnlo\_mc & -0.17 & -0.21 & 0.90\tabularnewline
PDF4LHC15\_nnlo\_100 & -0.13 & -0.17 & 0.91\tabularnewline
PDF4LHC15\_nnlo\_30 & -0.17 & -0.25 & 0.91\tabularnewline
\hline 
\end{tabular}
\par\end{centering}
\caption{\small \label{tab:correlation4} Same as Table~\ref{tab:correlation1}
  for the correlation coefficient of additional pairs
of LHC inclusive cross-sections.}
\end{table}

%% file: sec-recommendations.tex
\clearpage

\begin{table}[t]
  \centering
  \small
\begin{tabular}{l|c|c|c|l}
  \hline
      {\tt LHAPDF6} grid  & Pert order  & {\tt ErrorType}  & $N_{\rm mem}$  & $\alpha_s(m_Z^2)$ \\
      \hline
          {\bf \tt PDF4LHC15\_nnlo\_mc}  & NNLO  &  {\tt replicas}  &   100  & 0.118 \\
          {\bf \tt PDF4LHC15\_nnlo\_100}  & NNLO  &  {\tt symmhessian}  &   100  & 0.118 \\
          {\bf \tt PDF4LHC15\_nnlo\_30}  & NNLO  &  {\tt symmhessian}  &   30  & 0.118 \\
          {\bf \tt PDF4LHC15\_nnlo\_mc\_pdfas}  & NNLO  &  {\tt replicas+as}  &   102  & mem 0:100 $\to$ 0.118  \\
          &  &    &    & mem 101 $\to$ 0.1165  \\
          &  &    &    & mem 102 $\to$ 0.1195  \\
           {\bf \tt PDF4LHC15\_nnlo\_100\_pdfas}  & NNLO  &  {\tt symmhessian+as}  &   102  & mem 0:100 $\to$ 0.118  \\
          &  &    &    & mem 101 $\to$ 0.1165  \\
              &  &    &    & mem 102 $\to$ 0.1195  \\
          {\bf \tt PDF4LHC15\_nnlo\_30\_pdfas}  & NNLO  &  {\tt symmhessian+as}  &   32  & mem 0:30 $\to$ 0.118  \\
          &  &    &    & mem 31 $\to$ 0.1165  \\
              &  &    &    & mem 32 $\to$ 0.1195  \\
          {\bf \tt PDF4LHC15\_nnlo\_asvar}  & NNLO &  - & 1 & mem 0 $\to$ 0.1165 \\
           &   &    &  & mem 1 $\to$ 0.1195 \\
          \hline
\end{tabular}
\caption{\small Summary of the combined NNLO PDF4LHC15 sets with
$n_f^{\rm max}=5$
that
  are available from {\tt LHAPDF6}.
  The corresponding NLO sets
  are also available.
 Members 0 and 1 of {\bf \tt PDF4LHC15\_nnlo\_asvar} coincide
 with members 101 and 102 (31 and 32) of
 {\bf \tt PDF4LHC15\_nnlo\_mc\_pdfas} and
 {\bf \tt PDF4LHC15\_nnlo\_100\_pdfas} ({\bf \tt PDF4LHC15\_nnlo\_30\_pdfas}).
  Recall that in {\tt LHAPDF6} there is always a zeroth member,
  so that the total number of PDF members in a given set
  is always $N_{\rm mem}+1$.
  See text for more details. \label{tab:pdf4lhc15}
}
  \end{table}

\section{The PDF4LHC 2015 recommendations}
\label{sec:recommendations}

The 2015 PDF4LHC prescription has been motivated, and
its general principles spelled out,
in Sect.~\ref{sec:prescription}.
As discussed there, while in some cases individual
PDF sets should always be used, for other LHC applications
a combination of PDF sets is required.
For these cases, in Sect.~\ref{sec:frameworks}
we have constructed the statistical combination  of the
CT14, MMHT14 and NNPDF3.0 PDF sets using the Monte Carlo
method.
This strategy is  different from the envelope method employed in the
2010 recommendations, and allows a robust statistical
interpretation of the ensuing PDF uncertainties

The PDF4LHC15 combined sets are then delivered using three
different options, corresponding in all cases to the
same underlying prior combination.
The choice of which delivery method should be used depends
on {\it purely practical considerations}, such as whether a Hessian or Monte
Carlo representation is preferred, or on the trade-off between
computational speed and accuracy.
Explicit recommendations for the usage of each of these three 
delivery options are provided below.

In this section we first
present the final PDF4LHC15 sets which will be made
available in {\tt LHAPDF6} and provide
general guidelines for their usage.
Then we
review the formulae for the calculation of PDF and PDF+$\alpha_s$
uncertainties in each case.
Finally, we present the PDF4LHC15 combined sets to be used
in calculations in the $n_f=4$ scheme,
and summarise
the citation policy that should be used whenever these
combined PDF sets are used.

We emphasize that the present document
contains {\it recommendations},
not a {\it series of unique instructions}.
We believe that with the delivery options provided here
any PDF user will find the flexibility
to choose the strategy
that is better suited for each particular analysis.

\subsection{Delivery and guidelines}
\label{sec:guidelines}

The PDF4LHC15 combined
PDFs are based on an underlying   Monte Carlo
combination of CT14, MMHT14 and NNPDF3.0, denoted by MC900,
which is made publicly available in three different reduced delivery forms:
\begin{itemize}
\item {\bf PDF4LHC15\_mc}: a {\it Monte Carlo PDF set} with $N_{\rm rep}=100$ replicas.
\item {\bf PDF4LHC15\_30}: a {\it symmetric
  Hessian PDF set} with $N_{\rm eig}=30$ eigenvectors.
  \item {\bf PDF4LHC15\_100}: a {\it symmetric
  Hessian PDF set} with $N_{\rm eig}=100$ eigenvectors.
\end{itemize}
In the three cases, combined sets are available at NLO and at NNLO,
for the central value of $\alpha_s(m_Z^2)=0.118$.
In addition, we provide additional sets which contain the central
values for $\alpha_s(m_Z^2)=0.1165$ and $\alpha_s(m_Z^2)=0.1195$,
and that can be used for the computation of the combined PDF+$\alpha_s$
uncertainties, as explained in Sect.~\ref{sec:formulae}.
Finally, for ease of usage, the combined sets for $\alpha_s(m_Z^2)=0.118$
are also presented
bundled with the $\alpha_s$-varying sets in dedicated
grid files.
The specifications of each of the combined NNLO PDF4LHC15 sets that
are available from {\tt LHAPDF6} are summarized
in Table~\ref{tab:pdf4lhc15}; note that
 the corresponding NLO sets are also available.

\vspace{0.2cm}
\noindent
{\bf Usage of the PDF4LHC15 sets}.
As illustrated in Sect.~\ref{sec:frameworks}, the three
delivery options
provide a  reasonably accurate representation of the
original prior combination.
However, each of these methods has its own advantages and
disadvantages, which make them more suited in different specific 
contexts.
We now attempt to provide some general guidance about which of
the three PDF4LHC15 combined
sets should be used in specific  phenomenological
applications.

\begin{enumerate}

\item {\bf Comparisons between data and theory for Standard Model measurements}

{\bf Recommendations:} Use
  {\it individual PDF sets}, and, in particular, as many of the
  modern
  PDF sets~\cite{Alekhin:2013nda,Dulat:2015mca,Owens:2012bv,Jimenez-Delgado:2014twa,Abramowicz:2015mha,Harland-Lang:2014zoa,Ball:2014uwa} as possible.

{\bf Rationale:} Measurements
  such as jet production,  vector-boson single and pair production, or
  top-quark pair production,
have the power to constrain
PDFs, and this is best utilized and illustrated by comparing with many individual sets.

As a rule of thumb, {\it any measurement that potentially can be included
in PDF fits} falls in this category.

The same recommendation applies to the
{\it extraction of precision SM parameters},
  such as the strong coupling $\alpha_s(m_Z^2)$~\cite{CMS:2014mna,Khachatryan:2014waa},
  the $W$ mass $M_W$~\cite{Bozzi:2011ww},
  and
  the top quark mass $m_t$~\cite{Chatrchyan:2013haa} which are
 directly correlated to  the PDFs   used in the extraction.

\item {\bf Searches for Beyond the Standard Model phenomena}

{\bf Recommendations:} Use the {\tt PDF4LHC15\_mc} sets.

{\bf Rationale}: BSM searches, in particular for {\it new massive
  particles in the TeV
  scale}, often require the knowledge of PDFs in regions
  where available experimental constraints are limited, notably close
  to the hadronic threshold where $x
  \rightarrow 1$~\cite{Beenakker:2015rna}.
  In these extreme kinematical regions the PDF uncertainties are
  large, the {\it Monte Carlo combination of PDF sets
  is likely to be non-Gaussian}. {\it c.f.}
  Figs.~\ref{fig:nongaussian1} and~\ref{fig:nongaussian2}.

  This case also
  applies to the calculations of PDF uncertainties in related 
  {\it theoretical analysis}, such as when determining  exclusion limits for
  specific BSM scenarios.

  If it is necessary to use a Hessian representation
  of the PDF uncertainty, for example in order to express PDF errors
  as Gaussian systematic uncertainties, one can cross-check the
  {\tt PDF4LHC15\_mc} results with the two Hessian sets,
  {\tt PDF4LHC15\_30} and   {\tt PDF4LHC15\_100}, depending
  on the required accuracy.

\item {\bf Calculation of PDF uncertainties in situations when
computational speed is
needed, or a more limited number of error PDFs may be desirable}

{\bf Recommendations:} Use the {\tt PDF4LHC15\_30} sets.

{\bf Rationale}:
  In many situations, PDF uncertainties may affect the extraction of
  physics parameters.
  From the point of view of the statistical analysis, it might
  be useful in some cases
  to {\it limit the number of error PDFs}
  that need to be included in such 
  analyses.
In these cases, use of the
  {\tt PDF4LHC15\_30} sets  may be most suitable.

  In addition, the calculation of {\it acceptances,
  efficiencies or extrapolation factors}
  are affected by the corresponding PDF uncertainty.
  These quantities are only a moderate correction to
  the measured cross-section, and thus a mild loss of accuracy
  in the determination of PDF uncertainties in these corrections
  is acceptable, while computational speed can be an issue.
  In these cases, use of the
  {\tt PDF4LHC15\_30} sets is most suitable.


\item {\bf Calculation of PDF uncertainties in  precision observables}

{\bf Recommendation}: Use the {\tt PDF4LHC15\_100} sets.

{\bf Rationale}: For several
LHC phenomenological applications, the highest accuracy is
  sought for, with, in some cases, the need to {\it control PDF uncertainties
  to the percent level}, as currently allowed by the development of
  high-order computational techniques in the QCD and electroweak
  sectors of the Standard Model.

{\it Whenever the highest accuracy is desired}, the
    {\tt PDF4LHC15\_100} set is most suitable.

However, {\it calculations that have little impact
on the PDF dependence of
the precision measurement}, such as certain acceptances,
may also be computed according to
(3), using the {\tt PDF4LHC15\_30} set.

\end{enumerate}

Concerning the specific applications of the four cases
listed above, there are some important caveats to take into account.

\begin{itemize}

\item For the same process, {\it more than one of the user cases}
above might be applicable.
For instance, consider the total top quark production
cross-section $\sigma(t\bar{t})$: one should use either
(3) or (4) to estimate the {\it total PDF uncertainty}
(for instance to determine the overall compatibility of the SM theory
and data for this measurement), and at the same time
use (1) to gauge the sensitivity of this observable
to {\it constrain PDFs}.

Therefore, cases (1)--(4) above are {\it not exclusive}: one or
the other should be more adequate depending on the theoretical
interpretation of a given experimental measurement.

\item Since the three delivery methods
are based on the {\it same underlying PDF combination}, it
is perfectly consistent to use {\it different delivery options
in different parts of the same calculation}.

For instance, for the computation of the production cross-sections
of high-mass BSM resonances, one could use the {\tt PDF4LHC15\_mc}
sets to estimate the PDF uncertainty in the expected signal yield,
and at the same time the {\tt PDF4LHC15\_30} sets to estimate
the PDF uncertainties that affect the
corresponding acceptance calculation.

\item In the case of a {\it really  significant discrepancy}
between the experimental data and theoretical calculations,
it may be crucial to fully exclude the possibility 
that this discrepancy arises due to the PDFs.

In this case we {\it recommend to compare data with
a  broader variety of PDFs}, including PDF sets
without LHC data, or PDF sets based only on DIS data.

\end{itemize}

\subsection{Formulae for the calculation of PDF and PDF+$\alpha_s$ uncertainties}
\label{sec:formulae}

For completeness, we also collect
in this report the explicit formulae for
the calculation of PDF and combined PDF+$\alpha_s$ uncertainties
in LHC cross-sections when using the PDF4LHC15 combined sets.
Let us assume that
we wish to estimate the PDF+$\alpha_s$ uncertainty of 
given cross-section $\sigma$, which could be
a total
inclusive cross-section or any bin of a differential distribution.

First of all, to compute the PDF uncertainty, one has to evaluate this cross-section $N_{\rm mem}+1$ times, where $N_{\rm mem}$ is the number of error sets
(either symmetric eigenvectors or MC replicas) of the specific combined
set,
\be
\sigma^{(k)} \, , \quad k=0,\ldots,N_{\rm mem} \, ,
\ee
so in particular $N_{\rm mem}=30$ in {\tt PDF4LHC15\_30} and
$N_{\rm mem}=100$ in {\tt PDF4LHC15\_100} and {\tt PDF4LHC15\_mc}.

\vspace{0.2cm}
\noindent
{\bf PDF uncertainties for Hessian sets.} 
In the case of the Hessian sets, {\tt PDF4LHC15\_30} and
{\tt PDF4LHC15\_100},
the  master formula to evaluate the PDF uncertainty is given by
\begin{equation}
\label{eq:hessianerr}
\delta^{\rm pdf} \sigma=\sqrt{\sum_{k=1}^{N_{\rm mem}}\left(\sigma^{(k)}-\sigma^{(0)}\right)^{2}},
\end{equation}
This uncertainty is to be understood as a 68\% confidence level.
From this expression it is also easy to determine
the contribution
of each eigenvector $k$ to the total Hessian PDF uncertainty.

\vspace{0.2cm}
\noindent
{\bf PDF uncertainties for MC sets.} 
For the case of the Monte Carlo sets, {\tt PDF4LHC15\_mc},
PDF uncertainties can be computed in two ways.
First of all, one can use 
 the standard
deviation of the distribution
\be
\label{eq:mcerr1}
\delta^{\rm pdf}\sigma=\sqrt{\frac{1}{N_{\rm mem}-1} \sum_{k=1}^{N_{\rm mem}}\lp
\sigma^{(k)} - \la \sigma \ra \rp^2 } \ ,
\ee
where the mean value of the cross-section, $\la \sigma \ra$,
is computed as usual
\be
\label{eq:mcmean}
\la \sigma \ra = \frac{1}{N_{\rm mem}} \sum_{k=1}^{N_{\rm mem}}\sigma^{(k)} \, .
\ee
For Monte Carlo sets one should use always
Eq.~(\ref{eq:mcmean}) for the mean value of the cross-section,
though in many cases it is true that
$\la \sigma \ra \simeq \sigma^{(0)}$.

Alternatively, PDF uncertainties in a Monte Carlo set can be computed
from the 68\% confidence level.
This is achieved by reordering the $N_{\rm mem}=100$ values for the
cross-sections
in ascending order, so that we have
\be
\sigma^{(1)} \le \sigma^{(2)} \le \dots \le  \sigma^{(N_{\rm mem}-1)} \le
\sigma^{(N_{\rm mem})} \ ,
\ee
and then the PDF uncertainty computed as the 68\% CL interval of
will be given by
\be
\label{eq:mcerr2}
\delta^{\rm pdf}\sigma=\frac{\sigma^{(84)}-\sigma^{(16)}}{2} \, .
\ee
This definition is suitable wherever departures from the Gaussian
regime are sizable, since it gives the correct
statistical weight to outliers.
In general, it can be useful to compare results with the
two expressions, Eq.~(\ref{eq:mcerr1}) and Eq.~(\ref{eq:mcerr2}),
and whenever differences are found, use Eq.~(\ref{eq:mcerr2}).
In addition, in the non-Gaussian case the mean of the distribution
Eq.~(\ref{eq:mcmean}) is not necessarily the best choice
for the central value of the cross-section, and we recommend
to use instead the midpoint of the 68\% CL interval,
that is
\be
\bar{\sigma}= \frac{\sigma^{(84)}+\sigma^{(16)}}{2} \, .
\ee

\vspace{0.2cm}
\noindent
{\bf Combined PDF+$\alpha_s$ uncertainties.} 
Let us now turn to discuss the computation of the combined
PDF+$\alpha_s$ uncertainties.
The PDF4LHC15 combined are based on the
following value of $\alpha_s(m_Z^2)$ and of its associated
uncertainty,
\be
\label{eq:alphaserrPDG}
\alpha_s(m_Z^2)=0.1180 \pm 0.0015 \, ,
\ee
at the 68\% confidence level, and
both at NLO and at NNLO.
This choice is consistent with the current PDG average~\cite{Agashe:2014kda},
and reflects recent developments towards the updated 2015 PDG
average.\footnote{Preliminary results on the 2015 PDG average for
$\alpha_s(m_Z^2)$ have been presented in \url{https://indico.cern.ch/event/392530/contribution/1/attachments/1168353/1686119/alphas-CERN2015.pdf}.}
It is then recommended that PDF+$\alpha_s$ uncertainties are
determined by first computing the PDF uncertainty for the central
$\alpha_s$, using Eqs.~(\ref{eq:hessianerr}), (\ref{eq:mcerr1})
or (\ref{eq:mcerr2}),
then computing predictions for the upper and lower values
of $\alpha_s$, consistently using the corresponding PDF sets, and
finally adding results in quadrature.

Specifically, for the same cross-section $\sigma$ as before, the
$\alpha_s$ uncertainty can be computed as:
\be
\label{eq:aserr}
\delta^{\alpha_s} \sigma = \frac{\sigma(\alpha_s=0.1195)-\sigma(\alpha_s=0.1165)}{2} \, ,
\ee
corresponding to an uncertainty $\delta\alpha_s=0.0015$ at the 68\% confidence
level.
Note that 
Eq.~(\ref{eq:aserr}) is to be computed with the central values
of the corresponding PDF4LHC15 sets only.
Needless to say, the same value of $\alpha_s(m_Z^2)$ should always
be used in the partonic cross-sections and in the PDFs.
The combined PDF+$\alpha_s$ uncertainty is then computed as follows
\be
\label{eq:pdfaserr}
\delta^{\rm PDF+\alpha_s}\sigma = \sqrt{
  \lp \delta^{\rm pdf}\sigma \rp^2
  + \lp \delta^{\alpha_s}\sigma \rp^2 
} \, .
\ee

The result for any other value of
$\delta\alpha_s$,
as compared to the baseline Eq.~(\ref{eq:alphaserrPDG}),
can be obtained
from a trivial rescaling of Eq.~(\ref{eq:aserr}) assuming linear
error propagation.
That is, if we assume a different value for the uncertainty in $\alpha_s$,
\be
\tilde{\delta}\alpha_s = r \cdot \tilde{\delta}\alpha_s\, , \qquad
\delta \alpha_s=0.0015 \, ,
\ee
then the combined PDF+$\alpha_s$ uncertainty Eq.~(\ref{eq:pdfaserr})
needs to be modified as follows
\be
\label{eq:pdfaserr2}
\delta^{\rm PDF+\alpha_s}\sigma = \sqrt{
  \lp \delta^{\rm pdf}\sigma \rp^2
  + \lp r\cdot \delta^{\alpha_s}\sigma \rp^2 
} \, ,
\ee
everything else unchanged.
It is thus clear from Eq.~(\ref{eq:pdfaserr2}) that despite
combined PDF sets are provided for a range of $\delta\alpha_s=\pm 0.0015$
around the central value, any other choice for $\delta\alpha_s$
can be trivially implemented using the existing PDF4LHC15 sets.

\vspace{0.2cm}
\noindent
{\bf Implementation in LHAPDF6.} Starting from {\tt LHAPDF} v6.1.6,
it is possible to automatically compute the combined PDF+$\alpha_s(m_Z^2)$
uncertainties for the various cases listed above
using the corresponding built-in routines.\footnote{We thank Graeme Watt for the implementation of the combined PDF+$\alpha_s$ uncertainties in {\tt LHAPDF6}}

\vspace{0.2cm}
\noindent
{\bf PDF reweighting.} Many  NLO and NNLO
matrix-element calculators and event generators allow the computation
of PDF uncertainties without any additional CPU-time cost by means
of PDF reweighting techniques.
This functionality is
for example available, among others, in
{\tt MadGraph5\_aMC@NLO}~\cite{Alwall:2014hca},
{\tt POWHEG}~\cite{Alioli:2010xd}, {\tt Sherpa}~\cite{Gleisberg:2008ta},
{\tt FEWZ}~\cite{Gavin:2012sy} and {\tt RESBOS}~\cite{Balazs:1997xd, Landry:2002ix}.
These {\it (N)NLO reweighting methods can be used in the same way for
the PDF4LHC15 combined sets} as for the individual sets.

In addition, an {\it approximate LO PDF reweighting} is often used,
where event weights are rescaled by ratios of PDFs,
see for example Sect.~7
of the {\tt LHAPDF6} manual~\cite{Buckley:2014ana}.
This LO PDF reweighting can be applied to the PDF4LHC15
combined sets, with the caveat that it is only approximately correct:
(N)NLO PDF reweighting requires modifications of the LO reweighting
formula.
In particular, LO reweighting  misses potentially large contributions
from partonic channels that arise first at NLO.

Therefore, {\it exact NLO and NNLO PDF reweighting should be
used whenever possible}, and the approximate LO PDF
reweighting should be only used when the former is not
available.
The exception is of course LO event generators,
where  LO PDF reweighting is exact (except
for the dependence of the PDF in the parton shower - but
this is also true for NLO PDF reweighting).

\subsection{PDF4LHC15 combined sets in the $n_f=4$ scheme}

In addition to the combined sets listed in Table~\ref{tab:pdf4lhc15},
which are suitable for calculations in the $n_f=5$ scheme,
 PDF4LHC15 combined sets have also been made available in
the $n_f=4$ scheme.
These are 
required for consistent calculations where the
 partonic cross-sections are computed in the $n_f=4$ scheme,
 accounting for bottom quark mass effects~\cite{Maltoni:2012pa}.

The inputs for the $n_f=4$ PDF4LHC15 combination
are the $n_f=4$ versions of the CT14,
MMHT14 and NNPDF3.0 global fits.
Each of these is constructed~\cite{Ball:2011mu,Harland-Lang:2015qea,
Dulat:2015mca}
from the corresponding $n_f=5$ global
fits, using these as a boundary condition for $m_c \le Q_0 \le m_b$,
from which PDFs and $\alpha_s$ are obtained for $Q > Q_0$
using evolution equations with 
 only four active quark flavours.

As it is well known, in the $n_f=4$ scheme, $\alpha_s(m_Z^2)$ is rather
smaller than the global average value of 0.118 for $n_f=5$.
For example, at NLO, using $n_f=4$ active flavours
for the running of the strong couplings up to the $Z$ mass,
one finds that $\alpha_s^{(n_f=5)}(m_Z^2)=0.118$
corresponds to 
 $\alpha_s^{(n_f=4)}(m_Z^2) \simeq 0.113$.
The exact value depends on the choice of $m_b$, which is
slightly different in each PDF group, though these
differences are subdominant as compared to the
PDF uncertainties in the combination.\footnote{It is
also possible to produce PDF sets in a hybrid scheme
where $\alpha_s(Q)$ runs with five flavours,
but where the DGLAP equations include only four active
flavours, above the bottom threshold~\cite{Harland-Lang:2015qea,Bertone:2015gba}.}
When the exact matching conditions are unknown,
$\alpha_s^{(n_f=4)}(\mu)$ at an arbitrary renormalization scale $\mu$ 
can also be determined from the corresponding $\alpha_s^{(n_f=5)}(\mu)$
using the scheme transformation relations of
Ref.~\cite{Chetyrkin:2005ia}, currently known up to
four loops in QCD.

\begin{table}[h]
  \centering
  \small
\begin{tabular}{l|c|c|c|l}
  \hline
      {\tt LHAPDF6} grid  & Pert order  & {\tt ErrorType}  & $N_{\rm mem}$  & $\alpha_s^{(n_f=5)}(m_Z^2)$ \\
      \hline
      {\bf \tt PDF4LHC15\_nlo\_nf4\_100}  & NLO  &  {\tt symmhessian}  &   100  & 0.118 \\
          {\bf \tt PDF4LHC15\_nlo\_nf4\_30}  & NLO  &  {\tt symmhessian}  &   30  & 0.118 \\
          {\bf \tt PDF4LHC15\_nlo\_nf4\_100\_pdfas}  & NLO  &  {\tt symmhessian+as}  &   102  & mem 0:100 $\to$ 0.118  \\
          &  &    &    & mem 101 $\to$ 0.1165  \\
              &  &    &    & mem 102 $\to$ 0.1195  \\
          {\bf \tt PDF4LHC15\_nlo\_nf4\_30\_pdfas}  & NLO  &  {\tt symmhessian+as}  &   32  & mem 0:30 $\to$ 0.118  \\
          &  &    &    & mem 31 $\to$ 0.1165  \\
              &  &    &    & mem 32 $\to$ 0.1195  \\
          {\bf \tt PDF4LHC15\_nlo\_nf4\_asvar}  & NLO &  - & 1 & mem 0 $\to$ 0.1165 \\
           &   &    &  & mem 1 $\to$ 0.1195 \\
          \hline
\end{tabular}
\caption{\small
Same as Table~\ref{tab:pdf4lhc15} for the combined
PDF4LHC15 sets in the $n_f=4$ scheme.
We indicate the value of $\alpha_s^{(n_f=5)}(m_Z^2)$ in the
$n_f=5$ scheme, the actual value in the $n_f=4$ scheme is substantially
smaller, see text.
\label{tab:pdf4lhc15nf4}
}
  \end{table}

The available PDF4LHC15 combined sets in the $n_f=4$ scheme
are summarised in Table~\ref{tab:pdf4lhc15nf4}.
Only the NLO combination is required, since no NNLO calculations
in the $n_f=4$ scheme are yet available.
Also, only Hessian sets are produced, since $n_f=4$ calculations
are always in a region where the underlying PDF
combination is essentially Gaussian.

\subsection{Citation policy for the PDF4LHC recommendation}

The techniques and methods presented in this report are
the result of an intense collaborative efforts within the
PDF4LHC community.
The three reduction methods presented in Sect.~\ref{sec:frameworks},
the CMC-PDFs, META-PDFs and MC-H PDFs, have been substantially
improved and refined (and in some cases, even developed from scratch)
as a result of the fruitful discussions within the
PDF4LHC working group.
It is thus important to properly acknowledge this effort by providing
an accurate citation policy for the usage
of the PDF4LHC15 recommendations, which we spell out in some detail here:

\begin{itemize}

\item Whenever the PDF4LHC15 recommendations are used,
  this report should be cited.

  \item In addition, the individual PDF sets that enter the combination
  should also be cited:
  \begin{enumerate}
  \item CT14~\cite{Dulat:2015mca}: S. Dulat, T. J. Hou, J. Gao, M. Guzzi, J. Huston, P. Nadolsky, J. Pumplin, C. Schmidt, D. Stump and C. P. Yuan, {\it `The CT14 Global Analysis of Quantum Chromodynamics'}, arXiv:1506.07443.
  \item MMHT14~\cite{Harland-Lang:2014zoa}: L. A. Harland-Lang, A. D. Martin, P. Motylinski and R.S. Thorne, {\it `Parton distributions in the LHC era: MMHT 2014 PDFs'}, Eur. Phys. J. C75 (2015) 5, 204, arXiv:1412.3989.
    \item NNPDF3.0~\cite{Ball:2014uwa}: R. D. Ball, V. Bertone, S. Carrazza, C. S. Deans, L. Del Debbio, S. Forte, A. Guffanti, N. P. Hartland, J. I. Latorre, J. Rojo and M. Ubiali, {\it `Parton Distributions for the LHC Run II'}, JHEP 1504 (2015) 040, arXiv:1410.8849.
  \end{enumerate}
  In particular, citation of this report only without reference to the
 individual PDF sets that enter the combination is strongly
 discouraged.

\item When any of the two Hessian sets are used,
  {\tt PDF4LHC15\_30} or {\tt PDF4LHC15\_100}, the
  original publications where
  the Hessian reduction methods where developed should be cited:

  \begin{enumerate}
    \item META-PDFs~\cite{Gao:2013bia}: J. Gao and P. Nadolsky, {\it `A meta-analysis of parton distribution functions'}, JHEP(1407)035, arXiv:1401.0013.
    \item MCH-PDFs~\cite{Carrazza:2015aoa}: S. Carrazza, S. Forte, Z. Kassabov, J. I. Latorre and J. Rojo, {\it `An unbiased hessian representation for Monte Carlo PDFs'}, Eur.\ Phys.\ J.\ C 75, no. 8, 369 (2015),
      arXiv:1505.06736.
  \end{enumerate}

\item When the Monte Carlo sets {\tt PDF4LHC15\_mc} are used,
the   original publication where
  the Monte Carlo compression method was developed should be cited:
  \begin{enumerate}
\item CMC-PDFs~\cite{Carrazza:2015hva}: S. Carrazza, J. I. Latorre, J. Rojo  and G. Watt, {\it `A compression algorithm for the combination of PDF  sets'},
Eur.\ Phys.\ J.\ C 75, no. 10, 474 (2015),
arXiv:1504.06469.
  \end{enumerate}

\item In addition, when either of the two reduction methods
  are employed, one should cite the original publication where
  the Monte Carlo representation of Hessian sets was
  presented~\cite{Watt:2012tq}:
  \begin{enumerate}
  \item G. Watt and R. S. Thorne, {\it `Study of Monte Carlo approach to experimental uncertainty propagation with MSTW 2008 PDFs'}, 
JHEP {\bf 1208}, 052 (2012), 
arXiv:1205.4024.
    \end{enumerate}
  which is at the basis of the Monte Carlo method for the combination of
  PDF sets.
  \end{itemize}

%% file: sec-future.tex
\section{Future directions}
\label{sec:future}

We have presented in Sect.~\ref{sec:prescription} the new PDF4LHC
recommendation for the
computation of PDF and combined PDF+$\alpha_s$ uncertainties for
LHC applications.
While the general guideline remains to combine the individual
PDFs when such combination is appropriate (cf. Sect.~\ref{sec:combination}), 
both the way the combined uncertainties are
determined, and their form of delivery, have evolved substantially
since the last recommendation.

The main rationale for these changes is the observation that
the PDFs of several groups
have undergone significant evolution due to both new available data and
methodological 
improvements, and that, as a consequence, there is now
much better agreement between the PDF determinations based on the widest
available dataset.
The fact that the improvements are driven by increase of experimental
information and theoretical understanding suggests that they are not accidental.

As a consequence, it now appears 
advisable to recommend a statistical combination of the PDFs from
the three global analysis groups, 
with $\alpha_s$ uncertainties combined with PDF
uncertainties in the standard way (albeit with a conservative estimate
of the $\alpha_s$ uncertainty). In order to optimize and streamline
usage of this statistical combination, we have produced a combined
PDF4LHC15 set, both at NLO and NNLO.
The set is delivered in three
versions: as 100 Monte Carlo replicas, or as 
either 30 or 100  Hessian eigenvector sets. 
Guidelines for usage of the PDF4LHC15 set have been presented in 
Sect.~\ref{sec:recommendations}, though it should be borne in mind
that all its versions correspond to the same underlying information, and
the choice of a specific version is motivated by practical
considerations, such as speed versus accuracy.

Future updates of our current recommendation 
will require the release of new combined
PDF sets, to replace the current PDF4LHC15 sets.
In these future releases it might be 
desirable for all PDFs to use common values for heavy-quark
masses, as it is currently done for the strong coupling $\alpha_s(m_Z^2)$.
Such future updates are likely to be motivated by two
sets of considerations.

First, we expect the PDF sets which enter the current combination to undergo
various updates in the coming years, as a consequence of the availability
of LHC Run-II data, and also,
the availability of  full NNLO corrections for a variety of processes that
potentially provide  
important constraints on PDFs and that can be accurately measured at the
LHC. 

Second, while only the sets included in the current combination satisfy
the  requirements for inclusion in the combination spelled out 
in Sect.~\ref{sec:combination}, this may change in the future.
In particular, it may be possible to devise techniques for including
non-global sets, using weighted statistical combination of PDF sets
with uncertainties of different size or another method.

In the longer term, there are other important directions in the 
global PDF analysis that should be explored. 
Perhaps the most important one is a consistent estimation of theoretical
uncertainties.
While taking into account parametric uncertainties such as the value of
$\alpha_s(m_Z^2)$ and the heavy quark masses is feasible with the
current combination methods, estimating missing higher-order uncertainties 
is a more challenging problem, requiring investments by the individual PDF
collaborations. 

Another possible future avenue is to extend the PDF4LHC combinations beyond fixed-order
QCD NLO and NNLO global fits, and to perform combinations of PDF sets with improved theory such
as sets with QED corrections~\cite{Martin:2004dh,Ball:2013hta,Schmidt:2015zda}
(needed for consistent calculations when electroweak effects
are included) or of sets with threshold resummation~\cite{Bonvini:2015ira}
(required when soft-gluon resummation is
included in the partonic cross-sections).
Eventually, one might even want to further improve the accuracy of
global sets by using approximate N3LO calculations 
when available (for example for deep-inelastic coefficient
functions~\cite{Vermaseren:2005qc,Moch:2004xu}).

To summarize, we have presented a general, robust, and statistically consistent
procedure for combination of PDFs. It represents 
a significant advancement beyond the 2010 PDF4LHC 
recommendation ~\cite{Botje:2011sn}, and aims to meet
diverse needs of the LHC Run-II programme.
The advancements described here bring the modern PDFs and their
combinations to a novel level of accuracy, adequate for (N)NNLO QCD
computations and analysis of vast experimental information anticipated
in the near future.

\subsection*{Acknowledgments}
We are grateful to  Sergey Alekhin, Johannes Bl\"umlein,
Claire Gwenlan, Max Klein, Katerina Lipka,
Kristin Lohwasser, Sven Moch, Klaus Rabbertz and Reisaburo Tanaka
for their feedback on this report.
We are also grateful to
Richard Ball,  Andr\'e David, Lucian Harland-Lang, Maxime Gouzevitch,
Jan Kretzschmar, Jos\'e Ignacio Latorre, Alan Martin,
Patrick Motylinski, Ringaile Placakyte, Jon Pumplin,
Alessandro Tricoli, Dan Stump,
Graeme Watt, C.~P. Yuan, as well as to 
many other colleagues from the PDF4LHC Working Group
community for illuminating discussions about
the topics presented in this report.

S.~C. and S.~F. are supported in part by an Italian
PRIN2010 grant and by a European Investment Bank EIBURS
grant. S.~F. and Z.~K. are supported by the Executive Research Agency
(REA) of the European Commission under the Grant Agreement
PITN-GA-2012-316704 (HiggsTools). S.F. thanks Matteo Cacciari for hospitatly
at LPTHE, Universit\'e Paris VI, where part of this work was done,
supported by a Lagrange award.
S.~C. is also supported by the HICCUP ERC Consolidator grant (614577).
The research of J.~G. in the High
Energy  Physics  Division  at  Argonne National Laboratory is  supported  by  the  U.  S.  Department  of  Energy,  High  Energy
Physics, Office of Science, under Contract No.  DE-AC02-06CH11357.  The work of P. N. is supported
by the U.S. Department of Energy under grant DE-SC0013681.
J.~R. is supported by an STFC Rutherford Fellowship
and Grant ST/K005227/1 and ST/M003787/1, and
by an European Research Council Starting Grant ``PDF4BSM".
The work of R.~S.~T. is supported partly by the London Centre for
Terauniverse Studies (LCTS),
using funding from the European Research Council via the Advanced
Investigator Grant 267352.
R.~S.~T. thanks the
Science and Technology Facilities Council (STFC) for support via grant
awards ST/J000515/1 and ST/L000377/1.

%% file: PDF4LHC_Run2_recommendations.bbl
\providecommand{\href}[2]{#2}\begingroup\raggedright\endgroup